\newcommand*{\ATLASLATEXPATH}{}
\pdfoutput=1 
\documentclass[cernpreprint,texlive=2011,txfonts,UKenglish]{\ATLASLATEXPATH atlasdoc}

\usepackage{\ATLASLATEXPATH atlaspackage}
\usepackage{\ATLASLATEXPATH atlasbiblatex}

\usepackage[authblk=false]{\ATLASLATEXPATH atlascontribute}

\usepackage{\ATLASLATEXPATH atlasphysics}

\usepackage{multirow}
\usepackage{bigstrut}
\usepackage{longtable}
\ifpdf\usepackage{grffile}\else\fi

\graphicspath{{logos/}{figures/}}

\usepackage{atlas-document-defs}

\hypersetup{pdftitle={ATLAS draft},pdfauthor={The ATLAS Collaboration}}

\AtlasTitle{Search for an additional, heavy Higgs boson in the \htozz\ decay channel at $\sqrt{s} = 8\;\mbox{TeV}$ in $pp$ collision data with the ATLAS detector}

\author{The ATLAS Collaboration}

\AtlasRefCode{HIGG-2013-20}

\PreprintIdNumber{CERN-PH-EP-2015-154}

\AtlasJournal{EPJC}

\AtlasAbstract{%
  A search is presented for a high-mass Higgs boson in the \htollll, \htollvv, \htollqq, and \htovvqq\  
  decay modes using the ATLAS detector at the CERN
  Large Hadron Collider. The search uses proton--proton collision data at a
  centre-of-mass energy of 8~TeV corresponding to an integrated luminosity of 20.3~fb$^{-1}$. The
  results of the search are interpreted in the scenario of a heavy Higgs boson
  with a width that is small compared with the experimental mass resolution.
  The Higgs boson mass range considered extends up to $1\tev$ for all four decay modes and down to as low as $140\gev$, depending on the decay mode.
No significant excess of events over the Standard Model prediction is found.
A simultaneous fit to the four decay modes yields upper limits on the production
cross-section of a heavy Higgs~boson times the branching ratio
to $Z$~boson pairs.  
  95\% confidence level upper limits
  range from 0.53~pb at $\mH=195$~\gev\ to 0.008~pb at $\mH=950$~\gev\ for the
  gluon-fusion production mode and
  from 0.31~pb at $\mH=195$~\gev\ to 0.009~pb at $\mH=950$~\gev\ 
  for the vector-boson-fusion production mode.
  The results are also interpreted in the context
  of Type-I and Type-II two-Higgs-doublet models.
}

\AtlasCoverSupportingNote{\htollqq}{https://cds.cern.ch/record/1693159}
\AtlasCoverSupportingNote{\htollvv}{https://cds.cern.ch/record/1693161}
\AtlasCoverSupportingNote{\htollll}{https://cds.cern.ch/record/1693487}
\AtlasCoverSupportingNote{\htovvqq}{https://cds.cern.ch/record/1692942}
\AtlasCoverSupportingNote{Combination}{https://cds.cern.ch/record/1995509}

\AtlasCoverCommentsDeadline{25 June 2015}

\AtlasCoverAnalysisTeam{
S.~H.~Abidi,
C.~Anastopoulos,
S.~Angelidakis,
L.~Aperio Bella,
G.~Artoni,
M.~Baak,
P.~Clark,
F.~Conventi,
R.~Di~Nardo,
T.~Donszelmann Cuhadar,
N.~Edwards,
D.~Fassouliotis,
G.~Garrillo Montoya,
F.~Garay Walls,
C.~Gwilliam,
M.~Hoffmann,
R.~Harrington,
X.~Ju,
L.S.~Kaplan,
D.~Kyriazopoulos,
A.~Mehta,
E.~Mountricha,
R.~Nicolaidou,
S.~Oda,
H.~Okawa,
K.~Parker,
A.~Sanchez Pineda,
J.~Price,
M.~Rescigno,
C.~Savarese,
F.~Lo Sterzo,
S.~Snyder,
J.~Voosebeld,
C.~Wang,
S.M.~Wang,
S.L.~Wu,
L.~Xu,
L.~Zhang,
Z.~Zhao,
B.~Zhou,
G.~Zurzolo
}

\AtlasCoverEdBoardMember{M.~Vos~(chair)}
\AtlasCoverEdBoardMember{V.~Cavaliere}
\AtlasCoverEdBoardMember{P.~Maettig}
\AtlasCoverEdBoardMember{D.~Whiteson}

\AtlasCoverEgroupEditors{atlas-HIGG-2013-20-editors@cern.ch}

\AtlasCoverEgroupEdBoard{atlas-HIGG-2013-20-editorial-board@cern.ch}

\nolinenumbers
\begin{document}

\maketitle

\tableofcontents

\section{Introduction}
\label{sec:intro}
In 2012, a Higgs~boson $h$ with a mass of $125\gev$ was discovered by the ATLAS and CMS
collaborations at the LHC~\cite{HIGG-2012-27,CMS-HIG-12-028}.  One of the most important remaining
questions is whether the newly discovered particle is part of an extended scalar sector as
postulated by various extensions to the Standard Model (SM) such as the two-Higgs-doublet model
(2HDM)~\cite{Branco:2011iw} and the electroweak-singlet (EWS) model~\cite{Hill:1987ea}.  These
predict additional Higgs~bosons, motivating searches at masses other than $125\GeV$.

This paper reports four separate searches with the ATLAS detector for a heavy neutral scalar $H$
decaying into two SM $Z$ bosons, encompassing the decay modes $\zztollll$, $\zztollvv$, $\zztollqq$,
and $\zztovvqq$, where $\ell$ stands for either an electron or a muon.  These modes are referred to,
respectively, as $\llll$, $\llvv$, $\llqq$, and $\vvqq$.

It is assumed that additional Higgs bosons would be produced predominantly via the gluon fusion
(ggF) and vector-boson fusion (VBF) processes but that the ratio of the two production mechanisms is
unknown in the absence of a specific model. For this reason, results are interpreted separately for
ggF and VBF production modes.  For Higgs~boson masses below $200\gev$, associated production (\VH,
where $V$ stands for either a $W$~or a $Z$~boson) is important as well. In this mass range, only the
$\llll$ decay mode is considered. Due to its excellent mass resolution and high signal-to-background
ratio, the $\llll$ decay mode is well-suited for a search for a narrow resonance in the range
$140<\mH<500\gev$; thus, this search covers the $\mH$ range down to $140\gev$.  The $\llll$ search
includes channels sensitive to \VH production as well as to the VBF and ggF production modes. The
$\llqq$ and $\llvv$ searches, covering $\mH$ ranges down to $200\gev$ and $240\gev$ respectively,
consider ggF and VBF channels only. The $\vvqq$ search covers the $\mH$ range down to $400\gev$ and does
not distinguish between ggF and VBF production.  Due to their higher branching ratios, the $\llqq$,
$\llvv$, and $\vvqq$ decay modes dominate at higher masses, and contribute to the overall
sensitivity of the combined result.  The $\mH$ range for all four searches extends up to $1000\gev$.

The ggF production mode for the $\llll$ search is further divided into four channels based on lepton
flavour, while the $\llvv$ search includes four channels, corresponding to two lepton flavours for
each of the ggF and VBF production modes.  For the $\llqq$ and $\vvqq$ searches, the ggF production
modes are divided into two subchannels each based on the number of $b$-tagged jets in the event. For
Higgs~boson masses above $700\gev$, jets from $Z$~boson decay are boosted and tend to be
reconstructed as a single jet; the ggF $\llqq$ search includes an additional channel sensitive to
such final states.

For each channel, a discriminating variable sensitive to $\mH$ is identified and used in a
likelihood fit.  The $\llll$ and $\llqq$ searches use the invariant mass of the four-fermion system
as the final discriminant, while the $\llvv$ and $\vvqq$ searches use a transverse mass
distribution.  Distributions of these discriminants for each channel are combined in a simultaneous
likelihood fit which estimates the rate of heavy Higgs boson production and simultaneously the
nuisance parameters corresponding to systematic uncertainties.  Additional distributions from
background-dominated control regions also enter the fit in order to constrain nuisance parameters.
Unless otherwise stated, all figures show shapes and normalizations determined from this fit.  All
results are interpreted in the scenario of a new Higgs boson with a narrow width, as well as in
Type-I and Type-II 2HDMs.

The ATLAS collaboration has published results of searches
for a Standard Model Higgs boson decaying in the
\llll, \llqq, and \llvv\ modes with 4.7--4.8$~\ifb$ of data collected at
$\sqrt{s}=7\tev$~\cite{HIGG-2012-01,HIGG-2012-15,HIGG-2012-14}.  A heavy Higgs~boson with the width
and branching fractions predicted by the SM was excluded at the 95\% confidence level in the ranges
$182<\mH<233\GeV$, $256<\mH<265\gev$, and $268<\mH<415\gev$ by the $\llll$ mode; in the ranges $300
< \mH < 322\GeV$ and $353 < \mH < 410\GeV$ by the $\llqq$ mode; and in the range $319 < \mH <
558\GeV$ by the $\llvv$ mode. The searches in this paper use a data set of $20.3~\ifb$ of
$pp$ collision data collected at a centre-of-mass energy of $\sqrt{s}=8\tev$. Besides using
a larger data set at a higher centre-of-mass energy, these searches improve on the earlier results
by adding selections sensitive to VBF production for the $\llll$, $\llqq$, and $\llvv$ decay modes
and by further optimizing the event selection and other aspects of the analysis. In addition, the 
$\vvqq$ decay mode has been added; finally, results of searches in all four decay
modes are used in a combined search. The CMS Collaboration
has also recently published a search for a heavy Higgs~boson with SM width in $\htozz$ decays~\cite{cms-hzz}.
Since the searches reported here use a narrow width for each Higgs boson mass hypothesis
instead of the larger width corresponding to a SM Higgs boson, a direct comparison against earlier ATLAS
results and the latest CMS results is not possible.

This paper is organized as follows. After a brief description of the ATLAS detector in
\secref{sec:det}, the simulation of the background and signal processes used in this analysis is
outlined in \secref{sec:samples}.  \Secref{sec:objects} summarizes the reconstruction of the
final-state objects used by these searches.  The event selection and background estimation for the
four searches are presented in \secsref{sec:4l} to~\ref{sec:vvqq}, and \secref{sec:systs} discusses
the systematic uncertainties common to all searches.
\Secref{sec:combination} details the statistical combination of all
the searches into a single limit, which is given in \secref{sec:results}.  Finally,
\secref{sec:summary} gives the conclusions.

\section{ATLAS detector}
\label{sec:det}
ATLAS is a multi-purpose detector~\cite{PERF-2007-01} which provides nearly full
solid-angle coverage around the interaction point.\footnote{ATLAS uses a right-handed coordinate system
with its origin at the nominal interaction point (IP) in the centre of the detector and the $z$-axis
coinciding with the axis of the beam pipe. The $x$-axis points from the IP towards the centre of the
LHC ring, and the $y$-axis points upward. Cylindrical coordinates ($r$,$\phi$) are used in the
transverse plane, $\phi$ being the azimuthal angle around the $z$-axis. The pseudorapidity is defined
in terms of the polar angle $\theta$ as $\eta = - \ln \tan(\theta/2)$. 
The distance in ($\eta$,$\phi$) coordinates, $\Delta R = \sqrt{(\Delta\phi)^2 + (\Delta\eta)^2}$, is also
used to define cone sizes. Transverse momentum and energy are defined as $\pt = p \sin\theta$ and
$\et = E \sin \theta$, respectively.  }  It consists of a tracking system (inner detector or ID)
surrounded by a thin superconducting solenoid providing a 2 T magnetic
field, electromagnetic and hadronic calorimeters, and a muon
spectrometer (MS). The ID consists of pixel and silicon microstrip
detectors covering the pseudorapidity region $|\eta| < 2.5$,
surrounded by a transition radiation tracker (TRT), which improves
electron identification in the region $|\eta| < 2.0$. The sampling
calorimeters cover the region $|\eta| < 4.9$. The forward region ($3.2 <
|\eta| < 4.9$) is instrumented with a liquid-argon (LAr)
calorimeter for electromagnetic and hadronic measurements. In the
central region, a high-granularity lead/LAr electromagnetic
calorimeter covers $|\eta| < 3.2$.  Hadron calorimetry
is based on either steel absorbers with scintillator tiles
($|\eta| < 1.7$) or copper absorbers in LAr ($1.5 < |\eta| < 3.2$).
The MS consists of three large superconducting toroids
arranged with an eight-fold azimuthal coil symmetry around the
calorimeters, and a system of three layers of precision gas chambers
providing tracking coverage in the range $|\eta| < 2.7$, while
dedicated chambers allow triggering on muons in the region $|\eta| <
2.4$. The ATLAS trigger system~\cite{PERF-2011-02} consists of three
levels; the first (L1) is a hardware-based system, while the
second and third levels are software-based systems.

\section{Data and Monte Carlo samples}
\label{sec:samples}

\subsection{Data sample}
\label{sec:datasample}

The data used in these searches were collected by ATLAS at a centre-of-mass
energy of 8~TeV during 2012 and correspond to an integrated luminosity
of $20.3~\ifb$.

Collision events are recorded only if they are selected by the online trigger
system.  For the $\vvqq$ search this selection requires that
the magnitude $\met$ of the missing transverse momentum vector
(see \secref{sec:objects})
is above $80\gev$.  Searches with leptonic final states use a
combination of single-lepton and dilepton triggers in order to maximize acceptance.  The main single-lepton
triggers have a minimum $\pt$ (muons) or $\et$ (electrons) threshold of $24\gev$ and require that the
leptons are isolated.  They are complemented with triggers with higher thresholds
($60\gev$ for electrons and $36\gev$ for muons) and no isolation requirement in order to increase acceptance at
high $\pt$ and $\et$.  The dilepton triggers require two same-flavour leptons with a threshold of $12\gev$ for electrons and $13\gev$ for muons.
The acceptance in the $\llll$ search is increased further with an additional asymmetric
dimuon trigger selecting one muon with $\pt>18\gev$ and another one with
$\pt>8\gev$ and an
electron--muon trigger with thresholds of $\et^{e}>12\gev$ and $\pt^{\mu}>8\gev$.

\subsection{Signal samples and modelling}
\label{sec:signals}

The acceptance and resolution for the signal of a narrow-width 
heavy Higgs~boson decaying to a $Z$~boson pair are modelled using 
Monte Carlo (MC) simulation.
Signal samples are generated using
\POWHEG r1508~\cite{powheg4,powheg5}, which calculates separately the gluon and vector-boson-fusion
Higgs boson production processes up to
next-to-leading order (NLO) in $\alphas$.  The generated
signal events are hadronized with \PYTHIA~8.165 using the AU2 set of
tunable parameters for the underlying event~\cite{ATL-PHYS-PUB-2011-009,ATL-PHYS-PUB-2011-008}; \PYTHIA\ also
decays the $Z$ bosons into all modes considered in this search.  The contribution from
$Z$ boson decay to $\tau$~leptons is also included.  The NLO CT10~\cite{Lai:2010vv} parton
distribution function (PDF) is used. The associated production of Higgs bosons
with a $W$~or $Z$~boson ($WH$ and $ZH$) is significant for $\mH<200\gev$. It is
therefore included as a signal process for the \llll\ search for $\mH<400\gev$ and simulated using
\PYTHIA~8 with the LO CTEQ6L1 PDF set~\cite{Pumplin:2002vw} and the AU2 parameter set. 
These samples are summarized in Table~\ref{tab:MC}.

Besides model-independent results, a search in the 
context of a CP-conserving 2HDM~\cite{Branco:2011iw} is also 
presented.  This model has five physical Higgs bosons
after electroweak symmetry breaking: two CP-even, $h$ and $H$; one
CP-odd, $A$; and two charged, $H^{\pm}$.  
The model considered here has seven free parameters: the
Higgs boson masses ($m_h$, $m_H$, $m_A$, $m_{H^{\pm}}$), the ratio of
the vacuum expectation values of the two doublets ($\tan\beta$), the
mixing angle between the CP-even Higgs bosons ($\alpha$), and the
potential parameter $m_{12}^2$ that mixes the two Higgs doublets.  
The two Higgs doublets $\Phi_1$ and $\Phi_2$ can couple to
leptons and up- and down-type quarks in several ways.  In the Type-I model,
$\Phi_2$ couples to all quarks and leptons, whereas for Type-II,
$\Phi_1$ couples to down-type quarks and leptons and $\Phi_2$
couples to up-type quarks. The `lepton-specific' model is similar to
Type-I except for the fact that the leptons couple to $\Phi_1$,
instead of $\Phi_2$; the `flipped' model is similar to Type-II except
that the leptons couple to $\Phi_2$, instead of $\Phi_1$. In all these
models, the coupling of the $H$~boson to vector bosons is proportional to
$\cos(\beta-\alpha)$. In the limit $\cos(\beta-\alpha) \rightarrow 0$ the
light CP-even Higgs boson, $h$, is indistinguishable from a SM
Higgs~boson with the same mass.  
In the context of $\htozz$ decays there is no direct coupling of the
Higgs~boson to leptons, and so only the Type-I and -II interpretations are
presented.

The production cross-sections for
both the ggF and VBF processes are calculated using
\progname{SusHi}~1.3.0~\cite{Harlander:2012pb,Harlander:2002wh,Harlander:2003ai,Aglietti:2004nj,Bonciani:2010ms,Harlander:2005rq},
while the branching ratios are calculated with
\progname{2HDMC}~1.6.4~\cite{Eriksson:2009ws}. For the branching
ratio calculations it is assumed that $m_A = m_H = m_{H^{\pm}}$, $m_h
= 125\gev$, and $m_{12}^2 = m_A^2 \tan\beta/(1+\tan\beta^2)$. In the 2HDM parameter
space considered in this analysis, the cross-section times branching ratio for
$\htozz$ with $\mH=200\gev$ varies from
2.4~fb to 10~pb for Type-I and from 0.5~fb to 9.4~pb for Type-II.

The width of the heavy Higgs boson varies over the parameter space of the 2HDM
model, and may be significant compared with the experimental resolution. Since
this analysis assumes a narrow-width signal, the 2HDM interpretation is limited
to regions of parameter space where the width is less than 0.5\% of~$\mH$
(significantly smaller than the detector resolution).
In addition, the
off-shell contribution from the light Higgs boson and its interference with the
non-resonant $ZZ$ background vary over the 2HDM parameter space as the light
Higgs~boson couplings are modified from their SM values. Therefore
the interpretation is further limited
to regions of the parameter space where the light Higgs~boson
couplings are enhanced by less than a factor of three from their SM values;
in these regions the variation is found to have a negligible effect.

\subsection{Background samples}
\label{sec:bkgs}

Monte Carlo simulations are also used to model the shapes of distributions
from many of the sources of SM background to these searches.
Table~\ref{tab:MC} summarizes the simulated event samples
along with the PDF sets and underlying-event tunes used.
Additional  samples are also used to compute
systematic uncertainties as detailed in \secref{sec:systs}.  

\SHERPA~1.4.1~\cite{sherpa} includes the effects of heavy-quark masses
in its modelling of the production of $W$~and $Z$~bosons along with
additional jets ($V+\jets$).
For this reason it is used to model these backgrounds 
in the hadronic
$\llqq$ and $\vvqq$ searches, which are subdivided based on whether the $Z$~boson
decays into $b$-quarks or light-flavour quarks.  The \ALPGEN 2.14 $\Wjets$ and $Z/\gamma^{*}+\jets$
samples are generated with up to five hard partons and with the partons
matched to final-state particle jets~\cite{Mangano:2002ea,Mangano:2006rw}.
They are
used to describe these backgrounds in the other decay modes and
also in the VBF channel of the $\llqq$
search\footnote{The VBF channel is inclusive in quark flavour and hence dominated by the $Z$ +
light-quark jet background.} since the additional partons in the matrix element give a better
description of the VBF topology.  The \SHERPA (\ALPGEN)
$Z/\gamma^{*}+\jets$ samples have a dilepton invariant mass requirement of $\mll > 40\gev$ ($60\gev$)
at the generator level.

The background from the associated production of the $125\gev$ $h$~boson
along with a $Z$~boson
is non-negligible in the \llqq\ and \vvqq\ searches and is taken into account.
Contributions to $Zh$ from both $\qqbar$ annihilation 
and gluon fusion are included. The $\qqbar \to Zh$ samples take into account NLO electroweak corrections, including
differential corrections as a function of $Z$ boson
$\pt$~\cite{Ciccolini:2003jy,Denner:2011rn}. The Higgs~boson branching ratio is calculated using
\progname{hdecay}~\cite{Djouadi:1997yw}.  Further details can be found in Ref.~\cite{HIGG-2013-23}.

Continuum $\ZZstar$ events form the dominant background
for the \llll\ and \llvv\ decay modes;  this is modelled
with a dedicated $\qqbar \to \ZZstar$ sample.  This sample is corrected
to match the calculation described in Ref.~\cite{Cascioli:2014yka}, which is
next-to-next-to-leading order (NNLO) in $\alphas$, with a $K$-factor
that is differential in $\mZZ$.  Higher-order electroweak
effects are included following the calculation reported in
Refs.~\cite{Bierweiler:2013dja,Baglio:2013toa} by applying a $K$-factor based on the
kinematics of the diboson system and the initial-state quarks, using a procedure
similar to that described in Ref.~\cite{Gieseke:2014gka}. The off-shell SM ggF Higgs boson process, 
the $gg \to ZZ$ continuum, and their interference are considered as backgrounds.
These samples are generated at leading order (LO) in $\alphas$
using \progname{MCFM}~6.1~\cite{Campbell:2011bn} ($\llll$)
or \progname{gg2vv} 3.1.3~\cite{Kauer:2012hd,Kauer:2013qba} ($\llvv$)
but corrected to NNLO as a function of $\mZZ$~\cite{Passarino:2013bha} using
the same procedure as described in Ref.~\cite{HIGG-2012-15}. For the \llqq\ and \vvqq\ searches, the 
continuum $\ZZstar$ background is smaller so the $\qqbar \to \ZZstar$ sample is used alone. It is 
scaled to include the contribution from $gg \to \ZZstar$ using the $gg \to \ZZstar$ cross-section
calculated by \progname{MCFM}~6.1~\cite{Campbell:2011bn}. 

For samples in which the hard process is generated with
\ALPGEN or \progname{MC@NLO}~4.03~\cite{Frixione:2003ei}, \HERWIG6.520~\cite{Corcella:2000bw} is used
to simulate parton showering and fragmentation, with
\JIMMY~\cite{Butterworth:1996zw} used for the underlying-event simulation.
\PYTHIA6.426~\cite{pythia} is used for samples generated with
\progname{MadGraph}~\cite{madgraph} and \ACERMC~\cite{Kersevan:2004yg}, while \PYTHIA8.165~\cite{Pythia8} is used for
the  \progname{gg2vv} 3.1.3~\cite{Kauer:2012hd,Kauer:2013qba},
\progname{MCFM} 6.1~\cite{Campbell:2013una}, and \POWHEG samples.
 \SHERPA implements its own parton showering
and fragmentation model.

\begin{table}
\begin{center}
\scriptsize
\begin{tabular}{l l l l l l }
\toprule
Physics process & $H \to ZZ$ search & Generator & Cross-section & PDF set & Tune\\
                & final state   & & normalization & & \\

\midrule
\multicolumn{6}{c}{$W/Z$ boson + jets}\\
\midrule
\multirow{ 2}{*}{$Z/\gamma^{*}\rightarrow \ell^+ \ell^-/\nu\bar\nu$}  & $\ell\ell\ell\ell/\ell\ell\nu\nu$ &\ALPGEN
2.14~\cite{Mangano:2002ea} &NNLO \cite{Melnikov:2006kv} & CTEQ6L1 \cite{Pumplin:2002vw} & AUET2  \cite{ATL-PHYS-PUB-2011-008,ATL-PHYS-PUB-2011-014}\\

& $\ell\ell qq^{\dagger}/\nu\nu qq$ &  \SHERPA 1.4.1 \cite{sherpa} &NNLO \cite{DYNNLO1,DYNNLO2} &NLO CT10 & \SHERPA default\\

\bigstrut[t]
\multirow{ 2}{*}{$W \rightarrow \ell\nu$} & $\ell\ell\nu\nu$ &\ALPGEN 2.14 &NNLO \cite{Melnikov:2006kv} & CTEQ6L1 & AUET2\\

 & $\nu\nu qq$ & \SHERPA 1.4.1 & NNLO \cite{DYNNLO1,DYNNLO2} &NLO CT10 & \SHERPA default \\

\midrule
\multicolumn{6}{c}{Top quark}\\
\midrule

\multirow{ 2}{*}{$t\bar{t}$} & $\ell\ell\ell\ell/\ell\ell qq/\nu\nu qq$ &  \POWHEGBOX
r2129~\cite{powheg1,powheg2,powheg3} & 
NNLO+NNLL
& \multirow{ 2}{*}{NLO CT10} &\sc{Perugia2011C}~\cite{pythiaperugia}\\
& $\ell\ell\nu\nu$ & \progname{MC@NLO} 4.03~\cite{Frixione:2003ei} & \quad\cite{ttbarxsec1,ttbarxsec2}  & & AUET2\\

\bigstrut[t]
\multirow{ 2}{*}{$s$-channel and $Wt$} & $\ell\ell\ell\ell/\ell\ell qq/\nu\nu qq$ & \POWHEGBOX r1556 & 
NNLO+NNLL  
 & \multirow{ 2}{*}{NLO CT10} & \sc{Perugia2011C}\\ 
& $\ell\ell\nu\nu$ & \progname{MC@NLO} 4.03 & \quad\cite{Kidonakis:2010a,Kidonakis:2010b} & & AUET2\\

\bigstrut[t]
$t$-channel & all & \ACERMC 3.8 \cite{Kersevan:2004yg} &
NNLO+NNLL 
 & CTEQ6L1 & AUET2\\
&&&\quad\cite{Kidonakis:2011}\\

\midrule
\multicolumn{6}{c}{Dibosons}\\
\midrule

\multirow{ 3}{*}{$q\bar{q} \to ZZ(*)$} & $\ell\ell qq/\nu\nu qq$ & \POWHEGBOX r1508 \cite{Nason:2013ydw} & NLO \cite{diboson1,Campbell:2011bn} &NLO CT10 & AUET2\\
                                    & $\ell\ell\ell\ell/\ell\ell\nu\nu$ & \POWHEGBOX r1508 \cite{Nason:2013ydw} & NNLO QCD \cite{Cascioli:2014yka}  &NLO CT10 & AUET2\\
                                      & &  & NLO EW \cite{Bierweiler:2013dja,Baglio:2013toa} &  \\
\bigstrut[t]
EW $q\bar{q}~(\to h) \to$
   & $\ell\ell\ell\ell$ & \progname{MadGraph} 5 1.3.28~\cite{madgraph} && CTEQ6L1 & AUET2 \\
\quad$ZZ(*) + 2j$\\

\bigstrut[t]
\multirow{ 2}{*}{$gg~(\to h^*) \to ZZ$} & $\ell\ell\ell\ell$ & \progname{MCFM} 6.1 \cite{Campbell:2013una} &
 NNLO~\cite{Passarino:2013bha}
& NLO CT10 & AU2 \\
& $\ell\ell\nu\nu$ & \progname{GG2VV} 3.1.3 \cite{Kauer:2012hd,Kauer:2013qba} &
\quad(for $h \to ZZ$)
& NLO CT10 & AU2 \\

\bigstrut[t]
\multirow{ 2}{*}{$q\bar{q} \to WZ$} & $\ell\ell\nu\nu/\ell\ell qq/\nu\nu qq$ & \POWHEGBOX r1508
& \multirow{ 2}{*}{NLO \cite{diboson1,Campbell:2011bn}} & \multirow{ 2}{*}{NLO CT10} & AUET2\\
& $\ell\ell\ell\ell$ & \SHERPA 1.4.1 & & & \SHERPA default\\

\bigstrut[t]
$q\bar{q} \to WW$ & all &\POWHEGBOX r1508 & NLO \cite{diboson1,Campbell:2011bn} &NLO CT10 & AUET2\\

\midrule
\multicolumn{6}{c}{$\bigstrut$ $\mh=125\gev$ SM~Higgs~boson (background)$^{\ddagger}$}\\
\midrule
$q\bar{q} \to Zh \to$
 & $\ell\ell qq/\nu\nu qq$ &\PYTHIA 8.165 & NNLO~\cite{Ohnemus:1992bd,Baer:1992vx,Brein:2003wg} & CTEQ6L & AU2 \\
\quad$ \ell^+\ell^- b\bar b / \nu\bar\nu b\bar b$\\
\bigstrut[t]
$gg \to Zh \to$
 & $\ell\ell qq/\nu\nu qq$ &\POWHEGBOX r1508 & NLO~\cite{Altenkamp:2012sx} & CT10 & AU2 \\
\quad$ \ell^+\ell^- b\bar b / \nu\bar\nu b\bar b$\\
\midrule
\multicolumn{6}{c}{$\bigstrut$  Signal}\\
\midrule
$gg\ra H \ra ZZ(*)$
 & all & \POWHEGBOX r1508 & --- & NLO CT10 & AU2 \\
$q\bar{q}\ra H+2j$;
 & all & \POWHEGBOX r1508 & --- & NLO CT10 & AU2 \\
 \quad $H \ra ZZ(*)$ & & & \\
$q\bar{q}\ra (W/Z)H$;
 & $\ell\ell\ell\ell$ & \PYTHIA 8.163 & --- &  CTEQ6L1 & AU2 \\
 \quad $H \ra ZZ(*)$ & & & \\
\bottomrule
\end{tabular}
\caption{Details of the generation of simulated signal and background event samples. For each physics process, the table gives
  the final states generated, the $H \to ZZ$
  final states(s) for which they are used, the generator, the PDF
  set, and the underlying-event tune. For the background samples, the order in
  $\alphas$ used to normalize the event yield is also given; for the signal,
  the normalization is the parameter of interest in the fit.
  More details can be found in the text.
  \\\hspace{\textwidth}
  $^{\dagger}$The $\htollqq$ VBF search uses \ALPGEN instead.
  \\\hspace{\textwidth}
  $^{\ddagger}$For the $\htollll$ and $\htollvv$ searches, the SM $h\to ZZ$ boson contribution, along with
  its interference with the continuum $ZZ$ background, is included in the diboson samples.}
\label{tab:MC}
\end{center}
\end{table}

In the $\llqq$ and $\vvqq$ searches, which have jets in the final state,
the principal background is $V+\jets$, where $V$ stands for either
a $W$~or a $Z$~boson.  In simulations
of these backgrounds, jets are labelled according to which generated hadrons
with $\pt>5\gev$ are found within a cone of size $\Delta R = 0.4$ around the reconstructed jet axis.
If a $b$-hadron is found, the jet is labelled as a $b$-jet; if not and a charmed hadron is found, the
jet is labelled as a $c$-jet; if neither is found, the jet is labelled as a
light (i.e., $u$-, $d$-, or $s$-quark, or gluon) jet, denoted by `$j$'.
For $V+\jets$ events that pass the selections for these searches,
two of the additional jets are reconstructed as the hadronically-decaying
$Z$~boson candidate.  Simulated $V+\jets$ events are then categorized
based on the labels of these jets.
If one jet is labelled as a $b$-jet, the event belongs to the $V+b$
category; if not, and one of the jets is labelled as a $c$-jet, the event belongs to the $V+c$
category; otherwise, the event belongs to the $V+j$ category.  Further subdivisions are defined
according to the flavour of the other jet from the pair, using the same precedence order: $V+bb$,
$V+bc$, $V+bj$, $V+cc$, $V+cj$, and $V+jj$; the combination of $V+bb$, $V+bc$,
and $V+cc$ is denoted by $V$+hf.%

\subsection{Detector simulation}
\label{sec:detsim}

The simulation of the detector is performed with either a full ATLAS detector simulation
\cite{SOFT-2010-01} based on \GEANT~4 9.6~\cite{Agostinelli:2002hh} or a fast simulation\footnote{
  The background samples that use the parameterized fast simulation are: \SHERPA $W/Z+\jets$
  production with $\pT^{W/Z}<280\gev$ (for higher $\pt^{W/Z}$ the full simulation is used since it improves
  the description of the jet mass in the merged $\llqq$ search described
  in~\secref{sec:sel_merged}); \POWHEGBOX $\ttbar$, single top, and diboson production; and SM
  \PYTHIA $q\bar{q} \to Zh$ and \POWHEGBOX $gg \to Zh$ production with $h \to bb$.  The
  remaining background samples and the signal samples, with the exception of 
  those used for the $\vvqq$ search,
  use the full \GEANT~4 simulation.} based on a parameterization of the
performance of the ATLAS electromagnetic and hadronic calorimeters~\cite{ATL-PHYS-PUB-2010-013} and
on \GEANT~4 elsewhere.  All simulated samples are generated with a variable number of minimum-bias
interactions (simulated using \PYTHIA~8 with the MSTW2008LO PDF~\cite{Sherstnev:2007nd} and the A2
tune \cite{ATL-PHYS-PUB-2011-014}), overlaid on the hard-scattering event to account for additional
$pp$ interactions in either the same or a neighbouring bunch crossing (pile-up).

Corrections are applied to the simulated samples to account for differences between data
and simulation for the lepton trigger and reconstruction efficiencies, and for the efficiency and
misidentification rate of the algorithm used to identify jets containing $b$-hadrons ($b$-tagging).

\section{Object reconstruction and common event selection}
\label{sec:objects}

The exact requirements used to identify physics objects vary between the
different searches.  This section outlines features that are common
to all of the searches; search-specific requirements are given in the sections
below.

Event vertices are formed from tracks with $\pt>400\MeV$.
Each event must have an identified primary vertex, which is chosen
from among the vertices with at least three tracks as the one with
the largest $\sum\pt^2$ of associated tracks.

Muon candidates ('muons')~\cite{PERF-2014-05} generally consist
of a track in the ID matched with one in the MS.
However, in the forward region ($2.5<|\eta|<2.7$),
MS tracks may be used with no matching ID tracks; further,
around $|\eta|=0$, where there is a gap in MS coverage,
ID tracks with no matching MS track may be used if they match an
energy deposit in the calorimeter consistent with a muon.
In addition to quality requirements, muon tracks are
required to pass close to the reconstructed primary event vertex. The
longitudinal impact parameter, $z_0$, is required to be less than $10\mm$,
while the transverse impact parameter, $d_0$, is required to be less than $1\mm$
to reject non-collision backgrounds. This requirement is not applied in in the
case of muons with no ID track.

Electron candidates (`electrons')~\cite{PERF-2013-03,PERF-2013-05,ATLAS-CONF-2014-032}
consist of an energy cluster in the EM calorimeter 
with $|\eta|<2.47$ matched
to a track reconstructed in the inner detector.
The energy of the electron is measured from the energy of the calorimeter cluster,
while the direction is taken from the matching track.
Electron candidates
are selected using variables sensitive to the shape of the EM cluster,
the quality of the track, and the goodness of the match between
the cluster and the track.  Depending on the search, either a selection
is made on each variable sequentially or all the variables are combined
into a likelihood discriminant.

Electron and muon energies are calibrated from measurements
of $Z\ra ee/\mu\mu$ decays~\cite{PERF-2013-05,PERF-2014-05}.
Electrons and muons must be isolated from other tracks, using
$\pt^{\ell,\mathrm{isol}} / \pt^{\ell}<0.1$, where $\pt^{\ell,\mathrm{isol}}$
is the scalar sum of the transverse
momenta of tracks within a 
$\Delta R = 0.2$ cone around the electron or muon
(excluding the electron or muon track itself), and $\pt^{\ell}$ is the transverse
momentum of the electron or muon candidate.
The isolation requirement is not applied in the case of muons
with no ID track.
For searches with electrons or muons in the final state, the reconstructed
lepton candidates must match the trigger lepton candidates that
resulted in the events being recorded by the online selection.

Jets are reconstructed~\cite{Cacciari:2011ma} using the anti-$k_t$ algorithm~\cite{AntiKt} with
a radius parameter $R=0.4$ operating on massless calorimeter energy clusters
constructed using a nearest-neighbour algorithm. 
Jet energies and directions are calibrated
using energy- and $\eta$-dependent correction factors derived using
MC simulations, with an additional calibration applied to data samples
derived from in situ measurements~\cite{PERF-2012-01}.
A correction is also made for
effects of energy from pile-up.
For jets with $\pt<50\gev$ within the acceptance of the ID ($|\eta|<2.4$),
the fraction of the summed scalar $\pt$ of the tracks associated
with the jet (within  a $\Delta R=0.4$ cone around the jet axis) contributed
by those tracks originating from the primary vertex must be at least 50\%.
This ratio is called the jet vertex fraction (JVF), and this 
requirement reduces the number of jet candidates originating
from pile-up vertices~\cite{ATLAS-CONF-2012-064,ATLAS-CONF-2013-083}.

In the $\llqq$ search at large Higgs~boson masses, the decay products of
the boosted $Z$~boson may be reconstructed as a single anti-$k_t$ jet
with a radius of $R=0.4$.  Such configurations are identified using
the jet invariant mass, obtained by summing the momenta of the jet
constituents.  After the energy calibration, the jet masses are calibrated,
based on Monte Carlo simulations, as a function of jet $\pt$, $\eta$, and mass.

The missing transverse momentum, with magnitude \met, is the
negative vectorial sum of the transverse momenta from calibrated 
objects, such as identified electrons, muons, photons, hadronic decays 
of tau leptons, and jets~\cite{PERF-2011-07}. Clusters of calorimeter cells 
not matched to any object are also included.

Jets containing $b$-hadrons ($b$-jets) can be discriminated
from other jets (`tagged') based on the relatively long lifetime 
of $b$-hadrons.
Several methods are used to tag jets originating from the fragmentation of a $b$-quark, including
looking for tracks with a large impact parameter with respect to the primary event vertex, looking
for a secondary decay vertex, and reconstructing a $b$-hadron $\ra$ $c$~hadron decay chain.  For the $\llqq$ and
$\vvqq$ searches, this information is combined into a single neural-network discriminant (`MV1c').
This is a continuous variable that is larger for jets that are more like $b$-jets.  A selection is
then applied that gives an efficiency of about 70\%, on average, for identifying true $b$-jets,
while the efficiencies for accepting $c$-jets or light-quark jets are 1/5 and 1/140
respectively~\cite{HIGG-2013-23,ATLAS-CONF-2011-102,ATLAS-CONF-2012-043,ATLAS-CONF-2014-004,ATLAS-CONF-2012-040}.
The $\llvv$ search uses an alternative version of this discriminant,
`MV1'~\cite{ATLAS-CONF-2011-102}, to reject background due to top-quark production; compared with
MV1c it has a smaller $c$-jet rejection.  Tag efficiencies and mistag rates are calibrated using
data.  For the purpose of forming the invariant mass of the $b$-jets, $m_{bb}$, the energies of
$b$-tagged jets are corrected to account for muons within the jets and an additional $\pT$-dependent
correction is applied to account for biases in the response due to resolution effects.

In channels which require two $b$-tagged jets in the final state, the efficiency
for simulated events of the dominant $Z+\jets$ background to pass the $b$-tagging
selection is low.  To effectively increase the sizes of simulated samples,
jets are `truth tagged': 
each event is weighted by the flavour-dependent probability of the jets to actually
pass the $b$-tagging selection.

\section{\htollll\ event selection and background estimation}
\label{sec:4l}
\subsection{Event selection}
\label{sec:sel4l}
The event selection and background estimation for the $\htollll$ ($\llll$) search
is very similar to the analysis described in Ref.~\cite{HIGG-2013-21}.  More details
may be found there; a summary is given here.

Higgs boson candidates in the $\llll$ search must have two same-flavour,
opposite-charge lepton pairs. %
Muons
must satisfy $\pt>6\gev$ and $|\eta|<2.7$, while electrons are identified
using the likelihood discriminant corresponding to the `loose LH'
selection from Ref.~\cite{ATLAS-CONF-2014-032} and must satisfy $\pt>7\gev$.
The impact parameter requirements that are made for muons are also applied
to electrons, and electrons (muons) must also satisfy a requirement
on the transverse impact parameter significance, $|d_0|/\sigma_{d_0}  < 6.5$ (3.5).  
For this search, the track-based isolation requirement is relaxed to
$\pt^{\ell,\mathrm{isol}}/\pt^{\ell} < 0.15$ for both the electrons and muons.  
In addition, lepton candidates
must also be isolated in $\et^{\ell,\mathrm{isol}}$, the sum of the
transverse energies in calorimeter cells within a $\Delta R = 0.2$ cone around
the candidate (excluding the deposit from the candidate itself).
The requirement is $\et^{\ell,\mathrm{isol}} / \pt^{\ell}<0.2$ for electrons,
$<0.3$ for muons with a matching ID track, and $<0.15$ for other muons.
The three highest-$\pt$ leptons in the event
must satisfy, in order, $\pt > 20$, $15$, and $10\gev$.  To ensure
well-measured leptons, and reduce backgrounds containing electrons from
bremsstrahlung, same-flavour leptons
must be separated from each other by $\Delta R> 0.1$, and different-flavour
leptons by $\Delta R>0.2$.
Jets that are $\Delta R<0.2$ from electrons are removed.
Final states in this search are classified depending on the
flavours of the leptons present: $4\mu$, $2e2\mu$, $2\mu2e$, and $4e$.
The selection of lepton pairs is made separately for each of these
flavour combinations; the pair
with invariant mass closest to the $Z$~boson mass is called the leading
pair and its invariant mass, $\mleading$, must be in the range $50$--$106\gev$.
For the $2e2\mu$ channel, the electrons form the leading pair, while for the 
$2\mu2e$ channel the muons are leading.
The second, subleading, pair of each combination is the pair from the remaining
leptons with invariant mass $\msubleading$ closest to that of the $Z$~boson
in the range $\mmin < \msubleading < 115\gev$. Here 
$\mmin$ is $12\gev$ for $\mllll< 140\gev$, rises linearly to
$50\gev$ at $\mllll=190\gev$, and remains at $50\gev$ for 
$\mllll > 190\gev$. Finally, if more than one flavour combination passes the
selection, which could happen for events with more than four leptons,
the flavour combination with the highest expected signal acceptance is kept; i.e.,
in the order: $4\mu$, $2e2\mu$, $2\mu2e$, and $4e$. 
For $4\mu$ and $4e$ events, if an opposite-charge same-flavour dilepton pair
is found with $\mll$ below $5\gev$, the event is vetoed in order to reject
backgrounds from \jpsi\ decays.

To improve the mass resolution, the four-momentum of any reconstructed photon
consistent with having been radiated from one of the leptons in the leading pair
is added to the final state. Also, the four-momenta of the leptons in the leading
pair are adjusted by means of a kinematic fit assuming a $Z\ra\ell\ell$
decay; this improves the $\mllll$ resolution by up to 15\%, depending on \mH.
This is not applied to the subleading pair in order to retain sensitivity
at lower $\mH$ where one of the $Z$~boson decays may be off-shell.
For $4\mu$ events, the resulting mass resolution varies from 1.5\% at
$\mH=200\gev$ to 3.5\% at $\mH=1\tev$, while for $4e$ events
it ranges from 2\% at $\mH=200\gev$ to below 1\% at $1\tev$.

Signal events can be produced via ggF or 
VBF, or associated production (\VH, where $V$ stands
for either a $W$~or a $Z$~boson). In order  
to measure the rates for these processes separately, events passing the event
selection described above are classified into channels, either ggF, VBF, or \VH.
Events 
containing at least two jets with $\pt > 25\gev$ and $|\eta|<2.5$ or $\pt >
30\gev$ and $2.5<|\eta|<4.5$ and with the leading two such jets having
$\mjj>130\gev$ are classified as VBF events. 
Otherwise, if a jet pair
satisfying the same $\pt$ and $\eta$ requirements is present but with $40<\mjj<130\gev$, the
event is classified as \VH, providing it also passes a selection on a
multivariate discriminant used to separate the \VH\ and ggF signal.  The
multivariate discriminant makes use of $\mjj$, $\detajj$, the $\pt$ of the
two jets, and the $\eta$ of the leading jet.
In order to account for leptonic decays of the $V$ ($W$~or $Z$)~boson, events failing this selection
may still be classified as \VH\ if an additional lepton with $\pt>8\gev$
is present.  All remaining events are classified as {ggF}. Due to the
differing background compositions and signal resolutions, events in the
ggF channel are further classified into subchannels according to their final state:
\fe, \tetmu, \tmute, or \fmu.
The selection for VBF is looser than that used in the other searches;
however, the effect on the final results is small.
The $\mllll$ distributions for the three channels 
are shown in \figref{fig:llll_fit_m4l}. 

\begin{figure}[htp]
  \centering
\subfloat[ggF]{\includegraphics[width=0.48\textwidth]{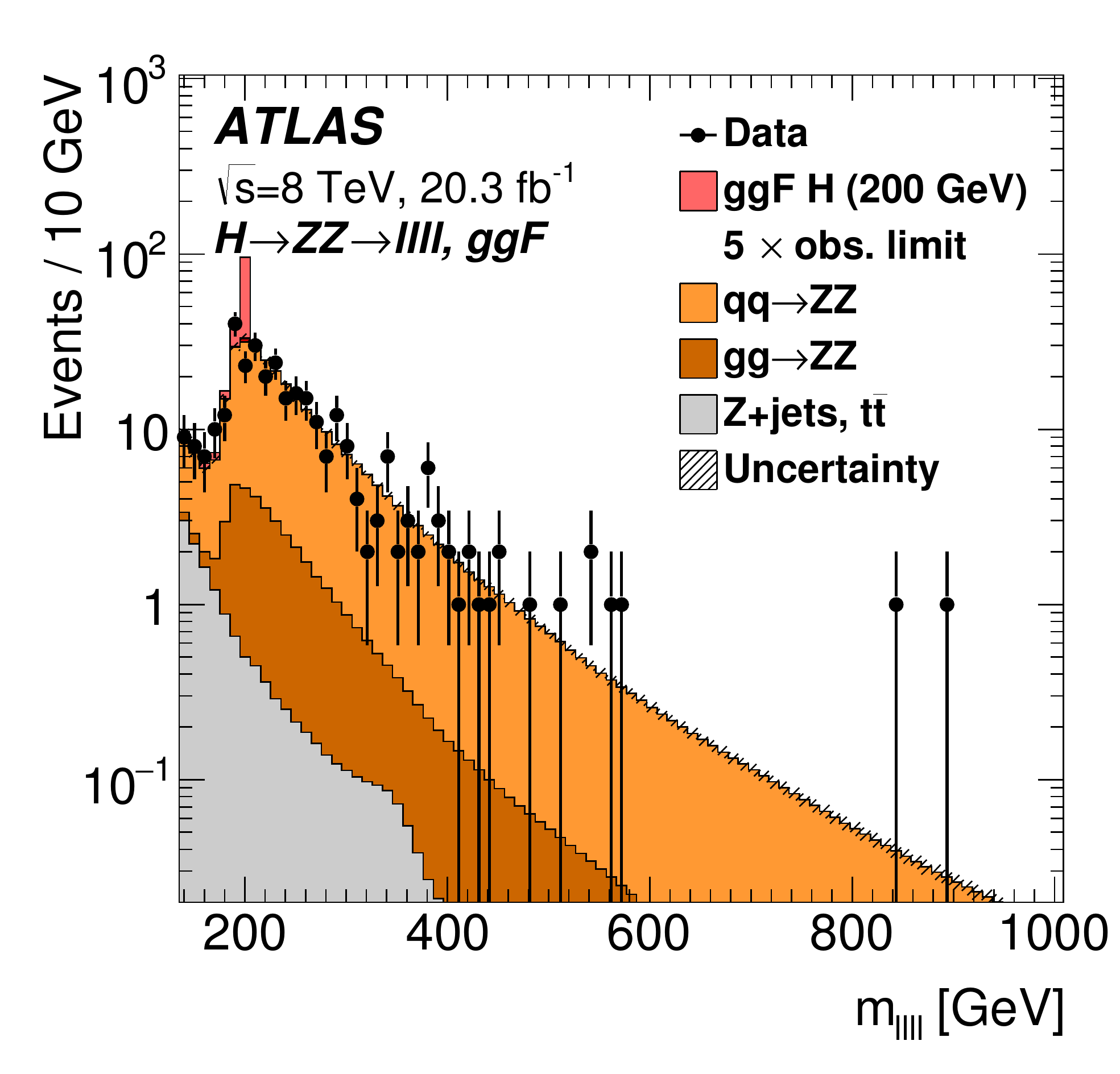}}
\subfloat[VBF]{\includegraphics[width=0.48\textwidth]{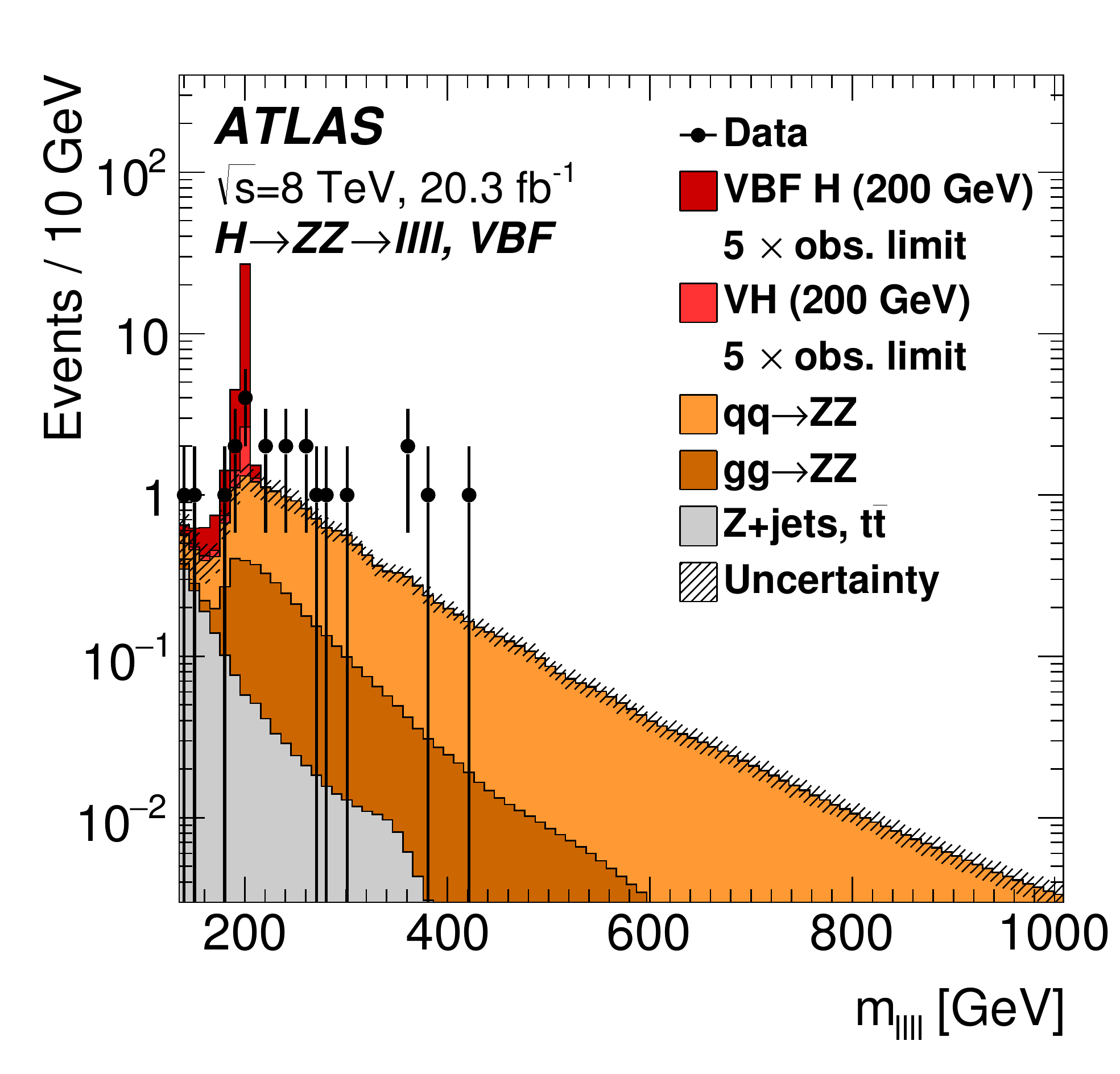}} \\
\subfloat[$VH$]{\includegraphics[width=0.48\textwidth]{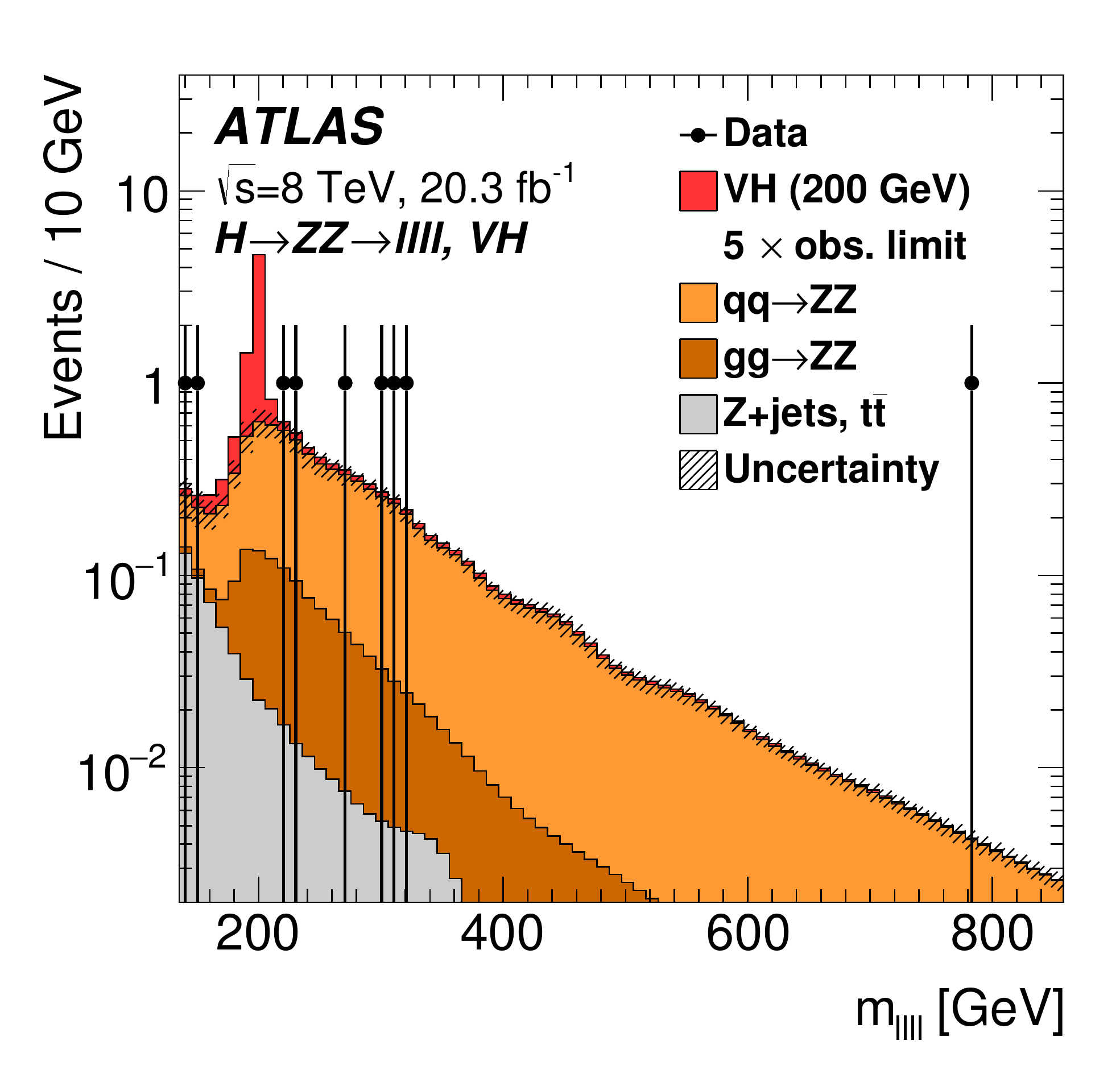}}
\caption{\label{fig:llll_fit_m4l} The distributions used in the likelihood
  fit of the four-lepton invariant mass $\mllll$ for the
  \htollll\ search in the (a)~ggF,  (b)~VBF, and (c)~$VH$ channels.
  The `$\Zjets$, $\ttbar$' entry includes all backgrounds other than $ZZ$,
  as measured from data.
  No events are observed beyond the upper limit of the plots.
  The simulated $\mH=200\GeV$ signal is normalized to a cross-section
  corresponding to five times the observed limit given in~\secref{sec:results}.
  Both the VBF and $VH$ signal modes are shown in~(b) as there is
  significant contamination of $VH$ events in the VBF category.
}
\end{figure}

\subsection{Background estimation}
\label{sec:sigbkg4l}
The dominant background in this channel is continuum \ZZ\
production. Its contribution to the yield is
determined from simulation using the samples described in \secref{sec:bkgs}.
Other background components are small and consist mainly of \ttbar\ and \Zjets\
events. These are difficult to estimate from MC simulations
due to the small rate at which
such events pass the event selection, and also because they
depend on details of jet fragmentation, 
which are difficult to model reliably in simulations.
Therefore, both the rate and composition of these backgrounds
are estimated from data. Since the composition of these
backgrounds depends on the flavour of the subleading dilepton pair, different
approaches are taken for the \llmumu\ and the \llee\ final states. 

The \llmumu\ non-$ZZ$ background comprises mostly \ttbar\ 
and $Z+b\bar{b}$ events, where in the latter the muons arise mostly from heavy-flavour semileptonic decays, and to a lesser
extent from $\pi$/\kaon~in-flight decays. The contribution from single-top
production is negligible.  The normalization of each component is
estimated by a simultaneous fit to the $m_{12}$ distribution in four control
regions, defined by inverting the impact parameter significance or isolation
requirements on the subleading muon, or by selecting a subleading $e\mu$ or
same-charge pair.  A small contribution
from $WZ$ decays is estimated using simulation.
The electron background contributing to the \llee\ final states comes mainly from jets
misidentified as electrons, arising in three ways: light-flavour hadrons misidentified as electrons, photon
conversions reconstructed as electrons, and non-isolated electrons from
heavy-flavour hadronic
decays. This background is estimated in a control region in which the three highest-\pt\
leptons must satisfy the full selection, with the third lepton 
being an electron.  For the lowest-$\pt$ lepton, which must also be an electron,
the impact parameter and isolation requirements are removed and the
likelihood requirement is relaxed.  In addition, it must
have the same charge as the other subleading electron in order to minimize 
the contribution from the \ZZ\ background. The yields of the
background components of the lowest-$\pt$ lepton are extracted with a fit to the 
number of hits in the innermost pixel layer
and the ratio of the number of high-threshold to low-threshold TRT hits
(which provides discrimination between electrons and pions).
For both backgrounds, the fitted yields in the control regions are extrapolated to the
signal region using efficiencies obtained from simulation. 

For the non-$ZZ$ components of the background, the
$m_{\llll}$ shape is evaluated for the 
\llmumu\ final states using simulated events, and from data for the \llee\ final states 
by extrapolating the shape from the \llee\ control region described 
above. The fraction of this background in each
channel (ggF, VBF, \VH) is evaluated using simulation. The non-$ZZ$ background contribution for
$\mllll>140\gev$ is found to be approximately 4\% of the total background. 

Major sources of uncertainty in the estimate of the non-$ZZ$
backgrounds include differences in the results when alternative
methods are used to estimate the background~\cite{HIGG-2013-21}, 
uncertainties in the transfer factors used to extrapolate from the
control region to the signal region, and the limited statistical 
precision in the control
regions.  For the \llmumu\ (\llee) background, the uncertainty is 21\% (27\%) in
the ggF channel, 100\% (117\%) in the VBF channel, and
62\% (79\%) in the \VH channel.  The larger uncertainty in the VBF channel
arises due to large statistical uncertainties on the fraction of \Zjets\ events
falling in this channel. Uncertainties in the expected $\mllll$ 
shape are estimated from differences in the shapes obtained using different methods for 
estimating the background.

\section{\htollvv\ event selection and background estimation}
\label{sec:llvv}
\subsection{Event selection}
\label{sec:selllvv}
The event selection for the \htollvv\ ($\llvv$) search starts with the reconstruction
of either a $Z\ra e^+e^-$ or $Z\ra\mu^+\mu^-$ lepton pair; the leptons must be
of opposite charge and must have invariant mass $76<\mll<106\gev$.
The charged lepton selection is tighter than that described in \secref{sec:objects}.
Muons must have matching tracks in the ID and MS and lie in the region $|\eta|<2.5$.
Electrons are identified using a series of sequential requirements
on the discriminating variables, corresponding
to the `medium' selection from Ref.~\cite{ATLAS-CONF-2014-032}.
Candidate leptons for the $Z\ra\ell^+\ell^-$~decay must have $\pt>20\gev$, and 
leptons within a cone of $\Delta R=0.4$ around jets are removed. Jets that lie
$\Delta R<0.2$ of electrons are also removed. 
Events containing a third lepton or muon with
$\pt>7\gev$ are rejected; for the purpose of this requirement, the `loose'
electron selection from Ref.~\cite{ATLAS-CONF-2014-032} is used.
To select events with neutrinos in the final state,
the magnitude of the missing transverse momentum must satisfy $\met > 70\gev$.

As in the \llll\ search, samples enriched 
in either ggF or VBF production are selected.
An event is classified as VBF
if it has at least two jets with $\pt>30\gev$ and $|\eta|<4.5$
with $\mjj>550\gev$ and $\detajj>4.4$.  Events failing to satisfy the VBF criteria and 
having no more than one jet with $\pt>30\gev$ and $|\eta|<2.5$ are classified as  
ggF\null. Events not satisfying either set of criteria are rejected. 

To suppress the Drell--Yan background, 
the azimuthal angle between the combined dilepton system and the missing transverse 
momentum vector $\Delta\phi(\ptll,\met)$
must be greater than 2.8 (2.7) for the ggF (VBF) channel 
(optimized for signal significance in each channel), 
 and the fractional $\pt$ difference, defined as 
$|\ptmetjet - \ptll|/\ptll$,
must be less than 20\%, where $\ptmetjet=\bigl|\metvec + \sum_{\mathrm{jet}}\ptvec^{\mathrm{jet}}\bigr|$. 
$Z$~bosons originating from the decay of a high-mass state are boosted;
thus, the 
azimuthal angle between the two leptons $\dphill$ must be less than 1.4.
Events containing a $b$-tagged jet with $\pt>20\gev$ and $|\eta|<2.5$ are 
rejected in order to reduce the background from top-quark production.
All jets in the event must have an azimuthal angle greater than 0.3
relative to the missing transverse momentum.

The discriminating variable used is the transverse mass $\mtzz$
reconstructed from the momentum of the 
dilepton system and the missing transverse momentum, defined by:
\begin{linenomath}
\begin{equation}\label{eq:mt}
(\mtzz)^2 \equiv 
        \left( \sqrt{\mZ^2+\left|\ptll \right|^2} +
               \sqrt{\mZ^2+\left|\met\right|^2} \right)^2 
- 
\left|\ptllvec + \metvec\right|^2.
\end{equation}
\end{linenomath}
The resulting resolution in $\mtzz$ ranges from 7\% at $\mH=240\gev$ to 15\% 
at $\mH=1\tev$.

\Figref{fig:llvv_fit_mT} shows the $\mtzz$ distribution in the 
ggF channel. The event yields in the VBF channel are very small
(see Table~\ref{tab:ZZllnn_yields_postfit}).

\begin{figure}[tbh]
  \centering
\includegraphics[width=0.47\textwidth]{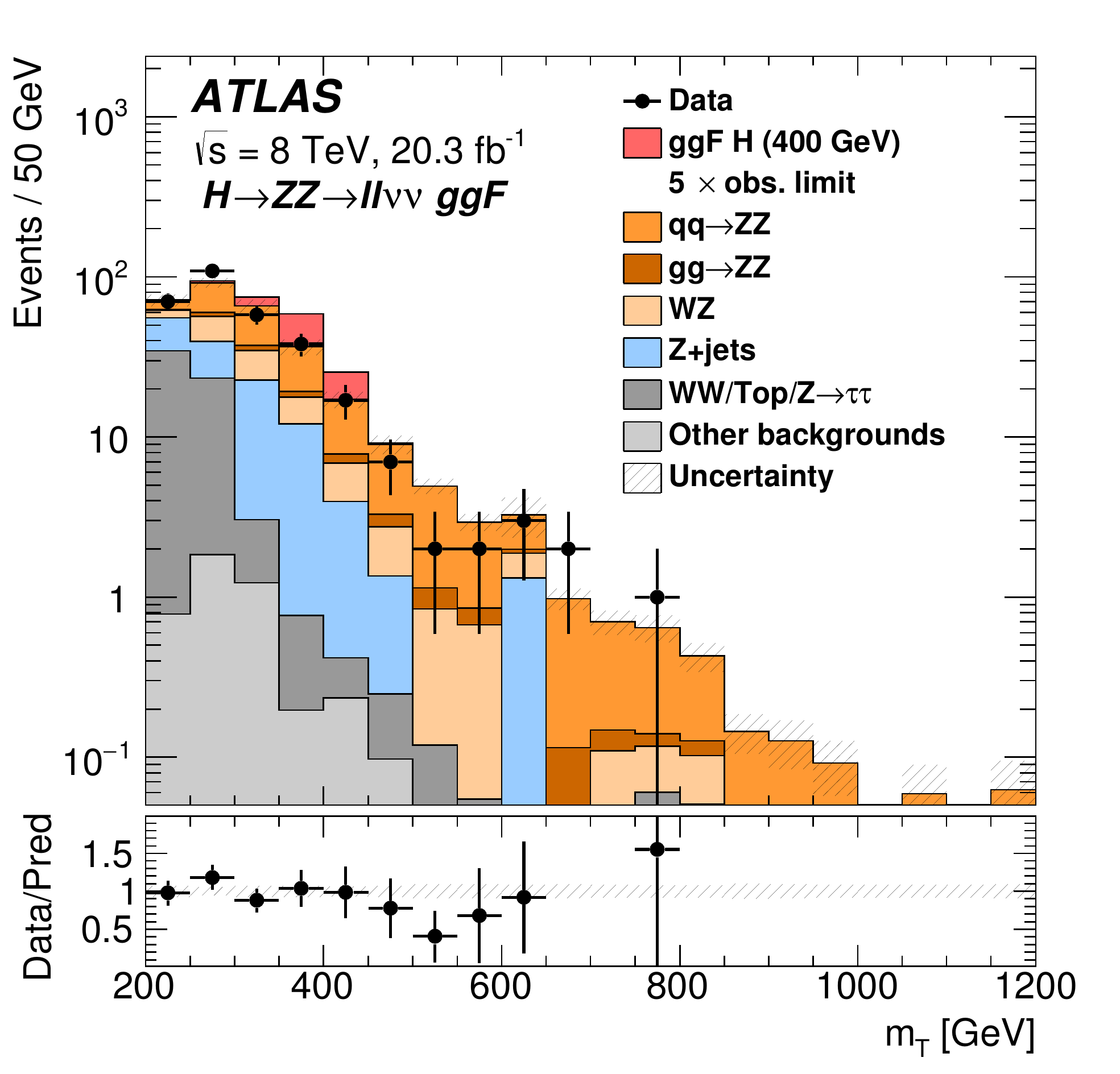}
\caption{\label{fig:llvv_fit_mT} The distribution used in the likelihood fit
  of the transverse mass
  $\mtzz$ reconstructed from the momentum of the dilepton system
  and the missing transverse momentum
  for the \htollvv\ search in the ggF channel. 
  The simulated signal is normalized to a cross-section
  corresponding to five times the observed limit given in~\secref{sec:results}.
  The contribution labelled as `Top' includes both the $\ttbar$ and single-top processes. 
  The bottom pane shows the ratio of the observed data
  to the predicted background.
}
\end{figure}

\subsection{Background estimation}
\label{sec:sigbkgllvv}
The dominant background is $ZZ$ 
production, followed by $WZ$~production. 
Other important backgrounds to this search include 
the $WW$, $\ttbar$, $Wt$, and $Z\to\tau^+\tau^-$~processes, and 
also the $\Zjets$ process with poorly reconstructed $\met$,
but these processes tend to yield final states with low $\mt$. 
Backgrounds from $\Wjets$, $\ttbar$, single top~quark ($s$- and $t$-channel),
and multijet processes with at least one jet misidentified as an electron or muon are 
very small.

The \POWHEG simulation is used to estimate the $ZZ$ background
in the same way as for the $\llll$ search.
The $WZ$ background is also estimated with \POWHEG\ and 
validated with data using a sample of events that pass the signal selection
and that contain an extra electron or muon in addition to the 
$Z\ra\ell^+\ell^-$~candidate.  

The $WW$, $\ttbar$, $Wt$, 
and $Z\to\tau^+\tau^-$~processes give rise to both
same-flavour as well as different-flavour lepton final states.  The total background from these
processes in the same-flavour final state can be estimated from control
samples that contain an electron--muon pair rather than a same-flavour
lepton pair by
\begin{eqnarray} 
\label{eq:ee_emubkg}
\begin{split}
N^{\rm bkg}_{ee} &=& \frac{1}{2} \times N^{\rm data,sub}_{e\mu}\times{f}, \\
N^{\rm bkg}_{\mu\mu} &=& \frac{1}{2} \times N^{\rm data,sub}_{e\mu}\times\frac{1}{f},
\end{split}
\end{eqnarray}
where $N^{\rm bkg}_{ee}$ and $N^{\rm bkg}_{\mu\mu}$ are the number of electron and
muon pair events in the signal region and 
$N^{\rm data,sub}_{e\mu}$ is the number of events in the $e\mu$ 
control sample with 
$WZ$, $ZZ$, and other small backgrounds ($\Wjets$, $\ttbar W/Z$, and triboson)
subtracted using simulation.   The factor of two arises because the
branching ratio to final states containing electrons and muons
is twice that of either $ee$ or $\mu\mu$.
The factor $f$ takes into account
the different efficiencies for electrons and muons and is measured 
from data as $f^2 = N_{ee}^{\mathrm{data}} / N_{\mu\mu}^{\mathrm{data}}$,
the ratio of the number of electron pair to muon pair events in the data
after the $Z$~boson mass requirement ($76 < \mll < 106\gev$). 
The measured value of $f$ is 0.94 with a systematic uncertainty 
of 0.04 and a negligible statistical uncertainty.   There is also
a systematic uncertainty from the background subtraction in the
control sample; this is less than 1\%.
For the VBF channel, no events remain in the $e\mu$ control sample
after applying the full selection.  In this case, the background estimate
is calculated after only the requirements on $\met$ and the number of jets;
the efficiencies of the remaining selections for this background are
estimated using simulation.

The $\Zjets$ background is estimated from data by comparing the signal region
(A) with regions in which one (B, C) or both (D) of the
$\dphill$ and $\Delta\phi(\ptll,\met)$ requirements are reversed.
An estimate of the number of background events in the signal region is then
$N_{\mathrm{A}}^{\mathrm{est}} = N_{\mathrm{C}}^{\mathrm{obs}}\times (N_{\mathrm{B}}^{\mathrm{obs}} / N_{\mathrm{D}}^{\mathrm{obs}})$,
where $N_X^{\mathrm{obs}}$ is the number of events observed
in region $X$ after subtracting non-$Z$~boson backgrounds.
The shape is estimated by taking $N_C^{\mathrm{obs}}$ (the region with
the $\dphill$ requirement reversed) bin-by-bin and applying a correction derived
from MC simulations to account for shape differences between regions~A and {C}.
Systematic uncertainties arise from differences in the shape
of the $\met$ and $\mtzz$ distributions among the four regions,
the small correlation between the two variables, and the subtraction
of non-$Z$~boson backgrounds.

The $\Wjets$ and multijet backgrounds are estimated from data using the
fake-factor method~\cite{HIGG-2013-03}.  This uses a control sample
derived from data
using a loosened requirement on $\met$ and several kinematic selections. 
The background
in the signal region is then derived using an efficiency factor from 
simulation to correct for the acceptance.  
Both of these backgrounds are found to be negligible.

Table~\ref{tab:ZZllnn_yields_postfit} shows the 
expected yields of the backgrounds and signal, and observed counts of data 
events. The expected yields of the backgrounds in the table 
are after applying the combined likelihood fit to the data, as explained in
Section~\ref{sec:combination}. 

\def\qqZZ{$q\bar{q}\rightarrow ZZ$}
\def\ggZZ{$gg\rightarrow ZZ$}
\def\Ztt{$Z\rightarrow \tau^+ \tau^-$}
\def\ZX{$Z(\rightarrow e^+e^-,\mu^+\mu^-)+$jets}
\def\WX{$W(\rightarrow e\nu/\mu\nu)+X$}
\def\hzzsev{$H\rightarrow ZZ~{m_{H}}=700 GeV$}
\def\hzzeig{$H\rightarrow ZZ~{m_{H}}=800 GeV$}
\def\hzznin{$H\rightarrow ZZ~{m_{H}}=900 GeV$}

\begin{table}[ht]
\centering
\def\ignore#1{}
\def\xroundl#1#2#3{\multicolumn{1}{S[table-format=#1, round-mode=figures, round-precision=#2]@{$\,\pm\,$}}{#3}}
\def\xroundr#1#2#3{\multicolumn{1}{S[table-format=#1, round-mode=figures, round-precision=#2]}{#3}}
\begin{tabular}{%
 l
 S[table-format=3.2, table-number-alignment=right,
   round-mode=figures, round-precision=2]@{$\,\pm\,$}
 S[table-format=1.2, table-number-alignment=right,
   round-mode=figures, round-precision=1]@{$\,\pm\,$}
 S[table-format=2.1, table-number-alignment=right,
   round-mode=figures, round-precision=1]
 S[table-format=2.2, table-number-alignment=right,
   round-mode=figures, round-precision=2]@{$\,\pm\,$}
 S[table-format=1.2, table-number-alignment=right,
   round-mode=figures, round-precision=1]@{$\,\pm\,$}
 S[table-format=1.2, table-number-alignment=right,
   round-mode=figures, round-precision=1]}
\toprule
Process & \multicolumn{3}{c}{ggF channel}  & \multicolumn{3}{c}{VBF channel} \\
\midrule
 \qqZZ            & 112   & 1   & 10      &  0.13 & 0.04 & 0.02    \\
 \ggZZ            &  10.5 & 0.1 &  4.9    &  0.12 & 0.01 & 0.05    \\
 $WZ$             &  47   & 1   &  5      &  0.10 & 0.05 & 0.10    \\
 $WW$/$\ttbar$/$Wt$/\Ztt
                  &  58   & 6   &  5      &  0.41 & 0.01 & 0.08    \\
 \ZX              &  74   & 7   & 19      &  \xroundl{2.2}{1}{0.8}  & 0.3  & 0.3     \\
 Other backgrounds&   4.5 & 0.7 &  0.5    & \multicolumn{3}{c}{---}\\
\midrule
 Total background & 307    & 9  & 44      &  1.6  & 0.3  & 0.5     \\
\midrule
 Observed         &\multicolumn{3}{c}{309}& \multicolumn{3}{c}{4}  \\
\midrule                                        
 ggF signal ($\mH=400 \GeV$)  
                  &  45.4 & \xroundl{1.2}{0,round-mode=places}{0.8} &  2.9    & \multicolumn{3}{c}{---}\\
 VBF signal ($\mH=400 \GeV$) 
                  &    \ignore{0.5}1& {$<0.1$}&1.6     & 10.3  & 0.5  & 1.1     \\
\bottomrule
\end{tabular}
\caption{\label{tab:ZZllnn_yields_postfit}
 Expected background yields and observed counts of data events after
 all selections for the ggF and VBF channels of the $\htollvv$ search.
 The first and second uncertainties correspond to the statistical and
 systematic uncertainties, respectively. 
}
\end{table}

\section{\htollqq\ event selection and background estimation}
\label{sec:llqq}
\subsection{Event selection}
\label{sec:selllqq}
As in the previous search, the event selection starts with the
reconstruction of a $\ztoll$ decay.  For the purpose of this search,
leptons are classified as either `loose', with $\pt>7\gev$, or
`tight', with $\pt>25\gev$.  Loose muons extend to $|\eta|<2.7$, while
tight muons are restricted to $|\eta|<2.5$ and must have tracks
in both the ID and the MS.  The transverse impact parameter requirement
for muons is tightened for this search to $|d_0|<0.1\mm$.
Electrons are identified using a likelihood
discriminant very similar to that used for the $\llll$ search, except
that it was tuned for a higher signal efficiency.  This selection is
denoted `very loose LH'~\cite{ATLAS-CONF-2014-032}.  To avoid double counting, the following
procedure is applied to loose leptons and jets.  First, 
any jets that lie $\Delta R < 0.4$ of an electron are removed.
Next, if a jet is within  a cone of $\Delta R = 0.4$ of a muon,
the jet is discarded if it has less than two matched tracks or if
the JVF recalculated without muons (see \secref{sec:objects}) is less
than 0.5, since in this case it is likely to originate from a muon
having showered in the calorimeter; otherwise the muon is discarded.
(Such muons are nevertheless included in the computation of the \met\
and in the jet energy corrections described in
\secref{sec:objects}.)  Finally, if an electron is within a cone of
$\Delta R = 0.2$ of a muon, the muon is kept unless it
has no track in the MS, in which case the electron is
kept.

Events must contain a same-flavour lepton pair with invariant mass satisfying
$83 < $\mll$ < 99\gev$.  At least one of the leptons must be tight, while the other may be either tight or loose.
Events containing any additional loose leptons are rejected.  The two muons in a pair are required
to have opposite charge, but this requirement is not imposed for electrons because larger energy
losses from showering in material in the inner tracking detector lead to higher charge
misidentification probabilities.

Jets used in this search to reconstruct the $\ztoqq$ decay, referred
to as `signal' jets, must have $|\eta|<2.5$ and 
$\pt>20\gev$; the leading signal jet must also have $\pt>45\gev$.  The
search for forward jets in the VBF production mode uses an alternative,
`loose', jet definition, which includes both signal jets and any
additional jets satisfying $2.5<|\eta|<4.5$ and $\pt>30\gev$.  Since
no high-$\pt$ neutrinos are expected in this search, the significance
of the missing transverse momentum,
$\met / \sqrt{\HT}$ (all quantities in GeV), where $\HT$ is the scalar
sum of the transverse momenta of the leptons and loose jets, must be
less than 3.5.  This requirement is loosened to 6.0 for the case of the
resolved channel (see \secref{sec:selllqq_ggf}) with two $b$-tagged jets due to the presence of
neutrinos from heavy-flavour decay.  The $\met$ significance requirement
rejects mainly top-quark background.

Following the selection of the $\ztoll$ decay, the search is divided into several channels: resolved
ggF, merged-jet ggF, and VBF, as discussed below.  %

\subsubsection{Resolved ggF channel}
\label{sec:selllqq_ggf}

Over most of the mass range considered in this search ($\mH\lesssim
700\gev$), the $\ztoqq$ decay results in two well-separated jets that
can be individually resolved.  Events in this channel should thus
contain at least two signal jets.  Since $b$-jets occur much more
often in the signal ($\sim21\%$ of the time) than in the dominant
$\Zjets$ background ($\sim2\%$ of the time), the
sensitivity of this search is optimized by dividing it into `tagged'
and `untagged' subchannels, containing events with exactly two and
fewer than two $b$-tagged jets, respectively.  Events with more than
two $b$-tagged jets %
are rejected.

In the tagged subchannel, the two $b$-tagged jets form the candidate $\ztoqq$ decay.  In the untagged
subchannel, if there are no $b$-tagged jets, the two jets with largest transverse momenta are used.
Otherwise, the $b$-tagged jet is paired with the non-$b$-tagged jet with the largest transverse momentum.
The invariant mass of the chosen jet pair $\mjj$ must be in the range
$70$--$105\gev$ in order to be consistent with $\ztoqq$ decay.
To maintain orthogonality, any events containing a VBF-jet pair as defined by the VBF channel (see
\secref{sec:vbf-selection}) are excluded from the resolved selection.

The discriminating variable in this search is the invariant mass of the $\lljj$ system, $\mlljj$; a
signal should appear as a peak in this distribution.  To improve the mass resolution,
the energies of the jets forming the dijet pair are scaled event-by-event by a single multiplicative
factor to set the dijet invariant mass $\mjj$ to the mass of the $Z$~boson ($m_Z$).  This
improves the resolution by a factor of 2.4 at $\mH=200\gev$.  The resulting $\mlljj$ resolution is
2--3\%, approximately independent of $\mH$, for both the untagged and tagged channels.

Following the selection of the candidate $\llqq$ decay, further requirements are applied in order to
optimize the sensitivity of the search.  For the untagged subchannel, the first requirement is on
the transverse momentum of the leading jet, $\ptj$, which tends to be higher for the signal than
for the background.  The optimal value for this requirement increases with increasing $\mH$.
In order to avoid having distinct selections for different $\mH$ regions, $\ptj$ is normalized by
the reconstructed final-state mass $\mlljj$; the actual selection is $\ptj> 0.1\times\mlljj$.
Studies have shown that the optimal requirement on $\ptj/\mlljj$ is nearly 
independent of the
assumed value of $\mH$.  Second, the total transverse momentum of the dilepton pair also increases
with increasing $\mH$.  Following a similar strategy, the selection is
$\ptll > \min[-54 \gev + 0.46\times\mlljj, 275\gev]$.  Finally, the azimuthal angle between the two leptons decreases with
increasing $\mH$; it must satisfy
$\dphill < (270\gev/\mlljj)^{3.5} + 1$.
For the
tagged channel, only one additional requirement is applied:
$\ptll > \min[-79\gev + 0.44\times\mlljj, 275\gev]$; the different
selection for $\ptll$ increases the sensitivity of the tagged channel
at low $\mH$.
\Figsref{fig:mlljj_ggf}(a) and~\ref{fig:mlljj_ggf}(b) show the $\mlljj$ distributions of the two
subchannels after the final selection.

\begin{figure}[tbh]
  \centering
\subfloat[]{\includegraphics[width=0.48\textwidth]{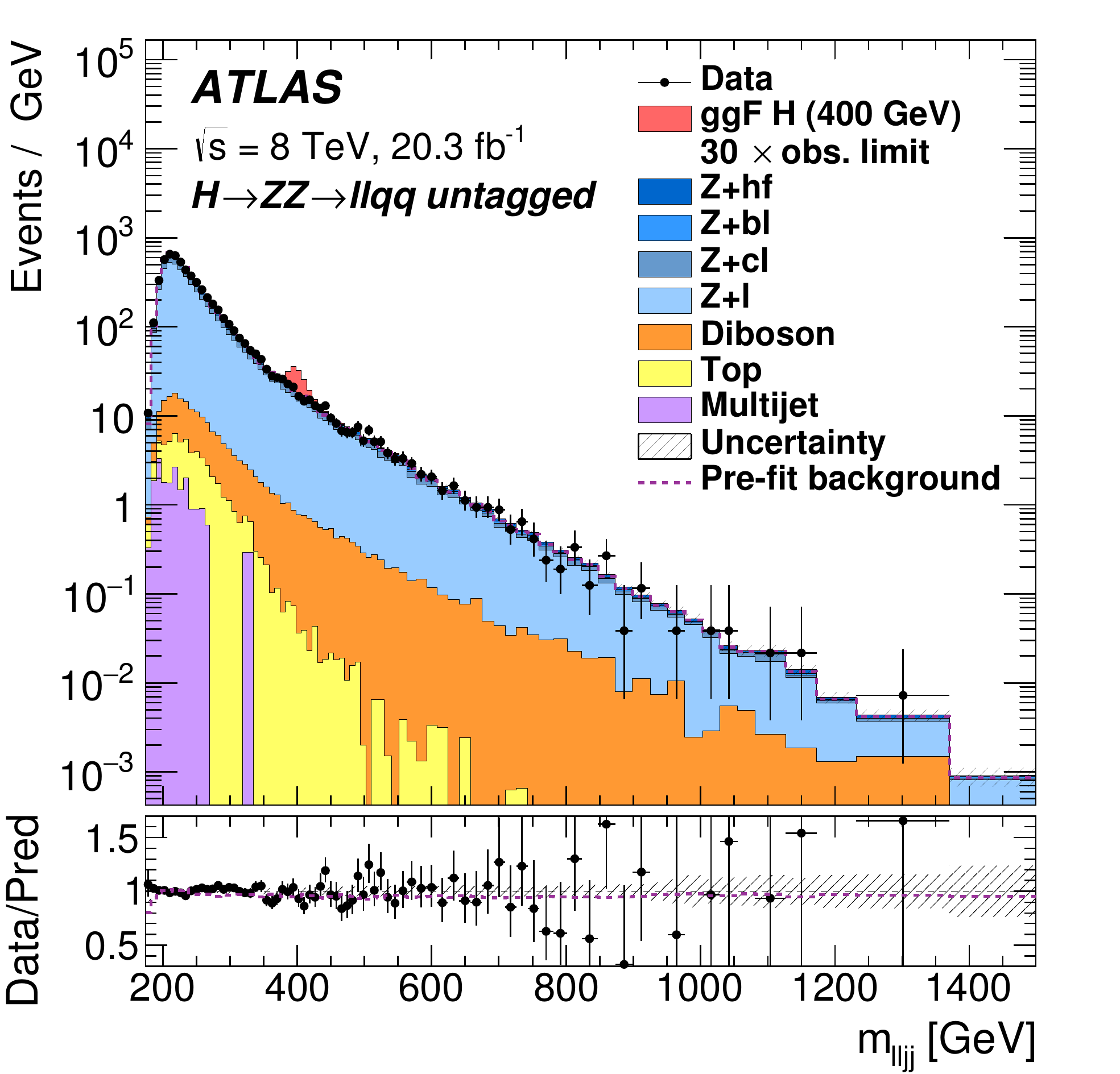}}
\subfloat[]{\includegraphics[width=0.48\textwidth]{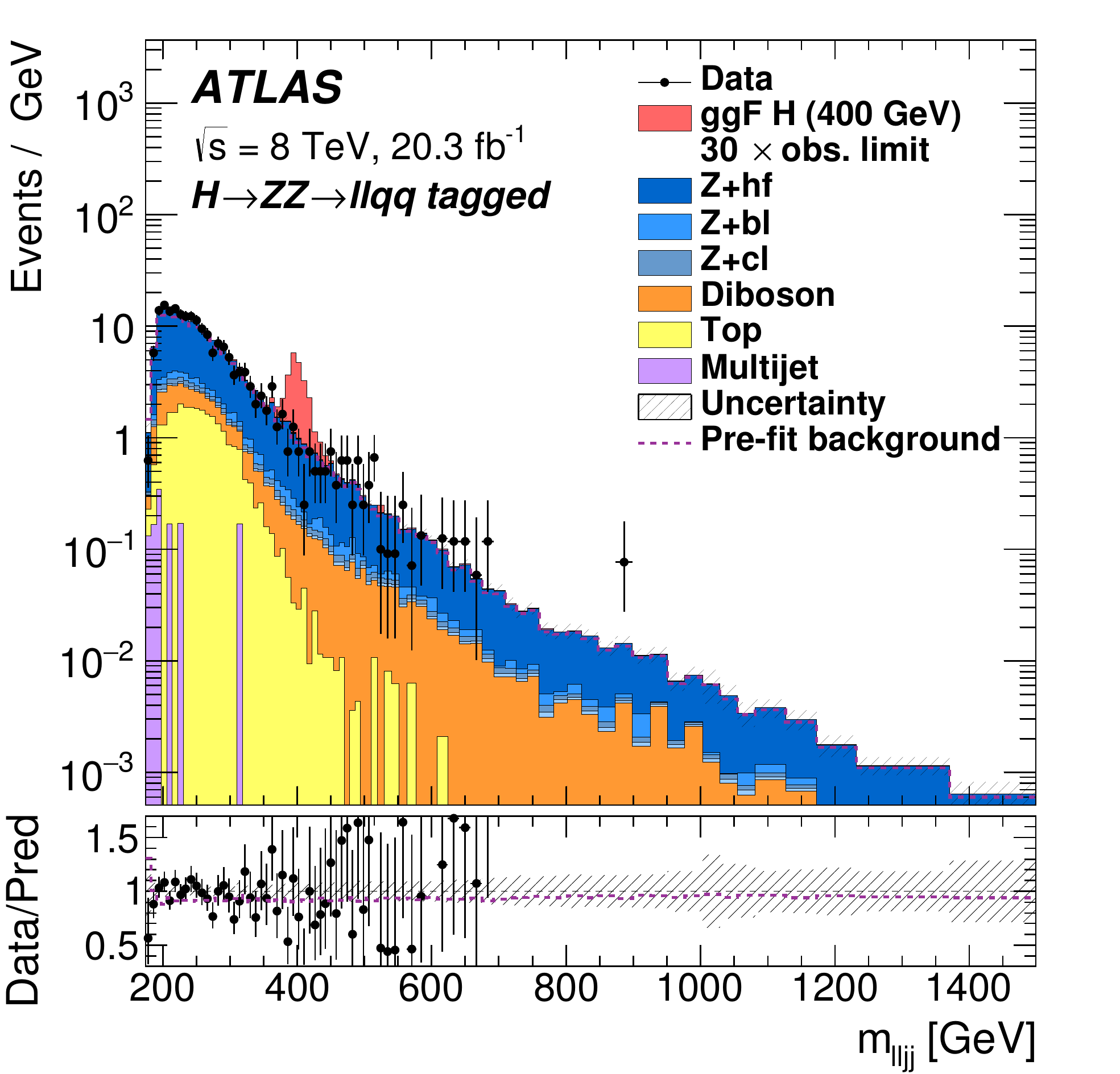}}\\
\caption{\label{fig:mlljj_ggf} The distributions used in the likelihood fit
  of the invariant mass of dilepton+dijet system $\mlljj$ 
  for the
  \htollqq\ search in the (a)~untagged and (b)~tagged resolved ggF subchannels.
  The dashed line shows the total background used as input to the fit.
  The simulated signal is normalized to a cross-section
  corresponding to thirty times the observed limit given in~\secref{sec:results}.
  The contribution labelled as `Top' includes both the $\ttbar$ and single-top processes.
  The bottom panes show the ratio of the observed data
  to the predicted background.
}
\end{figure}

\subsubsection{Merged-jet ggF channel}
\label{sec:sel_merged}

For very large Higgs boson masses, $\mH \gtrsim 700\gev$, the $Z$~bosons become highly boosted and
the jets from $\ztoqq$ decay start to overlap, causing the resolved channel to lose efficiency.  The
merged-jet channel recovers some of this loss by looking for a $\ztoqq$ decay that is reconstructed as a
single jet.

Events are considered for the merged-jet channel if they have exactly one signal jet, or if the selected
jet pair has an invariant mass outside the range $50$--$150\gev$ (encompassing both the signal
region and the control regions used for studying the background).  Thus, the merged-jet channel is
explicitly orthogonal to the resolved channel.

To be considered for the merged-jet channel, the dilepton pair must have
$\ptll>280\gev$.  
The leading
jet must also satisfy $\pt > 200\gev$ and $m / \pt > 0.05$, where
$m$ is the jet mass, in order to restrict the jet to the
kinematic range in which the mass calibration has been studied.  Finally, the invariant mass of the
leading jet must be within the range $70$--$105\gev$. 
The merged-jet channel is not split into subchannels based on the number
of $b$-tagged jets; as the sample size is small, this would not improve
the expected significance.

Including this channel increases the overall efficiency for the $\llqq$ signal
at $\mH=900\gev$ by about a factor of two.  \Figref{fig:merged}(a)
shows the distribution of the invariant mass of the leading jet after
all selections except for that on the jet invariant mass; it can be
seen that the simulated signal has a peak at the mass of the $Z$~boson, with a tail
at lower masses due to events where the 
decay products of the $Z$~boson are not fully contained in the jet cone.
The discriminating variable for this channel is the
invariant mass of the two leptons plus the leading jet, $\mllj$,
which has a resolution of 2.5\% for a signal with $\mH=900\gev$
and is shown in \figref{fig:merged}(b).

\begin{figure}
\centering
\subfloat[]{\includegraphics[width=0.48\textwidth]{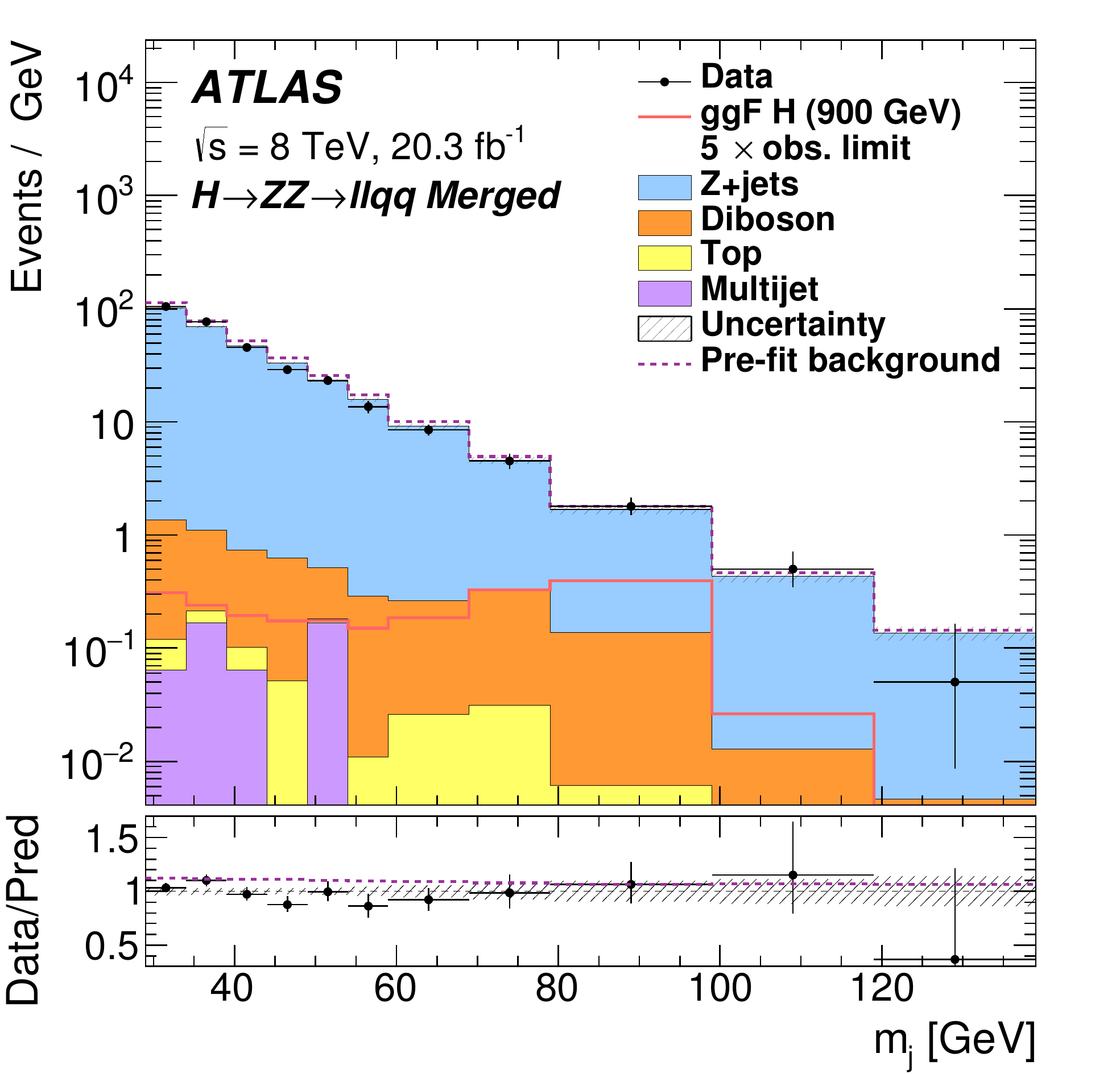}}
\subfloat[]{\includegraphics[width=0.48\textwidth]{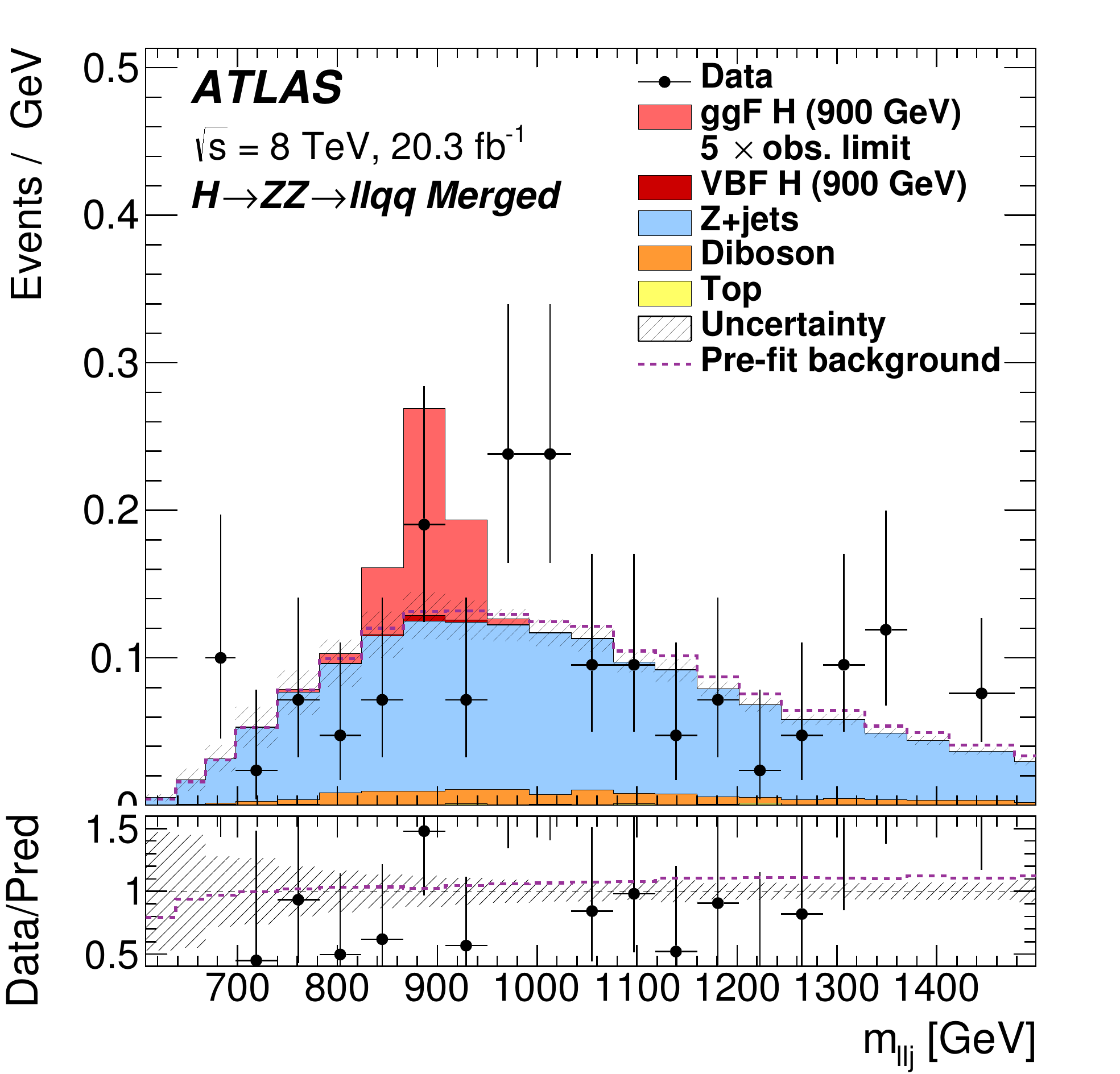}}
  \caption{\label{fig:merged} Distributions for the merged-jet channel
    of the $\htollqq$ search after the mass calibration.  
    (a)~The invariant mass of the leading jet,
    $\mj$, after the kinematic selection 
    for the $\llqq$ merged-jet channel.  (b)~The distribution used in the 
    likelihood fit of the invariant mass of the two leptons and the
    leading jet $\mllj$
    in the signal region. It is obtained requiring $70 < \mj < 105\gev$.
    The dashed line shows the total background used as input to the fit.
    The simulated signal is normalized to a cross-section
    corresponding to five times the observed limit given in~\secref{sec:results}.
  The contribution labelled as `Top' includes both the $\ttbar$ and single-top processes.
  The bottom panes show the ratio of the observed data
  to the predicted background.  The signal contribution is shown added
  on top of the background in~(b) but not in~(a).
}
\end{figure}

\subsubsection{VBF channel}
\label{sec:vbf-selection}

Events produced via the VBF process contain two forward jets in addition to the reconstructed
leptons and signal jets from $\zztollqq$ decay.  These forward jets are called `VBF jets'.  The
search in the VBF channel starts by identifying a candidate VBF-jet pair.  Events must have at least
four loose jets, two of them being non-$b$-tagged and pointing in opposite directions in $z$ (that
is, $\eta_1\cdot\eta_2 < 0$).  If more than one such pair is found, the one with the largest invariant
mass, $m_{jj,{\rm VBF}}$, is selected.
The pair must further satisfy $m_{jj,{\rm VBF}} > 500\gev$ and have a
pseudorapidity gap of $|\Delta\eta_{jj,{\rm VBF}}| > 4$.
The distributions of these two variables are shown in~\figref{fig:vbf}.  

Once a VBF-jet pair has been identified, the $\zztollqq$ decay is reconstructed in exactly the same
way as in the resolved channel, except that the jets used for the VBF-jet pair are excluded and no
$b$-tagging categories are created 
due to the small sample size.  The final
$\mlljj$ discriminant is shown in~\figref{fig:mlljj_vbf}.  Again, the resolution is improved by 
constraining the dijet mass to $m_Z$ as described in \secref{sec:selllqq_ggf}, resulting in a similar overall resolution of 2--3\%.

\begin{figure}
\subfloat[$m_{jj,{\rm VBF}}$]{ \includegraphics[width=0.48\textwidth]{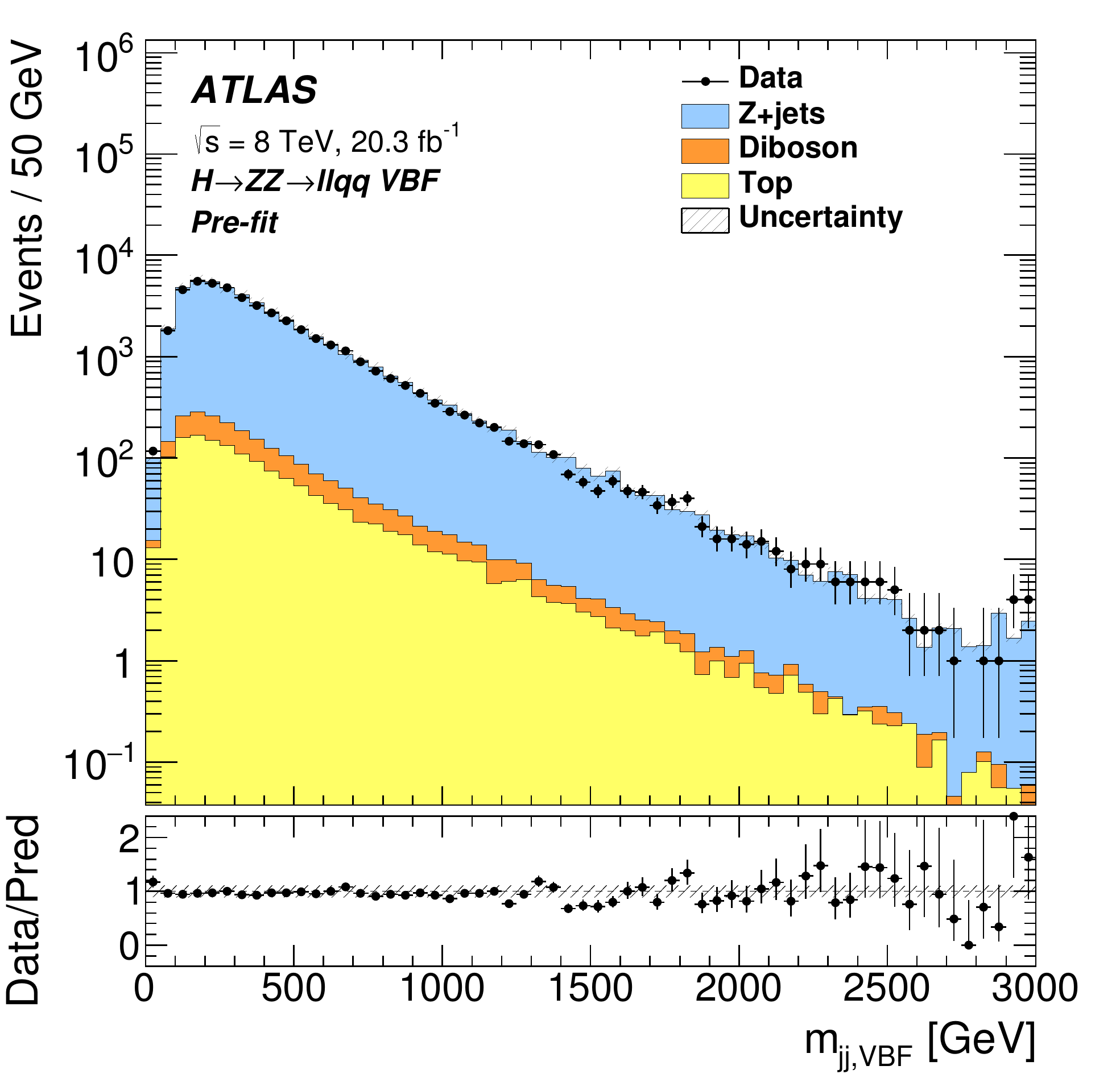}}
\subfloat[$\Delta\eta_{jj,{\rm VBF}}$]{ \includegraphics[width=0.48\textwidth]{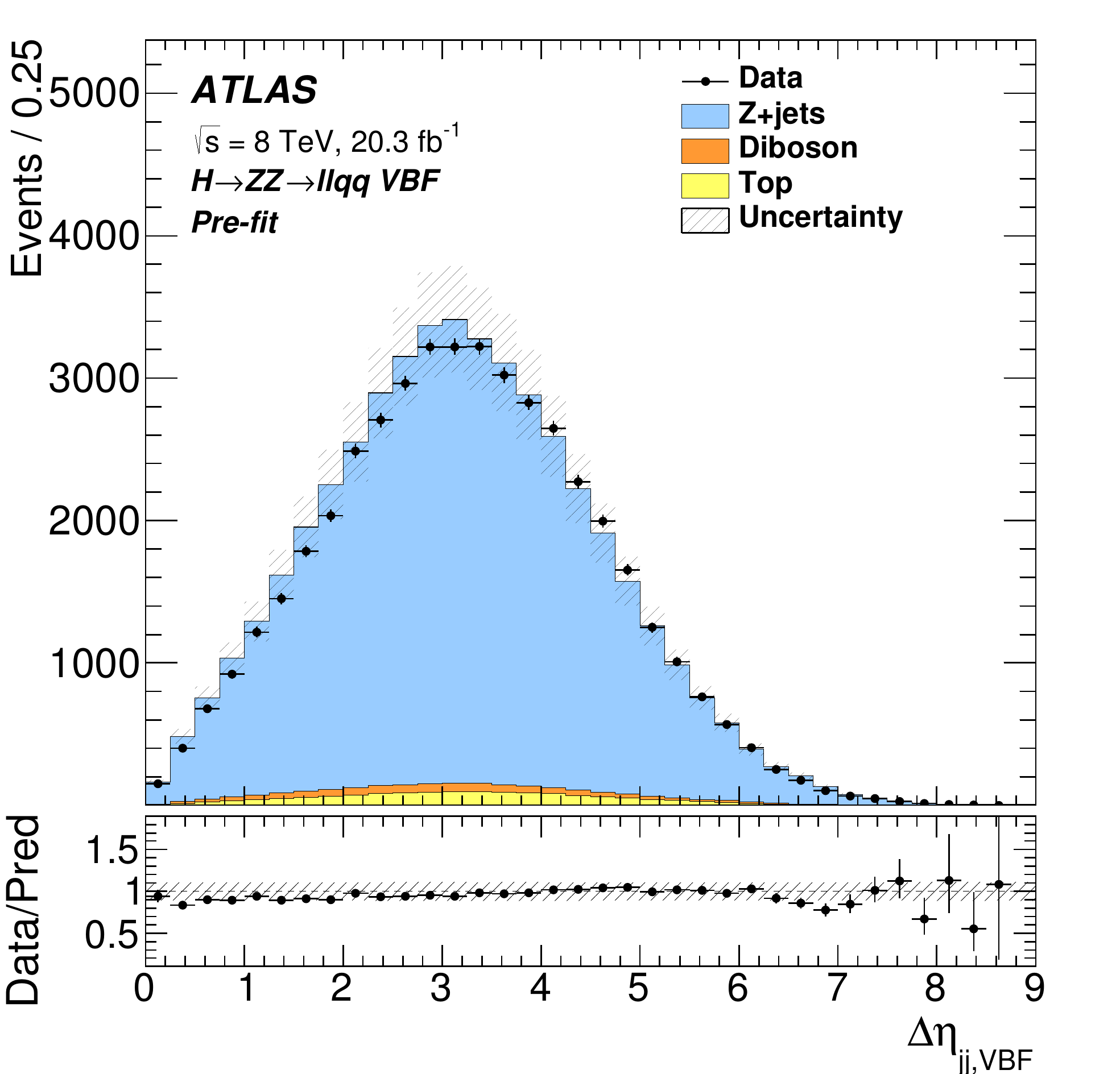}}
   \caption{\label{fig:vbf} Distribution of (a)~invariant mass and
     (b)~pseudorapidity gap for the VBF-jet pair in the VBF channel
     of the $\htollqq$ search before applying the requirements
     on these variables (and prior
     to the combined fit described in~\secref{sec:combination}).
     The contribution labelled as `Top' includes both the $\ttbar$ and single-top processes.
  The bottom panes show the ratio of the observed data
  to the predicted background.
}
\end{figure}

\begin{figure}
\centering
\includegraphics[width=0.48\textwidth]{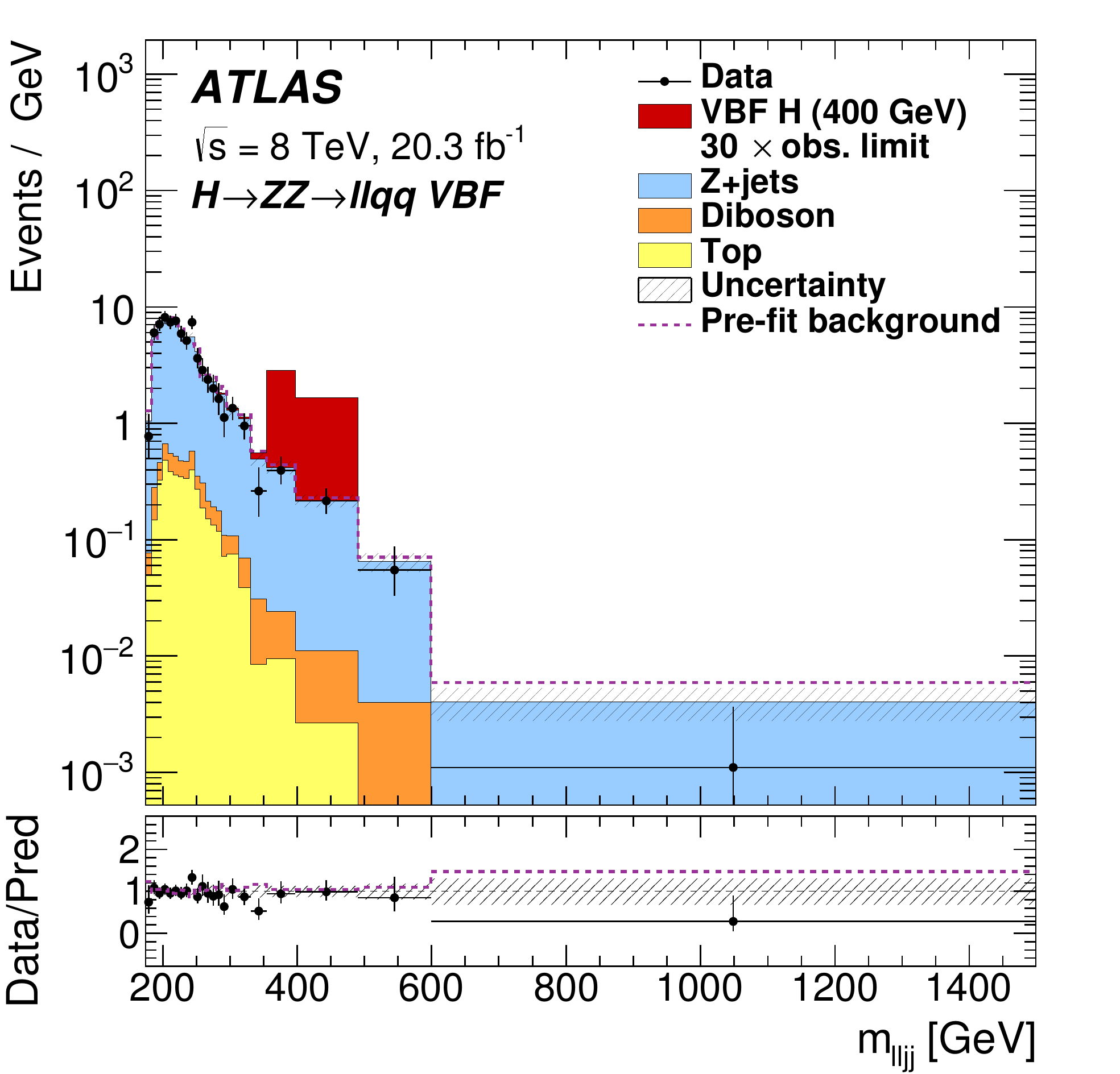}
\caption{\label{fig:mlljj_vbf} The distribution of $\mlljj$ used in the
  likelihood fit for the \htollqq\ search in the VBF channel. 
  The dashed line shows the total background used as input to the fit.
  The simulated signal is normalized to a cross-section
  corresponding to thirty times the observed limit given in~\secref{sec:results}.
  The contribution labelled as `Top' includes both the $\ttbar$ and single-top processes.
  The bottom pane shows the ratio of the observed data
  to the predicted background.
}
\end{figure}

\subsection{Background estimation}
\label{sec:sigbkgllqq}

The main background in the $\llqq$ search is $\Zjets$ production, with significant contributions
from both top-quark and diboson production in the resolved ggF channel, as well as a small
contribution from multijet production in all channels.
For the multijet background, the shape and normalization
is taken purely from data, as described below.  For the other background processes, the input is
taken from simulation, with data-driven corrections for $\Zjets$ and $\ttbar$ production.  The
normalizations of the $\Zjets$ and top-quark backgrounds are left free to float and are determined in
the final likelihood fit as described below and in~\secref{sec:combination}.

The $\Zjets$ MC sample is constrained using control regions that have the same selection as the
signal regions except that $\mjj$ ($\mj$ in the case of the merged-jet channel) lies in a 
region just outside of that selected by the signal $Z$~boson requirement.  For the resolved
channels, the requirement for the control region is $50<\mjj<70\gev$ or $105<\mjj<150\gev$; for the
merged-jet channel, it is $30<\mj<70\gev$.  In the resolved ggF channel, which is split into untagged
and tagged subchannels as described in~\secref{sec:selllqq_ggf}, 
the $\Zjets$ control region is further
subdivided into 0-tag, 1-tag, and 2-tag subchannels based on the number of $b$-tagged jets.  The sum of the
0-tag and 1-tag subchannels is referred to as the untagged control region, while the 2-tag subchannel is
referred to as the tagged control region.

The normalization of the $\Zjets$ background is determined by the final profile-likelihood fit as
described in~\secref{sec:combination}.  In the resolved ggF channel, the simulated $\Zjets$ sample is
split into several different components according to the true flavour of the jets as described
in~\secref{sec:bkgs}: $Z+jj$, $Z+cj$, $Z+bj$, and $Z+$hf.  
The individual normalizations for each of these four components are free to float in the fit and are
constrained by providing as input to the fit the distribution of the ``$b$-tagging category'' in the
untagged and tagged $\Zjets$ control regions.  The $b$-tagging category is defined by the
combination of the MV1c $b$-tagging discriminants of the two signal jets as described in
Appendix~\ref{app:mv1c}.  In the VBF and merged-jet ggF channels, which are not divided into $b$-tag
subchannels, the background is dominated by $Z+$light-jets.  Thus, only the inclusive $\Zjets$
normalization is varied in the fit for these channels.  Since these two channels probe very
different regions of phase space, each has a separate normalization factor in the fit; these are
constrained by providing to the fit the distributions of $\mlljj$ or $\mllj$ for the corresponding
$\Zjets$ control regions.

Differences are observed between data and MC simulation for the distributions
of the azimuthal angle between the two signal jets, $\dphijj$, and
the transverse momentum of the leptonically-decaying $Z$ boson, $\ptll$, for the
resolved region, and for the $\mlljj$ distribution in the VBF channel.
To correct for these differences, corrections are applied to the
\SHERPA $\Zjets$ simulation (prior to the likelihood fit) as described in
Appendix~\ref{app:llqqcorr}. The distributions of
$\mlljj$ or $\mllj$ in the various $\Zjets$ control regions are shown in \figref{fig:zcr_llqq}; it
can be seen that after the corrections (and after normalizing to the results of the
likelihood fit), the simulation provides a good description of the data.

The simulation models the $\mjj$ distribution well in the resolved ggF and VBF
channels.  An uncertainty is assigned by weighting each event of
the $\Zjets$ MC simulation by a linear function of $\mjj$ in
order to cover the residual difference between data and MC events in the control regions.

\begin{figure}[!ht]
\centering
\subfloat[Untagged resolved ggF channel]
{\includegraphics[width=0.48\textwidth]{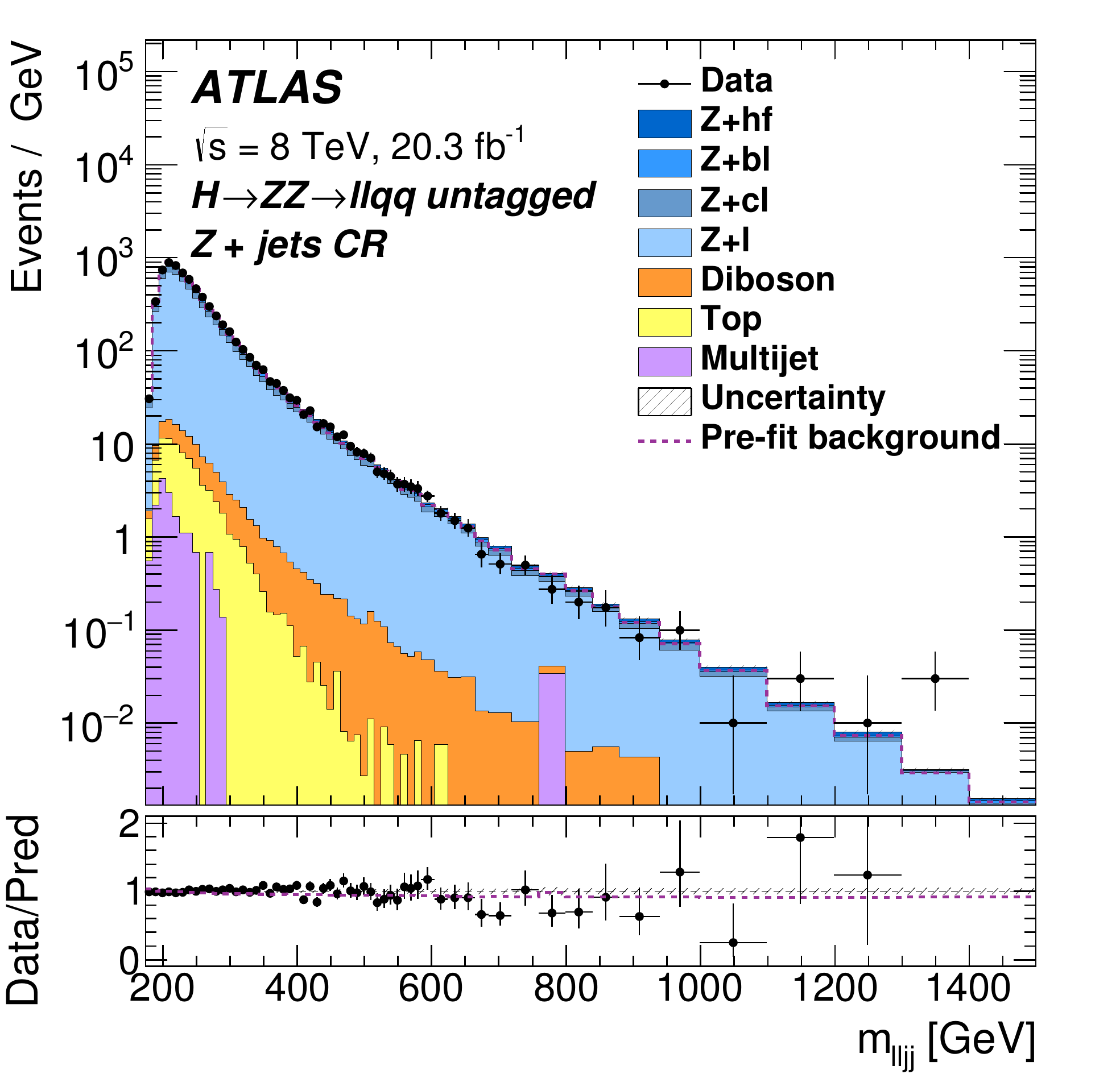}}
\subfloat[Tagged resolved ggF channel]
{\includegraphics[width=0.48\textwidth]{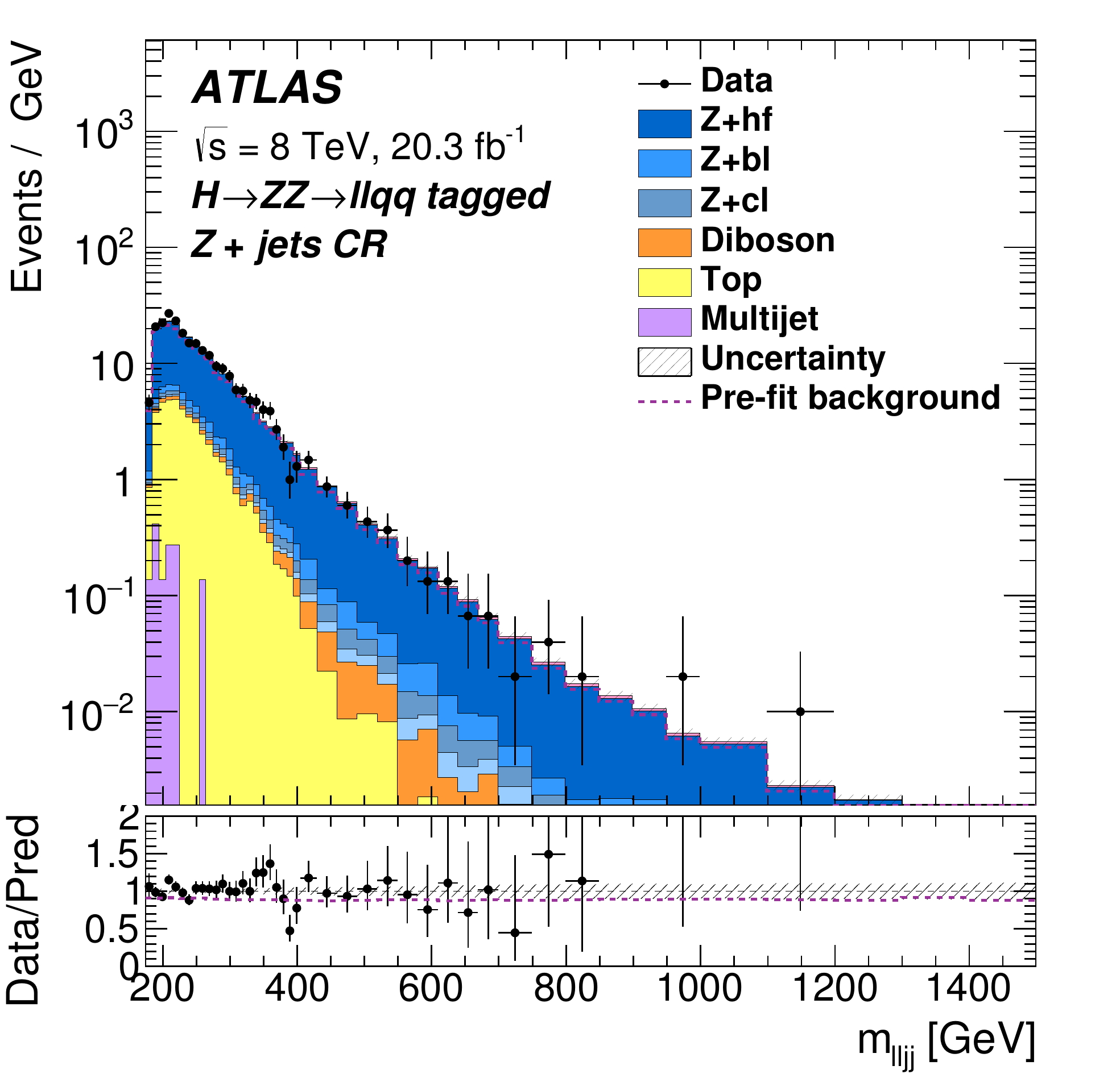}}\\
\subfloat[Merged-jet ggF channel]
{\includegraphics[width=0.48\textwidth]{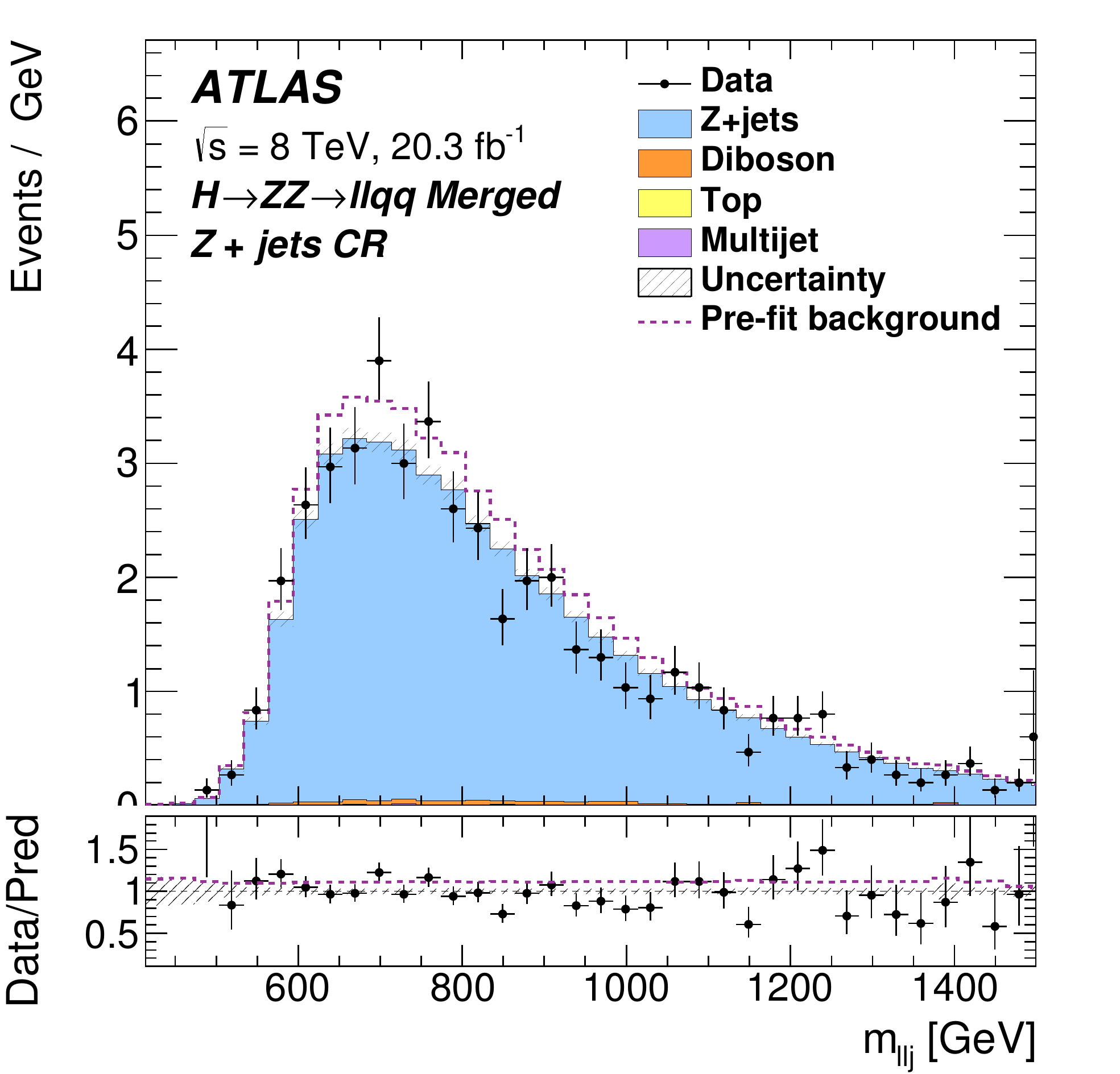}}
\subfloat[VBF channel]
{\includegraphics[width=0.48\textwidth]{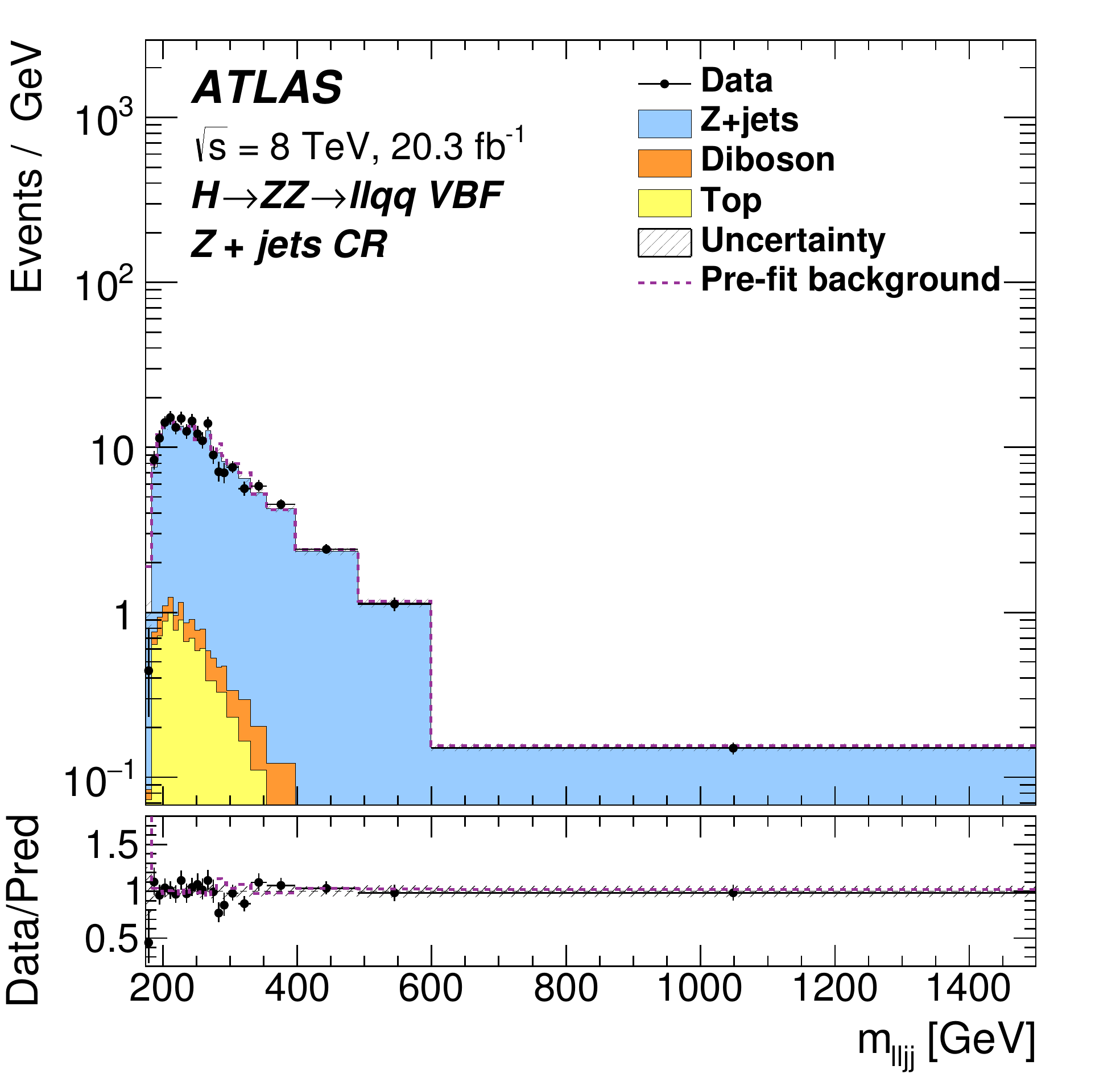}}
\caption{\label{fig:zcr_llqq} The distributions of $\mlljj$ or $\mllj$ in the
  $\Zjets$ control region of the $\htollqq$ search in the 
 (a)~untagged ggF, (b)~tagged ggF, (c)~merged-jet ggF, and (d)~VBF channels.
 The dashed line shows the total background used as input to the fit.
 The contribution labelled as `Top' includes both the $\ttbar$ and single-top processes.
  The bottom panes show the ratio of the observed data
  to the predicted background.
}
\end{figure}

Top-quark production is a significant background in the tagged subchannel of the resolved ggF
channel.  This background is predominantly ($> 97\%$) $t\bar{t}$ production with only a small
contribution from single-top processes, mainly $Wt$ production.
Corrections to the simulation to account for discrepancies in the $\pttt$ distributions are
described in Appendix~\ref{app:llqqcorr}.
The description of the
top-quark background is cross-checked and normalized using a control region with a selection
identical to that of the tagged ggF channel except that instead of two same-flavour leptons, events
must contain an electron and a muon with opposite charge.  The $\mlljj$ distribution in this control
region is used as an input to the final profile-likelihood fit, in which the normalization of the
top-quark background is left free to float (see~\secref{sec:combination}).
There are few events in the control region for the VBF and merged-jet ggF channels,
so the normalization is assumed to be the same across all channels, in which
the top-quark contribution to the background is very small.
\Figref{fig:topcr_llqq} shows that the data in the control region
are well-described by the simulation after the normalization.

\begin{figure}[tbh]
\centering
\includegraphics[width=0.48\textwidth]{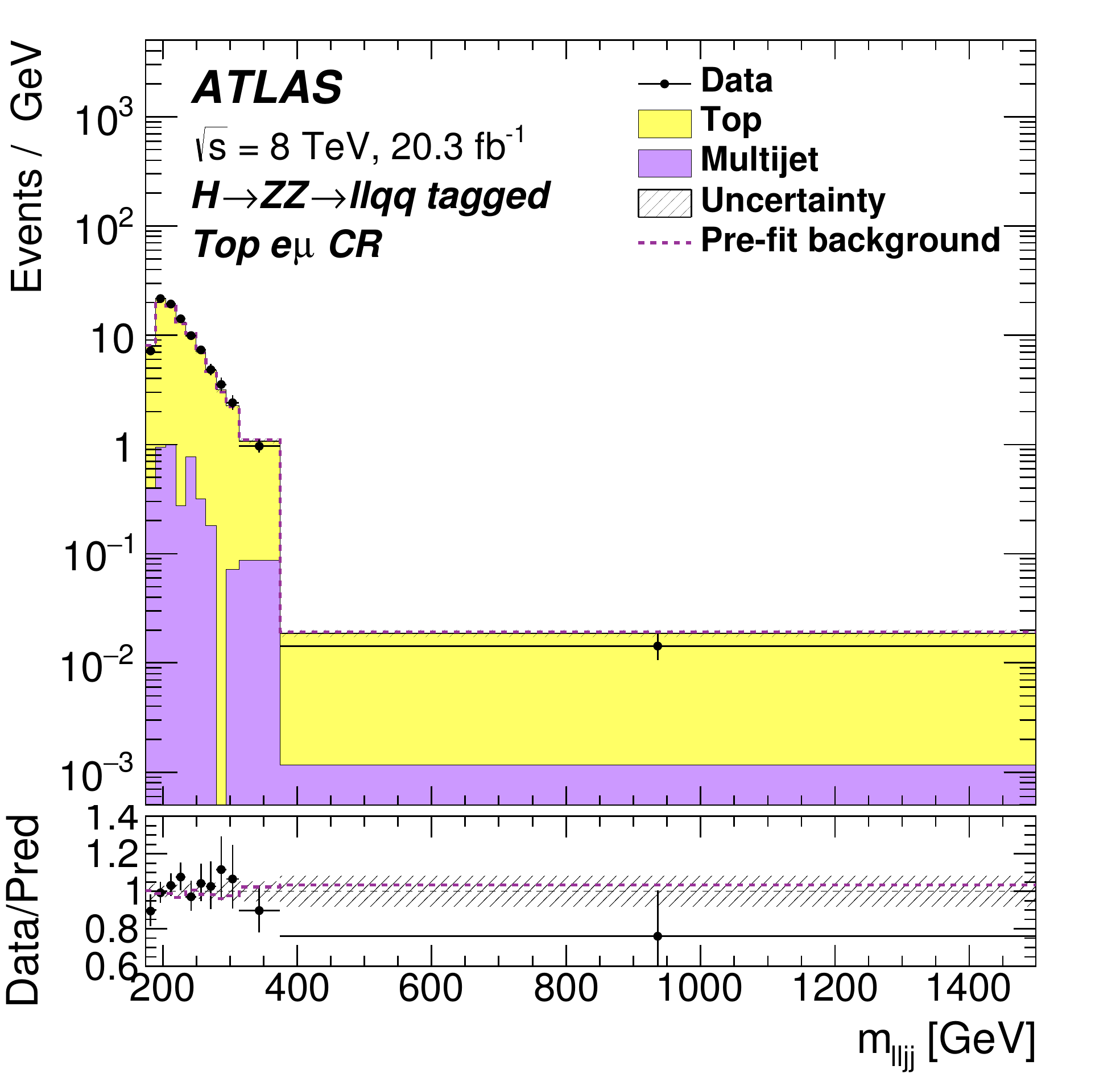}
\caption{\label{fig:topcr_llqq} The distribution of $\mlljj$ in the $e\mu$
  top-quark control region of the $\htollqq$ search in the tagged ggF channel.
  The dashed line shows the total background used as input to the fit.
 The contribution labelled as `Top' includes both the $\ttbar$ and single-top processes.
  The bottom pane shows the ratio of the observed data
  to the predicted background.
}
\end{figure}

Further uncertainties in the top-quark background arising from the parton showering and
hadronization models are estimated by varying the amount of parton showering in \ACERMC and also by
comparing with \progname{Powheg}+\HERWIG.  Uncertainties in the $\ttbar$ production matrix element
are estimated by comparing the leading-order MC generator \ALPGEN with the NLO generator
a\progname{MC@NLO}.  
Comparisons are also made with alternate PDF sets.
A similar procedure is used for single-top
production.  In addition, for the dominant $Wt$ single-top channel, uncertainties in the shapes of
the $\mjj$ and leading-jet $\pt$ distributions are evaluated by comparing results from \HERWIG to
those from \ACERMC.

The small multijet background in the $H \to ZZ \to eeqq$ decay mode is estimated from data by
selecting a sample of events with the electron isolation requirement inverted, which is then
normalized by fitting the $m_{ee}$ distribution in each channel.  In the $H \to ZZ \to \mu\mu qq$
decay mode, the multijet background is found to be negligible.  The residual multijet background in
the top-quark control region is taken from the opposite-charge $e\mu$ data events, which also accounts for
the small $W+\jets$ background in that region.  An uncertainty of 50\% is assigned to these two
normalizations, which are taken to be uncorrelated.

The diboson background, composed mainly of $ZZ$ and $WZ \to \ell\ell jj$ production, and
the SM $Zh \to \ell\ell bb$ background are taken directly from Monte Carlo simulation, as described
in \secref{sec:bkgs}.  The uncertainty in the diboson background is estimated by varying the
factorization and renormalization scales in an \progname{MCFM} calculation~\cite{Campbell:2011bn}.
The method described in Refs.~\cite{stmethodA,stmethodB} is used to avoid underestimating the uncertainty due
to cancellations.  Differences due to the choice of alternate PDF sets
and variations in the value of $\alphas$ are included in the
normalization uncertainty.  Additional shape uncertainties in the $\mjj$ distribution are obtained
by comparing results from \HERWIG, an LO simulation,
with those from \progname{Powheg}+\PYTHIA, an NLO simulation.

The rate of the SM $Vh (V=W/Z, h\ra bb)$ process, relative to the SM expectation, has been measured
by ATLAS as $\mu = \sigma/\sigma_{SM} = 0.52 \pm 0.32~{\rm (stat.)} \pm 0.24~{\rm (syst.)}$~\cite{HIGG-2013-23}.  
Since this is compatible with the SM expectation, the small $Zh (h\ra bb)$ background
in this channel is normalized to the SM cross-section and a 50\% uncertainty is assigned to cover
the difference between the prediction and the measured mean value.

\section{\htovvqq\ event selection and background estimation}
\label{sec:vvqq}
\subsection{Event selection}
\label{sec:selvvqq}
Events selected for this search must contain no electrons or muons as defined
by the `loose' lepton selection of the $\llqq$ search.
To select events with neutrinos in the final
state, the magnitude of the missing transverse momentum vector
must satisfy $\met>160\gev$;
the trigger is 100\% efficient in this range.
Events must have at least two jets with $\pt>20\gev$ and $|\eta|<2.5$;
the leading jet must further satisfy $\pt>45\gev$.  To select a candidate
$\ztoqq$ decay, the invariant mass of the leading two jets must satisfy
$70<\mjj<105\gev$.

The multijet background, due mainly to the mismeasurement of
jet energies, is suppressed using a track-based missing transverse
momentum, $\ptmissvec$, defined as the negative vectorial
sum of the transverse momenta of all good-quality inner detector tracks.
The requirements are $\ptmiss$ $>30\gev$, 
the azimuthal angle between the directions of $\metvec$
and $\ptmissvec$ satisfy $\Delta\phi(\metvec, \ptmissvec)<\pi/2$,
and the azimuthal angle between the directions of  $\metvec$ and the
nearest jet satisfy $\Delta\phi(\metvec, j)>0.6$.

As in the resolved ggF channel of the \llqq\ search,
this search is divided into `tagged' (exactly two $b$-tagged jets)
and `untagged' (fewer than two $b$-tagged jets) subchannels.
Events with more than two $b$-tags are rejected.

The sensitivity of this search is improved by adding a
requirement on the jet transverse momenta.  As in  the $\llqq$ search,
the optimal threshold depends on $\mH$.  However, due to the 
neutrinos in the final state, this decay mode does not provide a good
event-by-event measurement of the mass of the diboson system, $\mZZ$.  
So, rather than having a single requirement on the jet transverse
energy which is a function of the measured $\mZZ$, instead there is
a set of requirements, based on the generated $\mH$, with the background
estimated separately for each of these separate jet requirements.
The specific requirement is found by rounding the generated $\mH$ to
the nearest $100\gev$; this is called $\mHbin$.  Then the subleading
jet must satisfy $\ptjtwo > 0.1\times\mHbin$ in events with no $b$-tagged jets,
and $\ptjtwo > 0.1\times\mHbin - 10\gev$ in events with at least one
$b$-tagged jet.

The discriminating variable for this search is the transverse mass of the
$\vvqq$ system, shown in \figref{fig:vvqq_fit_sr}, defined as in Eq.~(\ref{eq:mt}) with 
$\ptjj$ replacing $\ptll$. To improve the transverse mass resolution,
the energies of the leading two jets are scaled 
event-by-event by a multiplicative factor to set the dijet invariant mass $\mjj$
to the $Z$~boson mass, in the same manner as in the \llqq\ search. This improves
the transverse mass resolution by approximately 20\% at $\mH=400$~\gev\ and by
approximately 10\% at $\mH=1$~\tev.
The resulting resolution in $\mt$ ranges from about 9\% at $\mH=400\gev$
to 14\% at $\mH=1\tev$.

\begin{figure}[!h]
  \centering
\subfloat[Untagged, $\mH=400\gev$]{\includegraphics[width=0.48\textwidth]{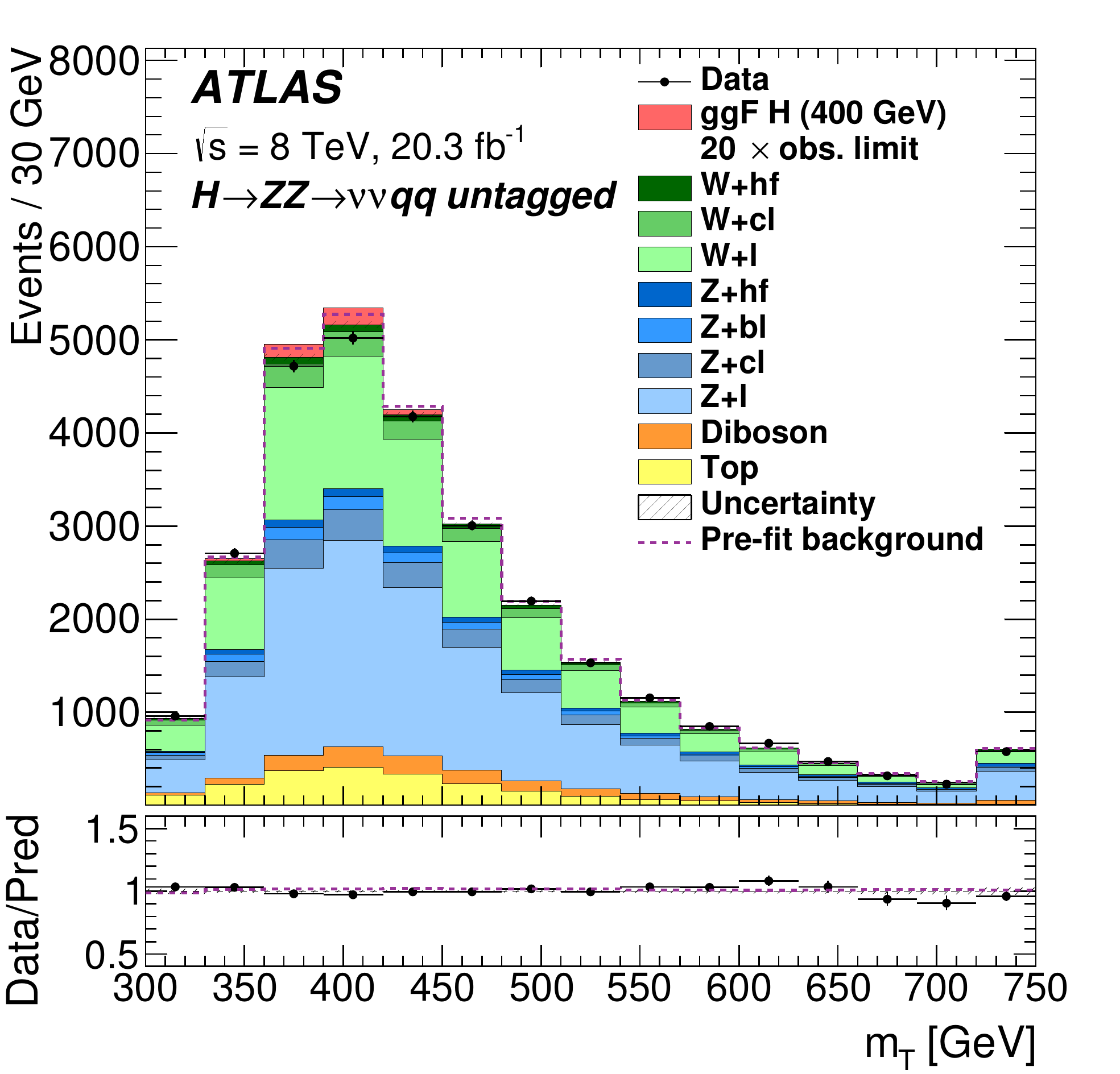}}
\subfloat[Tagged, $\mH=400\gev$]{\includegraphics[width=0.48\textwidth]{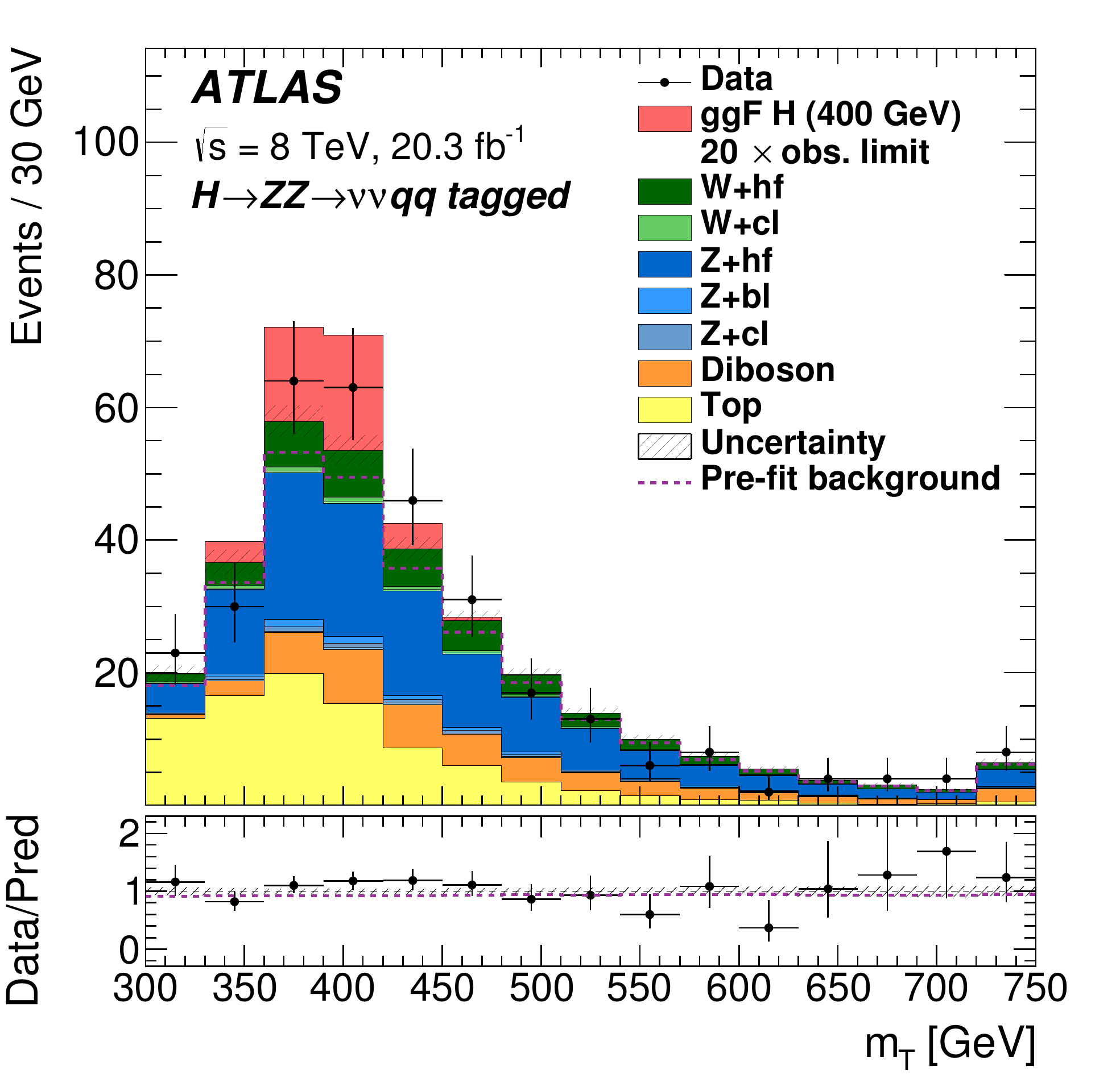}}\\
\subfloat[Untagged, $\mH=900\gev$]{\includegraphics[width=0.48\textwidth]{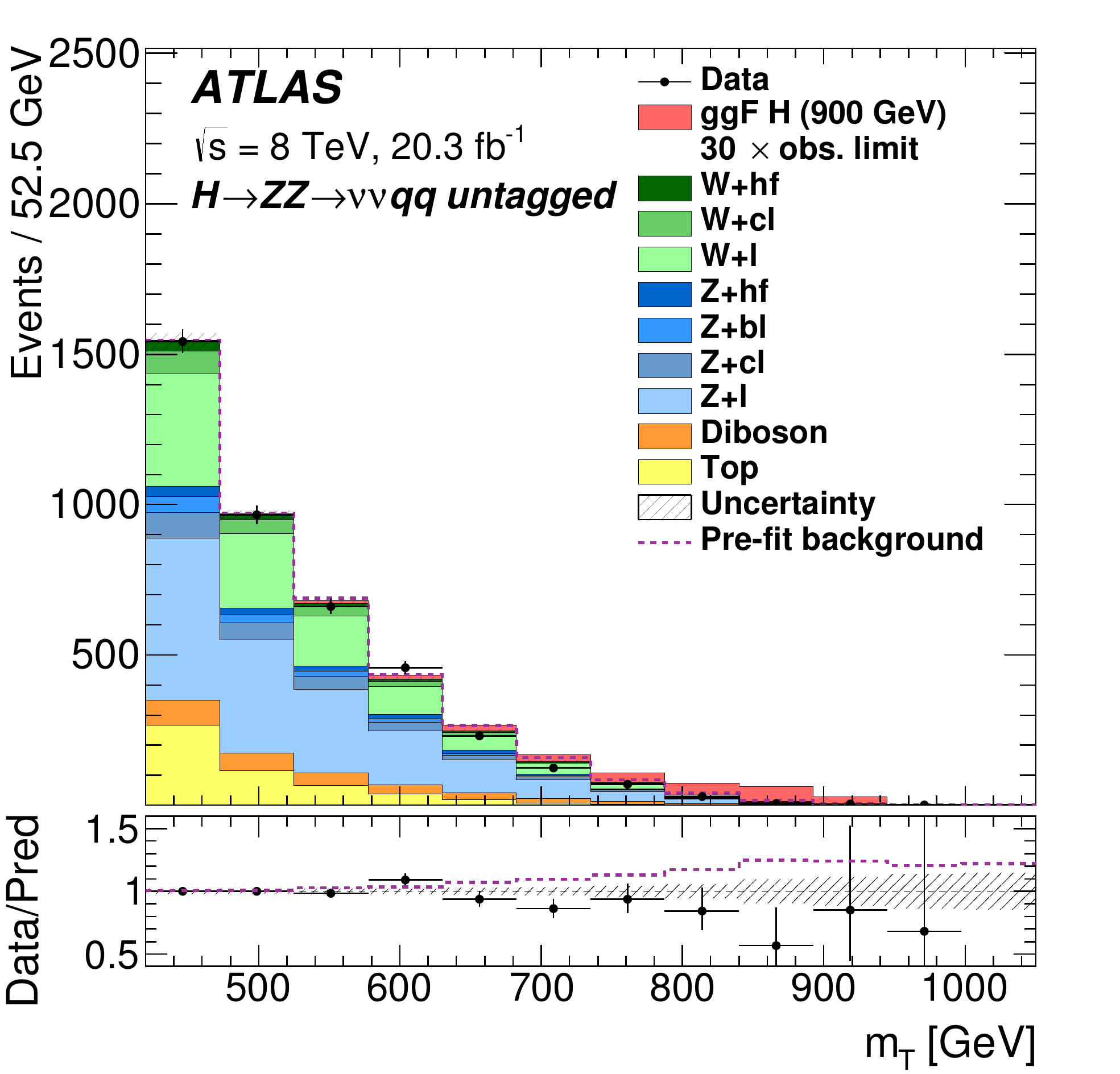}}
\subfloat[Tagged, $\mH=900\gev$]{\includegraphics[width=0.48\textwidth]{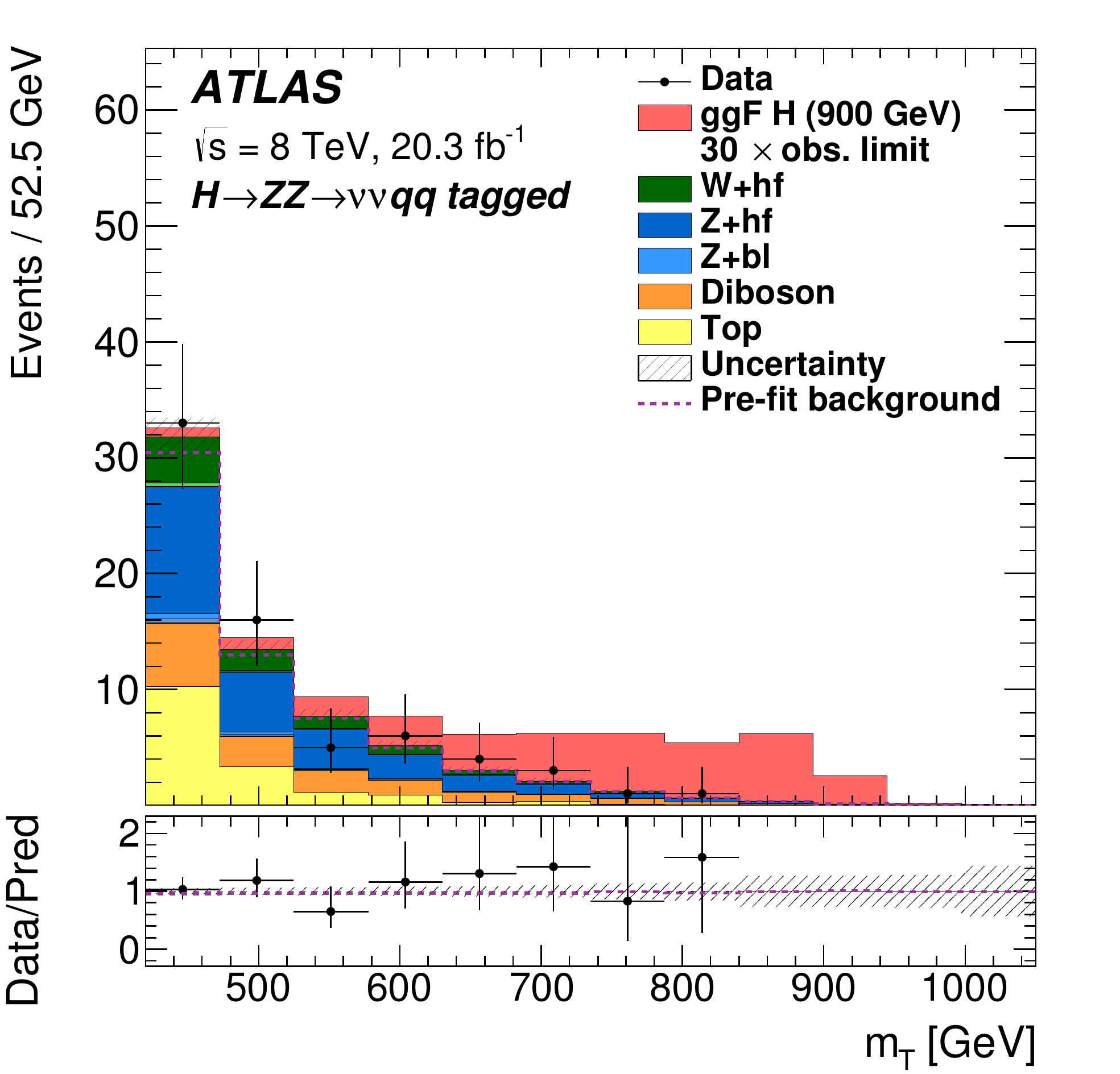}}\\
\caption{\label{fig:vvqq_fit_sr} 
The distributions of $\mt$, the transverse mass of the $Z(\nu\nu)Z(jj)$
system, used in the likelihood fit for the
  \htovvqq\ search in the (a,~c)~untagged and (b,~d)~tagged channels, 
  for Higgs~boson mass hypotheses of (a,~b)~$\mH=400\gev$
  and (c,~d)~$\mH=900\gev$.
  The dashed line shows the total background used as input to the fit.
  For the $\mH=400\gev$ hypothesis (a,~b) the simulated signal is normalized to a cross-section
  corresponding to twenty times the observed limit given
  in~\secref{sec:results}, while for the $\mH=900\gev$ hypothesis (c,~d) it is
  normalized to thirty times the observed limit.
  The contribution labelled as `Top' includes both the $\ttbar$ and single-top processes.
  The bottom panes show the ratio of the observed data
  to the predicted background.
}
\end{figure}

\subsection{Background estimation}
\label{sec:sigbkgvvqq}
The dominant backgrounds for this search are $\Zjets$, $\Wjets$, and $\ttbar$ production.  The
normalization of the $\Zjets$ background is determined using the $\Zjets$ control region from the
$\llqq$ channel in the final profile-likelihood fit as described in~\secref{sec:combination}. To
check how well this background is modelled after the $\vvqq$ selection, an alternative $\Zjets$ control region is
defined in the same way as the signal sample for $\mHbin = 400$~GeV except that events must contain exactly two loose
muons.  The $\met$ is calculated without including the muons and must satisfy the same requirement
as for the signal: $\metnomuon>160\gev$. The $\Zjets$ MC simulation is corrected as a function of
$\dphijj$ and $\ptll$ in the same manner as in the resolved ggF channel of the $\llqq$ search, as
described in \secref{sec:sigbkgllqq} and Appendix~\ref{app:llqqcorr}.

The $\Wjets$ background estimate similarly uses a control sample with the same selection as the
signal sample for $\mHbin = 400$~GeV except that there must be exactly one loose muon and the $\met$ requirement is again
on $\metnomuon$.  
The simulated $\Wjets$ sample is also split into several different flavour components,
as in the case of $\Zjets$.
The normalization of the $W+jj$ and $W+cj$ components are free to float in the final profile-likelihood fit, 
and are constrained by providing as input to the fit the distribution of the MV1c $b$-tagging category,
described in Appendix~\ref{app:mv1c}, in the 0-$b$-tag and 1-$b$-tag control regions.
Unlike the $\Zjets$ case, the 2-$b$-tag control region is not used in the final profile-likelihood fit to constrain
the $W+bj$ and $W+$hf background components  since it is highly dominated by $\ttbar$ production.
Their normalizations are instead taken from the NNLO cross section predictions with 
an uncertainty of 50\%. The uncertainty is determined by comparing the nominal fit value from 
the profile-likelihood fit with the value when including the 2-$b$-tag control region,
where $W+bj$ and $W+$hf are free to float; this uncertainty also covers the normalization determined
in Ref.~\cite{HIGG-2013-23}.
Following Ref.~\cite{HIGG-2013-23}, the agreement between simulation and data for this background is
improved by applying a correction to $\dphijj$ for $W+jj$ and $W+cj$, with half the correction
assigned as a systematic uncertainty; in the case of $W+bj$ and $W+$hf, no correction is applied,
but a dedicated systematic uncertainty is assigned.

Even after these corrections, the simulation does not accurately describe the data in 
the $\Zjets$ and $\Wjets$ control sample with no $b$-tagged jets (which is dominated 
by $Z/W+jj$) for important kinematic distributions such as $\met$ and jet transverse
momenta.  Moreover, because the resolution of the transverse mass of the
$\zztovvqq$ system is worse than that of $\mlljj$, the $\vvqq$ search
is more sensitive to $\met$ (i.e. $Z/W$~boson $\pt$) than the $\llqq$ search.
Therefore, a further correction is applied, as a linear function of $\met$, derived from measuring 
the ratio of the $\met$ distributions from simulation and data in the control sample 
with no $b$-tagged jets after non-$Z/W+jj$ backgrounds have been subtracted.
An uncertainty of 50\% is assigned to this correction.
Following this correction, there is good agreement between simulation and data, as shown 
in \figsref{fig:Zjet0tag} and~\ref{fig:Wjet0tag}. 
For higher $\mHbin$ signal samples, which have tighter selections on kinematic variables than the control
sample, the $\met$ correction is somewhat underestimated, leading to some remaining difference
between data and pre-fit simulation at high $m_T$, as can be seen in~\figref{fig:vvqq_fit_sr}(c).
However, the profile-likelihood-ratio fit (\secref{sec:combination}) is able to correct this
residual mismodelling, leading to reasonable agreement between the data and simulation.

\begin{figure}[htbp]
\centering
\subfloat[$\met$]{\includegraphics[width=0.47\textwidth]{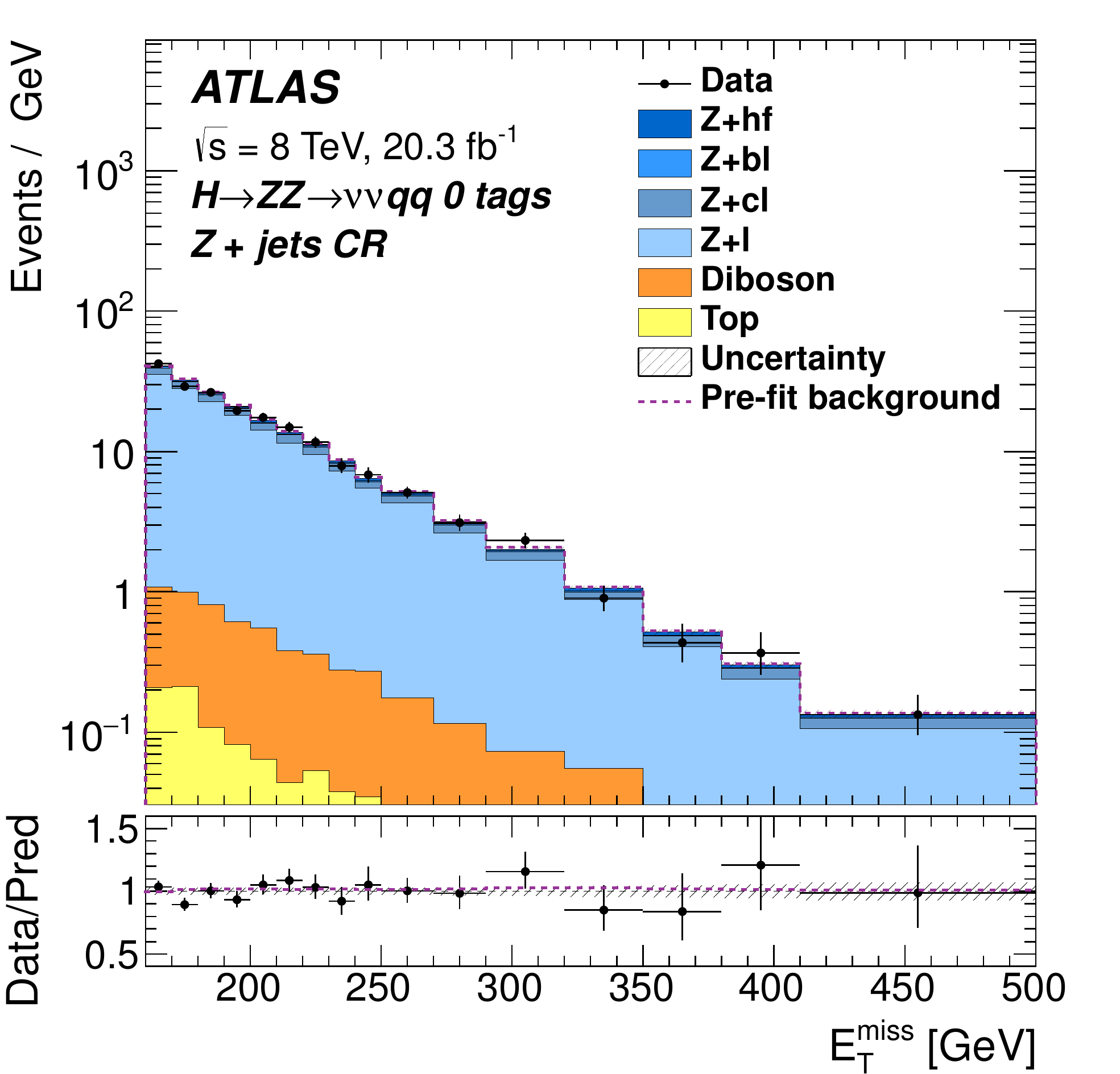}}
\subfloat[Leading-jet $\pt$]{\includegraphics[width=0.47\textwidth]{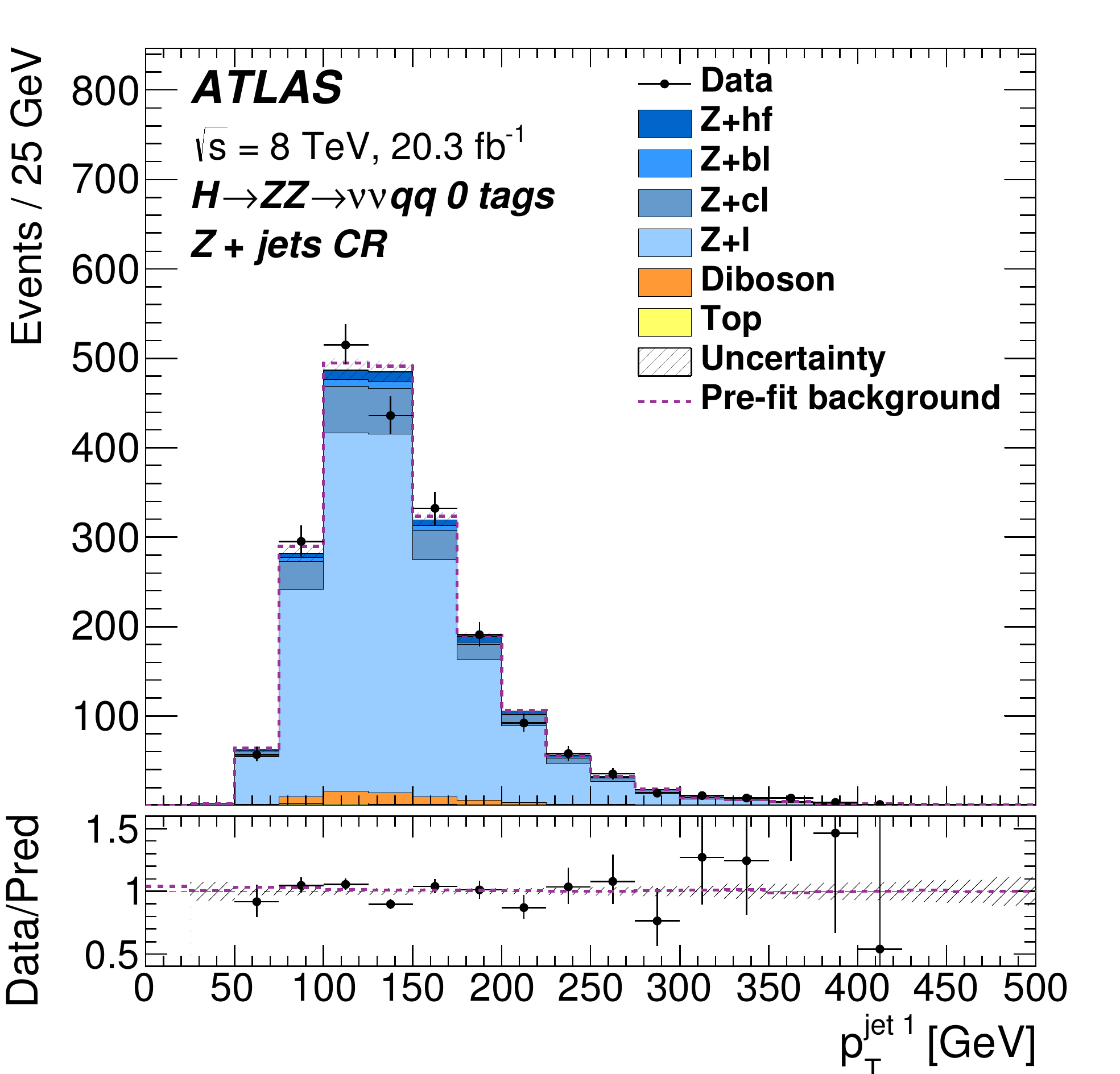}}
\caption
{\label{fig:Zjet0tag}
The distributions of (a)~missing transverse momentum $\met$ and (b)~leading-jet $\pt$ from the
untagged $(Z\ra\mu\mu) + \jets$ control sample of the $\htovvqq$ search.
The dashed line shows the total background used as input to the fit.
The contribution labelled as `Top' includes both the $\ttbar$ and single-top processes.
  The bottom panes show the ratio of the observed data
  to the predicted background.
}
\end{figure}

\begin{figure}[htbp]
\centering
\subfloat[$\met$]{\includegraphics[width=0.47\textwidth]{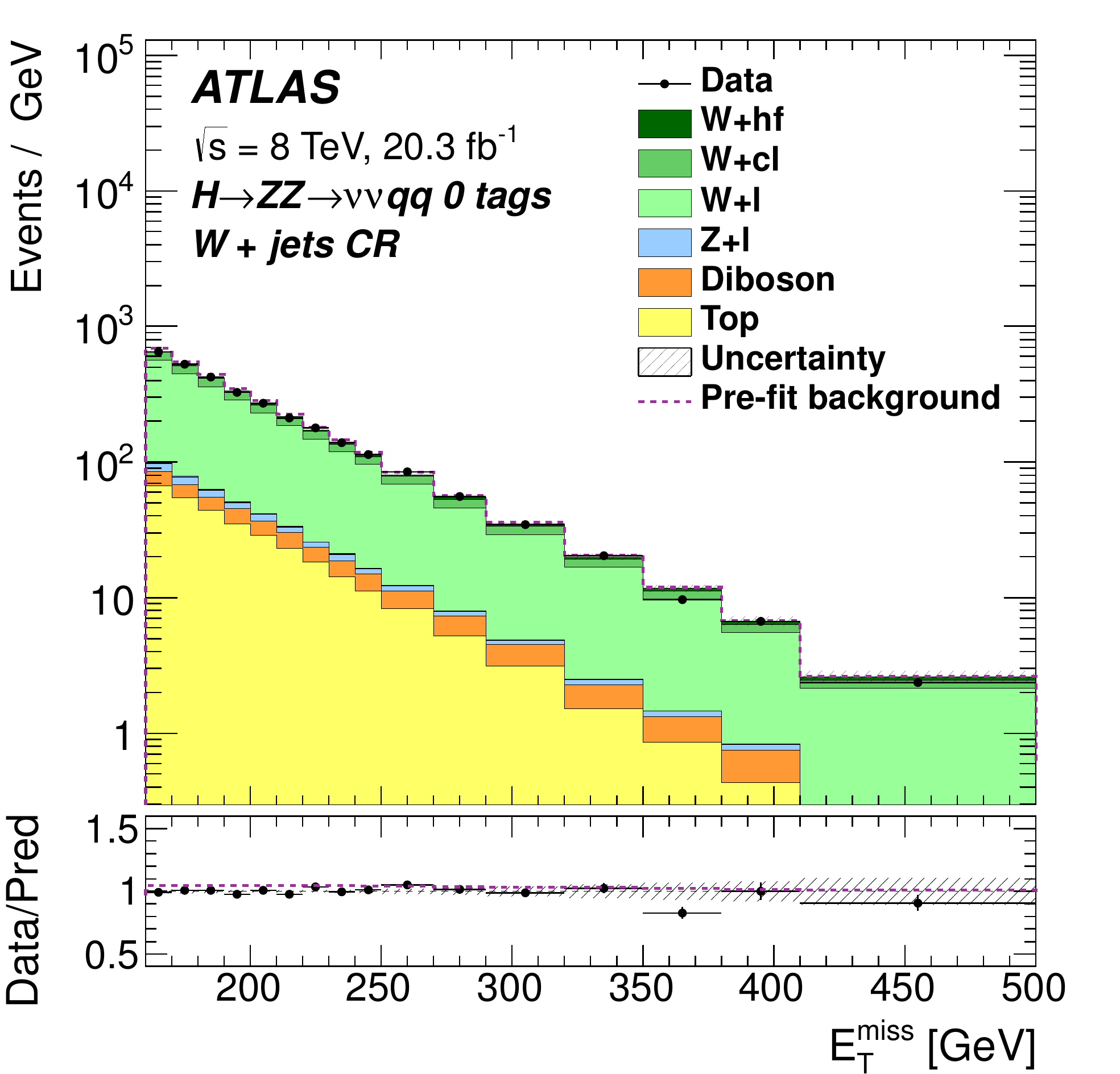}}
\subfloat[Leading-jet $\pt$]{\includegraphics[width=0.47\textwidth]{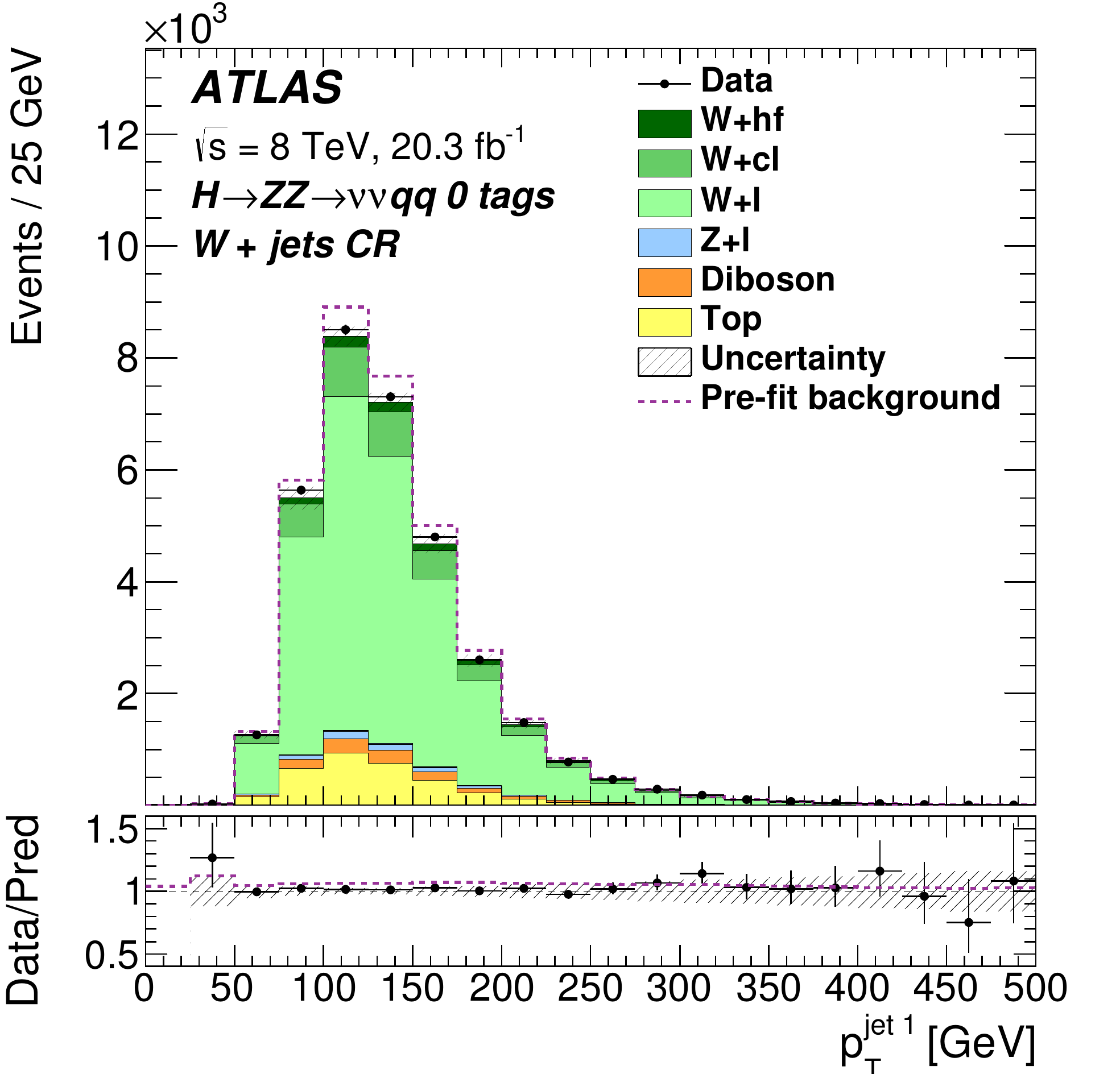}}
\caption
{\label{fig:Wjet0tag}
The distributions of (a)~$\met$ and (b)~leading-jet $\pt$ from the
untagged $(W\ra\mu\nu) + \jets$ control sample of the $\htovvqq$ search.
The dashed line shows the total background used as input to the fit.
The contribution labelled as `Top' includes both the $\ttbar$ and single-top processes.
The bottom panes show the ratio of the observed data
to the predicted background.
}
\end{figure}

The $\ttbar$ background is treated in the same manner as in the $\llqq$ search; 
in particular, $\pttt$ is corrected in the same way and the normalization is determined 
by $\ttbar$ control region from $\llqq$ channel in the final profile-likelihood fit.

Backgrounds from diboson and single-top production are estimated directly from MC simulations, 
both for shapes and normalization. The multijet background is estimated
using a method similar to that used for the $\Zjets$ background in 
the $\llvv$ search (\secref{sec:sigbkgllvv}), except that the variables used are
$\Delta\phi(\metvec,\ptmissvec)$ and $\Delta\phi(\metvec, j)$~\cite{HIGG-2013-23}.
It is found to be negligible.

\section{Systematic uncertainties}
\label{sec:systs}
The systematic uncertainties can be divided into three categories:
experimental uncertainties, related to the detector or to the reconstruction
algorithms, uncertainties in the modelling of the signal,
and uncertainties in the estimation of the backgrounds.
The first two
are largely common to all the searches and are treated as
fully correlated.  The uncertainties in the estimates of most backgrounds
vary from search to search, and are summarized in the background
estimation sections above.
The estimation of the uncertainty of the $\ZZ$ background is outlined
in \secref{sec:systszzbg}.

\subsection{Experimental uncertainties}
\label{sec:systs-common}
The following detector-related systematic uncertainties are common to all
the searches unless otherwise stated.

The uncertainty in the integrated luminosity is determined to be 2.8\%
in a %
calibration following the methodology detailed in Ref.~\cite{DAPR-2011-01}
using beam-separation scans performed in November 2012. This uncertainty is
applied to the normalization of the signal and also to backgrounds for which the
normalization is derived from MC calculations, and is correlated between all
of the searches.
There is also an uncertainty of 4\% in the average number of interactions per
bunch crossing, which leads to an uncertainty on distributions sensitive to
pile-up.

There are small systematic uncertainties of $O(1\%)$ in the reconstruction
and identification efficiencies for electrons and
muons~\cite{PERF-2014-05,PERF-2013-03,PERF-2013-05,ATLAS-CONF-2014-032}.
For the $\vvqq$ search, the uncertainty is instead in the efficiency
of the lepton veto, and is also $O(1\%)$.  Uncertainties in the lepton
energy scale and resolution are also taken into account. These uncertainties are
treated as uncorrelated between all of the searches due to differences in lepton selections 
optimized for each search. 

The uncertainty in the jet energy scale has several sources,
including uncertainties in the in situ calibration analysis,
corrections for pile-up, and the flavour composition of
the sample~\cite{PERF-2011-03,PERF-2012-01}.  These
uncertainties are decomposed into independent components.  For central jets,
the total relative uncertainty on the jet energy scale ranges from about 3\%
for jets with a $\pt$ of $20\gev$ to about 1\% for a $\pt$ of $1\tev$.
The calibration of the $b$-jet transverse energy has an additional uncertainty of 1--2\%.
There is also an uncertainty in the jet energy resolution~\cite{PERF-2011-04},
which ranges from 10--20\% for jets with a $\pt$ of $20\gev$ to less than
5\% for jets with $\pt>200\gev$.  The uncertainty associated with the
pile-up rejection requirement (\secref{sec:objects}) is evaluated by varying the nominal
value of 50\% between 47\% and 53\%~\cite{ATLAS-CONF-2013-083}.
The jet energy scale uncertainties are correlated between the \llqq\ and \vvqq\
searches, and separately between the \llll\ and \llvv\ searches. 
They are not correlated between the two pairs of searches because although
the $\llqq$ and $\vvqq$ control regions have the power to constrain
the jet energy scale uncertainties, these constraints do not necessarily apply
to the $\llll$ and $\llvv$ searches due to differences
 in the jet kinematics and composition.

Uncertainties on the lepton and jet energy scales are propagated into
the uncertainty on $\met$.  A contribution to $\met$ also comes from
energy deposits that are not associated with any identified physics object;
uncertainties on the energy calibration (8\%) and resolution (3\%) 
of the sum of
these deposits are also propagated to the uncertainty on $\met$~\cite{ATLAS-CONF-2013-082}.

Uncertainties in the efficiency for tagging $b$-jets and in the rejection factor 
for light jets are determined from $\ttbar$ and dijet control samples
~\cite{ATLAS-CONF-2012-043,ATLAS-CONF-2014-004,ATLAS-CONF-2012-040}. Additional
uncertainties account for differences in $b$-tagging efficiency between 
simulated samples generated with \progname{sherpa} and \progname{pythia}
and for differences observed between standard $b$-tagging and truth
tagging (defined at the end of \secref{sec:objects}) for close-by jets~\cite{HIGG-2013-23}.

The efficiencies for the lepton triggers in events with reconstructed
leptons are nearly 100\%, and hence the related 
uncertainties are negligible.  
For the selection used in the \vvqq\ search, the efficiency for the $\met$
trigger is also close to 100\% with negligible associated uncertainties.

The merged-jet channel of the $\llqq$ search relies on measuring single-jet
masses.  To estimate the uncertainty in this measurement, jets reconstructed as described in \secref{sec:objects}
 are compared with jets constructed using the
same clustering algorithm but using as input charged-particle tracks rather
than calorimeter energy deposits.  The uncertainty is found 
using a procedure similar to that described in Ref.~\cite{PERF-2012-02} by
studying
the double ratio of masses of jets found by both the calorimeter- and
track-based algorithms:
 $R^m_{\mathrm{trackcalo}} = r^{m,\mathrm{data}}_{\mathrm{trackcalo}} / r^{m,\mathrm{MC}}_{\mathrm{trackcalo}}$,
where $r^{m,X}_{\mathrm{trackcalo}} = m^X_{\mathrm{calo}} / m^X_{\mathrm{track}}$,
$X = $ data or MC simulation, and $m$ is the jet mass.  The uncertainty is taken
as the deviation of this quantity from unity.
Studies performed on dijet samples yield a constant value of
$10\%$ for this uncertainty.
Applying the jet mass calibration derived from single jets in
generic multijet samples to merged jets originating from boosted $Z$~bosons
results in a residual topology-dependent miscalibration.
This effect can be bounded by an additional uncertainty of $10\%$. Adding these two
effects in quadrature gives a total uncertainty on the jet mass scale
of $14\%$.
The uncertainty on the jet mass resolution has a negligible effect
on the final result.

\subsection{Signal acceptance uncertainty}
\label{sec:systs_signal}
The uncertainty in the experimental acceptance for the Higgs~boson
signal due to the modelling of Higgs~boson production is estimated
by varying parameters in the generator and re-applying the signal
selection at generator level.  The renormalization and factorization
scales are varied up and down both independently and coherently
by a factor of two; the amounts of initial- and final-state radiation
(ISR/FSR)
are increased and decreased separately; and the PDF set used is changed
from the nominal CT10 to either MSTW2008 or NNPDF23.

\subsection{\ZZ\ background uncertainties}
\label{sec:systszzbg}
Uncertainties on the \ZZ\ background are treated as correlated
between the \llll\ and \llvv\ searches. 

Uncertainties in the PDF and in $\alphas$ are taken
from Ref.~\cite{LHCHiggsCrossSectionWorkingGroup:2012vm} and are derived
separately for the \qqZZ\ and \ggZZ\ backgrounds,  using
the envelope of the CT10, MSTW, and NNPDF error sets following the PDF4LHC
prescription given in Refs.~\cite{botje:2011sn,Alekhin:2011sk}, giving an uncertainty
parameterized in $\mZZ$.  These uncertainties amount to 3\% for
the \qqZZ\ process and 8\% for the \ggZZ\ process and are found to be 
anti-correlated between the two processes; this is taken into account in the fit.
The QCD scale uncertainty for the \qqZZ\ process is also taken
from Ref.~\cite{LHCHiggsCrossSectionWorkingGroup:2012vm} and is based on
varying the factorization and renormalization scales up and down by a factor of
two, giving an uncertainty parameterized in $m_{ZZ}$ amounting to 4\% on average. 

The deviation of the NLO electroweak $K$-factor from unity is varied up and down by 100\%
in events with high QCD activity or with an off-shell $Z$~boson, as described in
Ref.~\cite{Aad:2015xua}; this leads to an additional overall uncertainty of 1--3\%
for the \qqZZ\ process.

Full NLO and NNLO QCD calculations exist for the $gg\ra h^{*} \ra \ZZ$ 
process, but not for the \ggZZ\ continuum process.  However, Ref.~\cite{Bonvini:2013jha}
showed that higher-order corrections affect
$gg\ra WW$ and $gg\ra h^{*} \ra WW$ similarly, within a 30\% uncertainty
on the interference term.  This yields about a 60\% uncertainty
on the $gg\ra WW$ process.  Furthermore, Ref.~\cite{Bonvini:2013jha} states
that this conclusion also applies to the \ZZ\ final state, 
so the $gg$-induced part of the off-shell light Higgs boson $K$-factor
from Ref.~\cite{Passarino:2013bha} is applied to the \ggZZ\ background.
The uncertainty on this $K$-factor depends on \mZZ\ and is
about 30\%.  An additional uncertainty of 100\% is assigned to this
procedure; this covers the 60\% mentioned above.  This
uncertainty corresponds to the range considered
for the \ggZZ\ background $K$-factor in the ATLAS off-shell Higgs boson
signal-strength measurement described in Ref.~\cite{Aad:2015xua}.

Acceptance uncertainties for the ggF and VBF (and
\VH\ for \llll) channels due
to the uncertainty on the $\leq$1-jet and 2-jet cross-sections are estimated 
for the \qqZZ\ background by comparing the acceptance upon varying the factorization and renormalization
scales and changing the PDF set. For \llll
this leads to uncertainties of 4\%, 8\%,
and 3\% on the ggF, VBF, and \VH\ channels, respectively, where the
uncertainty is fully anti-correlated between the ggF channel and the VBF and
\VH\ channels. For the \ggZZ\
process where only LO generators are available, the VBF jets are simulated only
in the parton shower, and so the acceptance uncertainty is estimated by taking
the difference between the acceptances predicted by \progname{MCFM+Pythia8}
and \SHERPA, which have different parton shower simulations; this amounts to 90\% for the
\VH channel.

\section{Combination and statistical interpretation}
\label{sec:combination}
The statistical treatment of the data is similar to that described in
Refs.~\cite{HIGG-2012-17,ATL-PHYS-PUB-2011-011,Moneta:2010pm,HistFactory,ROOFIT}, and uses a
simultaneous profile-likelihood-ratio fit to the data from all of the searches. The parameter of interest
is the cross-section times branching ratio for heavy Higgs~boson production, assumed
to be correlated between all of the searches. It is assumed that an additional
Higgs~boson would be produced predominantly via the ggF and VBF processes
but that the ratio of the two production mechanisms is
unknown in the absence of a specific model. For this reason, fits
for the ggF and VBF production processes are done separately, and in each case the other
process is allowed to float in the fit as an additional nuisance
parameter. The \VH\ production mechanism is included in the fit for the $\llll$
search and is assumed to scale with the VBF signal since both the \VH\ and VBF production
mechanisms depend on the coupling of the Higgs~boson to vector bosons.

The simultaneous fit proceeds as follows.  For each channel of each search, there is a distribution
of the data with respect to some discriminating variable; these distributions are fitted to a sum of
signal and backgrounds.  The particular variables used are summarized in Table~\ref{tab:variables}.
The distributions for the $\llll$ search are unbinned, since the resolution of $\mllll$ is very
good, while other searches have binned distributions.  For the VBF channels of the $\llvv$ search,
only the overall event counts are used, rather than distributions, as the sample sizes are very
small.  The $\llqq$ and $\vvqq$ searches include additional distributions in control regions in
order to constrain the background, using either distributions of the mass variable or of the 
MV1c $b$-tagging category. 
The details of the specific variables used and the definitions of the
signal and control regions are discussed in \secsref{sec:4l} to~\ref{sec:vvqq}.

\begin{table}                                                                  
 \begin{center}                                                               
 \begin{tabular}{lllllll}
\toprule
Search    & \multicolumn{2}{c}{Channel}      & SR  & $Z$ CR & $W$ CR & Top CR\\
\midrule
\multirow{4}{*}{$\llll$}   
          & \multirow{2}{*}{ggF} &&  $m_{eeee}$, $m_{\mu\mu\mu\mu}$,  \\
          &  &&   $m_{ee\mu\mu}$, $m_{\mu\mu ee}$ \\
          & VBF &              &  $\mllll$ \\
          & \VH  &              &  $\mllll$ \\
\midrule
\multirow{2}{*}{$\llvv$}
          & $\bigstrut[b]$ ggF &    & $\mT^{ee}$, $\mT^{\mu\mu}$     \\
          & $\bigstrut[t]$ VBF &    & $N_{\mathrm{evt}}^{ee}$, $N_{\mathrm{evt}}^{\mu\mu}$     \\
\midrule
\multirow{4}{*}{$\llqq$}
          & \multirow{3}{*}{ggF} & untagged     & $\mlljj$  & MV1c cat. \\
          &  & tagged       & $\mlljj$  & MV1c cat. & & $\mlljj$ \\
          &  & merged-jet    & $\mllj$   & $\mllj$ & \\
          & VBF &              & $\mlljj$  & $\mlljj$ & \\
\midrule
\multirow{2}{*}{$\vvqq$}
          & \multirow{3}{*}{ggF} & \multirow{2}{*}{untagged}     & \multirow{2}{*}{$\mT$}     &  &          MV1c cat. (0 $b$-tags) \\
          &  &      &      &  & MV1c cat. (1 $b$-tag) \\
          &  & tagged       & $\mT$     &  \\
\bottomrule
    \end{tabular}                                                              
  \end{center}
  \caption{Summary of the distributions entering the likelihood fit for
    each channel of each search, both in the signal region (SR) and the various
    control regions (CR) used to constrain the background.  Each entry
    represents one distribution; some channels have several
    distributions for different lepton flavours.  MV1c cat.\ refers to the MV1c $b$-tagging event
  category. The distributions
    are unbinned for the $\llll$ search and binned elsewhere.  The VBF
    channels of the $\llvv$ search use only the overall event counts.
    See the text for the definitions of the specific variables used
    as well as for the definitions of the signal and control regions.
    \label{tab:variables}}
\end{table}

As discussed in \secref{sec:systs}, the signal acceptance uncertainties, and many of the
background theoretical and experimental uncertainties, are treated as fully correlated
between the searches. 
A given correlated uncertainty is modelled in the fit by using
a nuisance parameter common to all of the searches.
The mass hypothesis for the
heavy Higgs~boson strongly affects which sources of systematic uncertainty
have the greatest effect on the result.
At lower masses, the
\ZZ\ background theory uncertainties, the \Zjets modelling uncertainties, and the uncertainties on the jet energy
scale dominate. At higher masses, uncertainties in the \llvv\ non-$ZZ$
background, the jet mass scale, and the \Zjets\ background in the merged-jet regime
dominate.  The contribution to the uncertainty on the best-fit signal cross-section from the dominant
systematic uncertainties is shown in Table~\ref{tab:ranking}.

\begin{table}[ht]
\small
\centering
\begin{tabular}{%
 l
 S[table-format=2.1, table-number-alignment=right,
   round-mode=figures, round-precision=2] 
 |
 l
 S[table-format=2.1, table-number-alignment=right,
   round-mode=figures, round-precision=2]
}
\toprule
\multicolumn{2}{c|}{ggF mode} &  \multicolumn{2}{c}{VBF mode} \\
\multicolumn{2}{l|}{Systematic source    \hfill Effect [\%]}   & \multicolumn{2}{l}{Systematic source \hfill Effect [\%]}\\
\midrule
\multicolumn{4}{c}{$\bigstrut$  $\mH=200\gev$}\\
\midrule
$gg\rightarrow ZZ$ $K$-factor uncertainty & 26.5 & $gg\rightarrow ZZ$ acceptance & 13.4 \\ 
$Z+$hf $\Delta\phi$ reweighting & 5.3 & Jet vertex fraction ($\llqq$/$\vvqq$) & 13.4 \\ 
Luminosity & 5.2 & $gg\rightarrow ZZ$ $K$-factor uncertainty & 12.9 \\ 
Jet energy resolution ($\llqq$/$\vvqq$) & 3.9 & $\Zjets$ $\Delta\phi$ reweighting & 7.9 \\ 
QCD scale $gg\rightarrow ZZ$ & 3.7 & Jet energy scale $\eta$ modelling ($\llqq$/$\vvqq$) & 5.3 \\ 
\midrule
\multicolumn{4}{c}{$\bigstrut$  $\mH=400\gev$}\\
\midrule
$qq\rightarrow ZZ$ PDF & 20.8 & $\Zjets$ estimate ($\llvv$) & 33.8 \\ 
QCD scale $qq\rightarrow ZZ$ & 13.2 & Jet energy resolution ($\llll$/$\llvv$) & 6.5 \\ 
$\Zjets$ estimate ($\llvv$) & 12.6 & VBF $\Zjets$ $\mlljj$ & 5.5 \\ 
Signal acceptance ISR/FSR ($\llll$/$\llvv$) & 7.8 & Jet flavour composition ($\llll$/$\llvv$) & 5.3 \\ 
$Z+b\bar{b}$, $Z+c\bar{c}$, $\ptll$ & 5.6 & Jet vertex fraction ($\llqq$/$\vvqq$) & 4.8 \\ 
\midrule
\multicolumn{4}{c}{$\bigstrut$  $\mH=900\gev$}\\
\midrule
Jet mass scale ($\llqq$) & 7 & $\Zjets$ estimate ($\llvv$) & 19.2 \\ 
$Z+jj$ $\pt^Z$ shape ($\vvqq$) & 5.6 & Jet mass scale ($\llqq$) & 8.7 \\ 
$qq\rightarrow ZZ$ PDF & 4.3 & $Z+jj$ $\ptll$ shape & 7.3 \\
QCD scale $qq\rightarrow ZZ$ & 3.5 & Jet energy resolution ($\llll$/$\llvv$) & 4.4 \\ 
Luminosity & 2.6 & Jet flavour composition ($VV$/Signal) & 2.6 \\ 
\bottomrule
\end{tabular}
\caption{\label{tab:ranking}
The effect of the leading systematic uncertainties on the best-fit
signal cross-section uncertainty, expressed as a percentage of the total
(systematic and statistical) uncertainty,
for the ggF (left) and VBF (right) modes at $\mH=200$, 400, and $900\gev$.
The uncertainties are listed in decreasing order of their effect on the total uncertainty;
additional uncertainties with smaller effects are not shown.
}
\end{table}

As no significant excess is observed, exclusion limits are
calculated with a modified frequentist method~\cite{cls}, also known as $CL_s$,
using the $\tilde{q}_{\mu}$ test statistic in the asymptotic
approximation~\cite{Cowan:2010js,Cowan-erratum}.  The observed limits can be compared with
expectations by generating `Asimov' data sets, which are
representative event samples that provide both the median expectation for an
experimental result and its expected statistical variation in the asymptotic
approximation, as described in Refs.~\cite{Cowan:2010js,Cowan-erratum}.
When producing the Asimov data set for the expected limits, the background-only
hypothesis is assumed and the cross-sections for both ggF and VBF production of the heavy Higgs~boson
are set to zero. The remaining nuisance parameters are set to the value that
maximizes the likelihood function for the observed data (profiled). When using the
asymptotic procedure to calculate limits it is necessary to generate an Asimov data set both for the
background-only hypothesis and for the signal hypothesis. When setting the
observed limits, the cross-section for the other production mode not under
consideration is profiled to data before generating the background-only 
Asimov data set.

\section{Results}
\label{sec:results}

Limits on the cross-section times branching ratio
from the combination of all of the searches are shown in \figref{fig:CLAll}.
Also shown are expected limits from the \llll, \llvv\ and the combined
\llqq+\vvqq searches (the latter two searches are only shown in combination as they share control
regions). At low mass the \llll\ search has the best sensitivity while at high
mass the sensitivity of the combined \llqq+\vvqq search is greatest, with the
sensitivity of the \llvv\ channel only slightly inferior.
In the mass range considered for this search 
the 95\% confidence level (CL) upper limits on the cross-section times branching ratio for heavy Higgs~boson
production vary 
between 0.53~pb at $\mH=195$~\gev\ and 0.008~pb at $\mH=950$~\gev\ in the ggF channel and
between 0.31~pb at $\mH=195$~\gev\ and 0.009~pb at $\mH=950$~\gev\ in the VBF
channel.
The excursions into the $2\sigma$ band around the expected limit originate from 
local deviations in the input distributions. For example, the excess occurring
around $200\gev$ and the deficit occurring around $300\gev$ arise from the $\llll$ (see
\figref{fig:llll_fit_m4l}) search. Deficits at higher 
mass are driven by fluctuations in the $\llqq$ search (see
\figsref{fig:mlljj_ggf} and \ref{fig:mlljj_vbf}).

\Figref{fig:2hdm_200} shows exclusion limits in the
$\cos(\beta - \alpha)$ versus $\tan \beta$ plane for Type-I and Type-II 2HDMs,
for a heavy Higgs boson with mass $\mH=200$~\gev.  This \mH\ value is chosen so
the assumption of a narrow-width Higgs~boson is valid over most of the parameter
space, and  the experimental sensitivity is at a maximum.
As explained in \secref{sec:signals}, the 
range of $\cos(\beta - \alpha)$ and $\tan \beta$ explored
is limited to the region where the assumption of a heavy narrow-width Higgs~boson
with negligible interference is valid.
When calculating the limits at a given choice of $\cos(\beta-\alpha)$ and
$\tan{\beta}$, the relative rate of ggF and VBF 
production in the fit is set according to the prediction of the 2HDM for that parameter
choice. \Figref{fig:2hdm_mH} shows exclusion limits as a function of the heavy
Higgs boson mass $\mH$ and the parameter $\tan{\beta}$ for $\cos(\beta-\alpha)=-0.1$.
The white regions in the exclusion plots indicate regions of parameter space
not excluded by the present analysis; in these regions the cross-section predicted by the
2HDM is below the experimental sensitivity.
Compared with recent studies of indirect limits~\cite{Aad:2015pla},
the exclusion presented here is considerably more stringent for
Type-I with $\cos(\beta-\alpha)<0$ and $1<\tan\beta<2$, and for 
Type-II with $0.5<\tan\beta<2$. 

\begin{figure}[htbp]
\centering
\subfloat[ggF]{\includegraphics[scale=0.5]{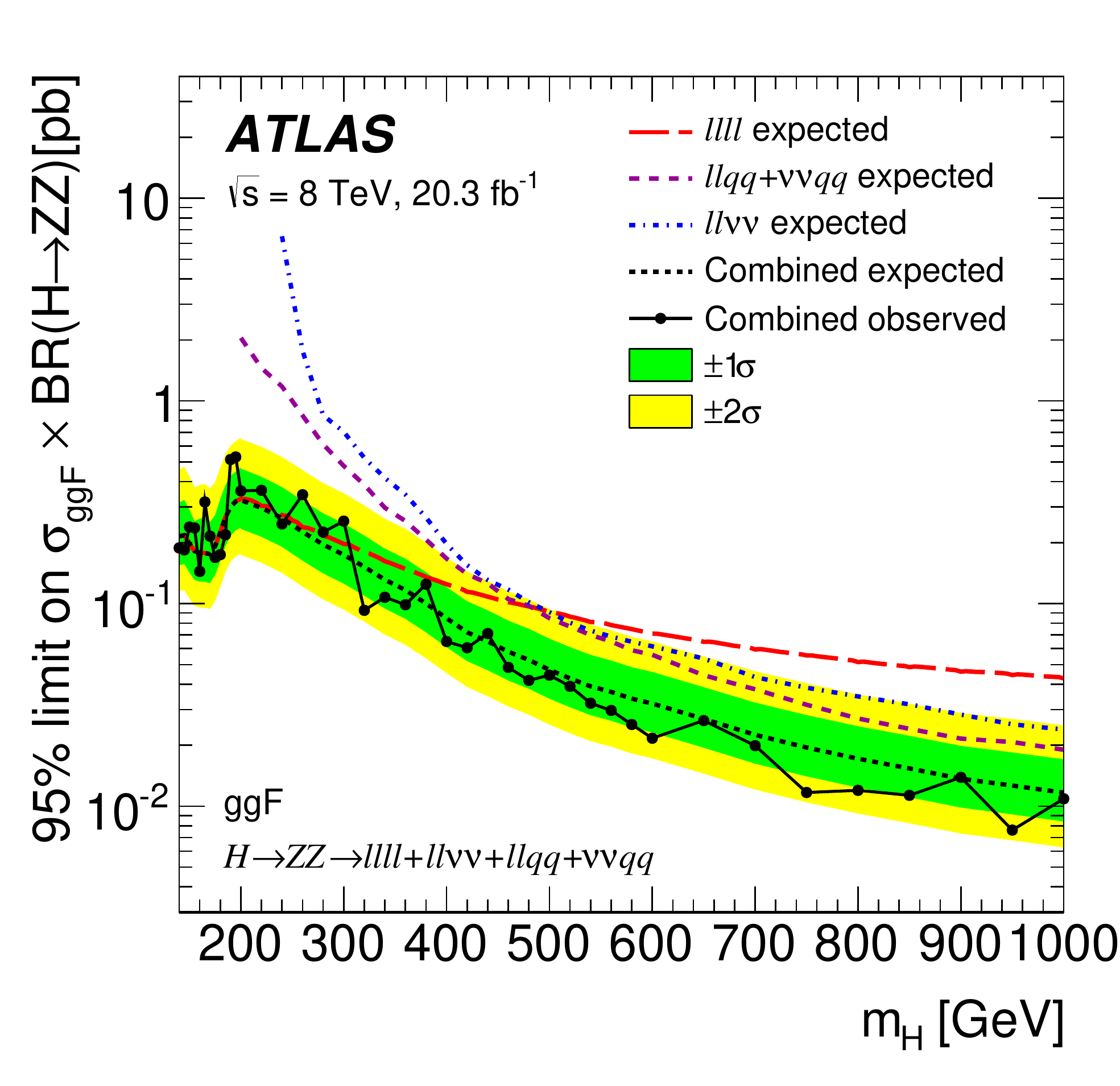}} \\
\subfloat[VBF]{\includegraphics[scale=0.5]{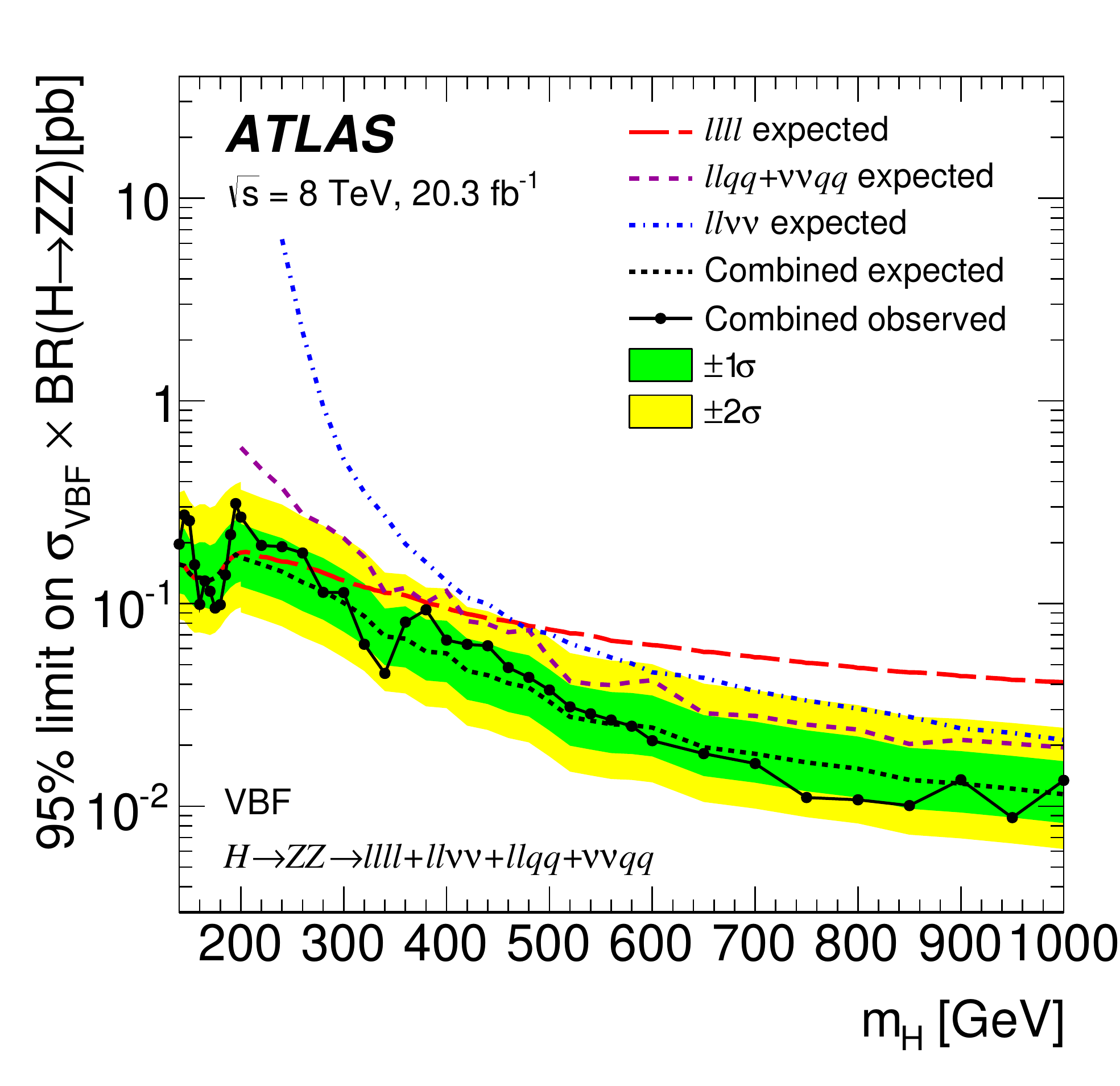}}
\caption{
95\% CL upper limits on $\sigma \times {\mathrm{BR}}(H \rightarrow ZZ)$ as a
function of \mH,
resulting from the combination of all of the searches
in the (a)~ggF and (b)~VBF channels.
The solid black line and points indicate the observed limit.
The dashed black line indicates the expected limit and the
bands the 1-$\sigma$ and 2-$\sigma$ uncertainty ranges about the expected limit.
The dashed coloured lines indicate the expected limits obtained from the
individual searches; for the \llqq\ and \vvqq\ searches, only the combination of
the two is shown as they share control regions.
\label{fig:CLAll}
}
\end{figure}

\begin{figure}[htbp]
\centering
\subfloat[Type-I]{\includegraphics[scale=0.35]{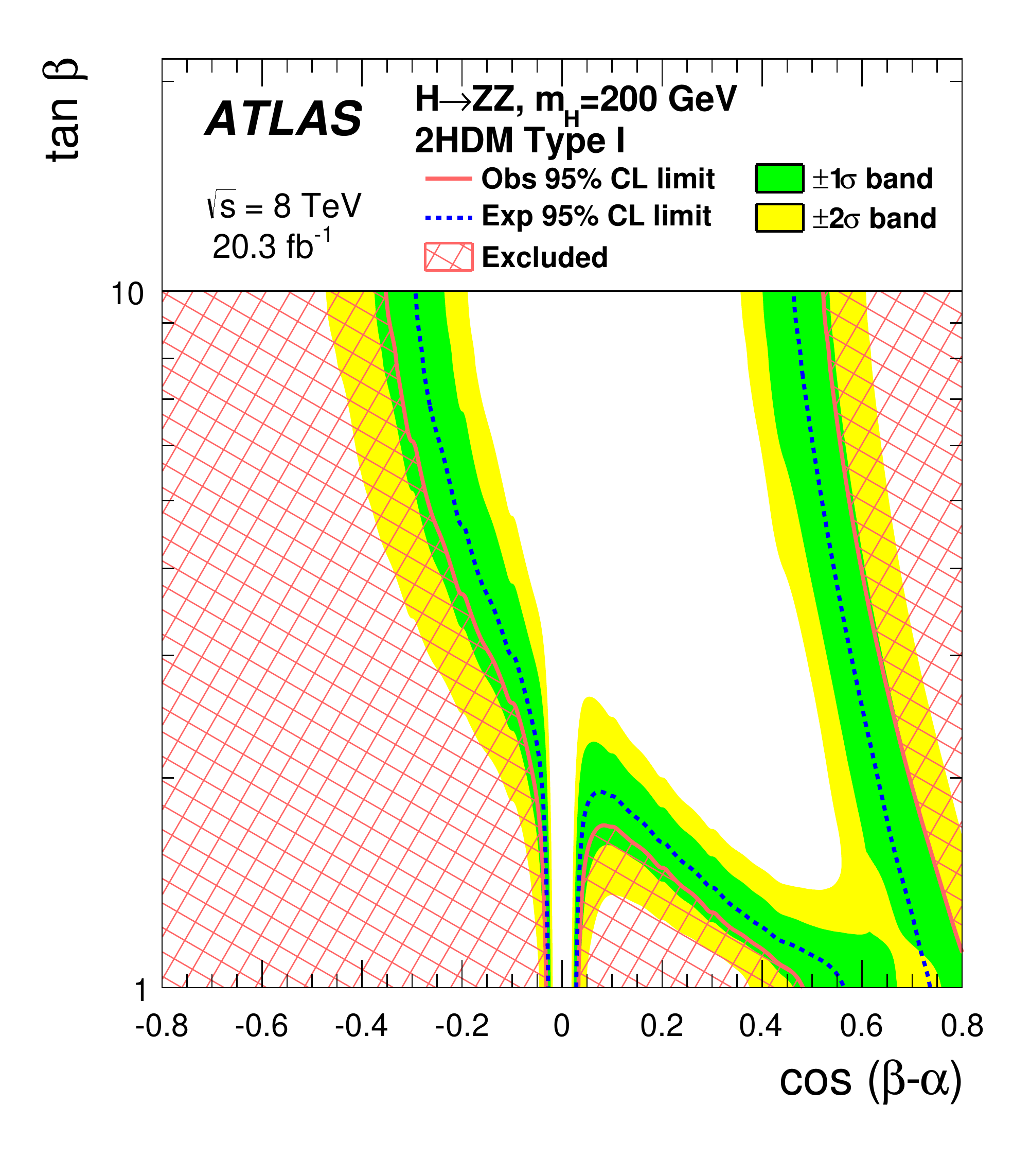}}
\subfloat[Type-II]{\includegraphics[scale=0.35]{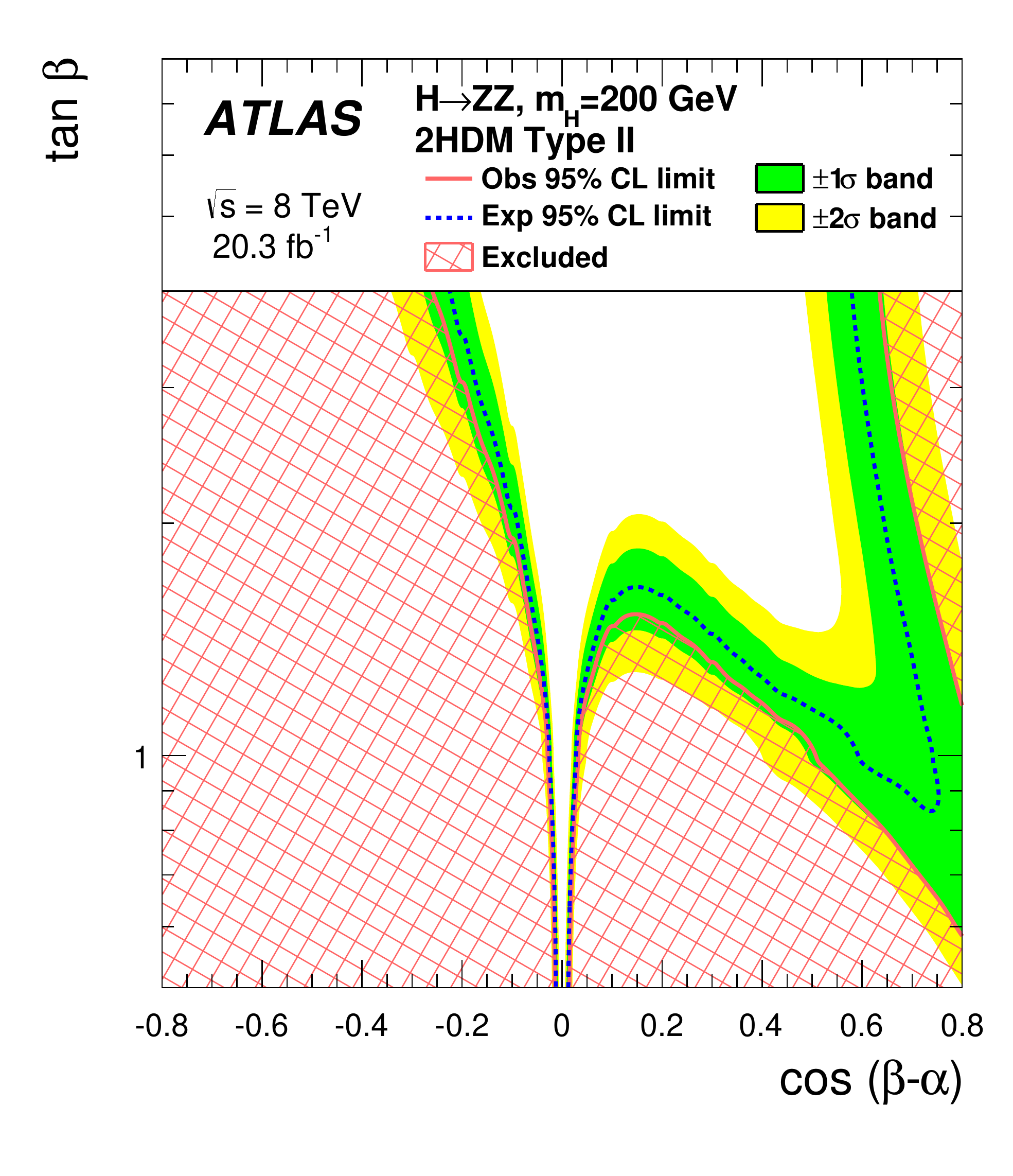}}
\caption{
95\% CL exclusion contours in the 2HDM (a)~Type-I and (b)~Type-II
models for $\mH=200\gev$, shown as a
function of the parameters $\cos(\beta-\alpha)$ and $\tan{\beta}$. The red hashed
area shows the observed exclusion, with the solid red line denoting the edge of the
excluded region. The dashed blue line represents the expected exclusion contour and the
shaded bands the 1-$\sigma$ and 2-$\sigma$ uncertainties on the
expectation. The vertical axis range is set such that regions where the light
Higgs couplings are enhanced by more than a factor of three from their SM values
are avoided.
\label{fig:2hdm_200}
}
\end{figure}

\begin{figure}[htbp]
\centering
\subfloat[Type-I]{\includegraphics[scale=0.35]{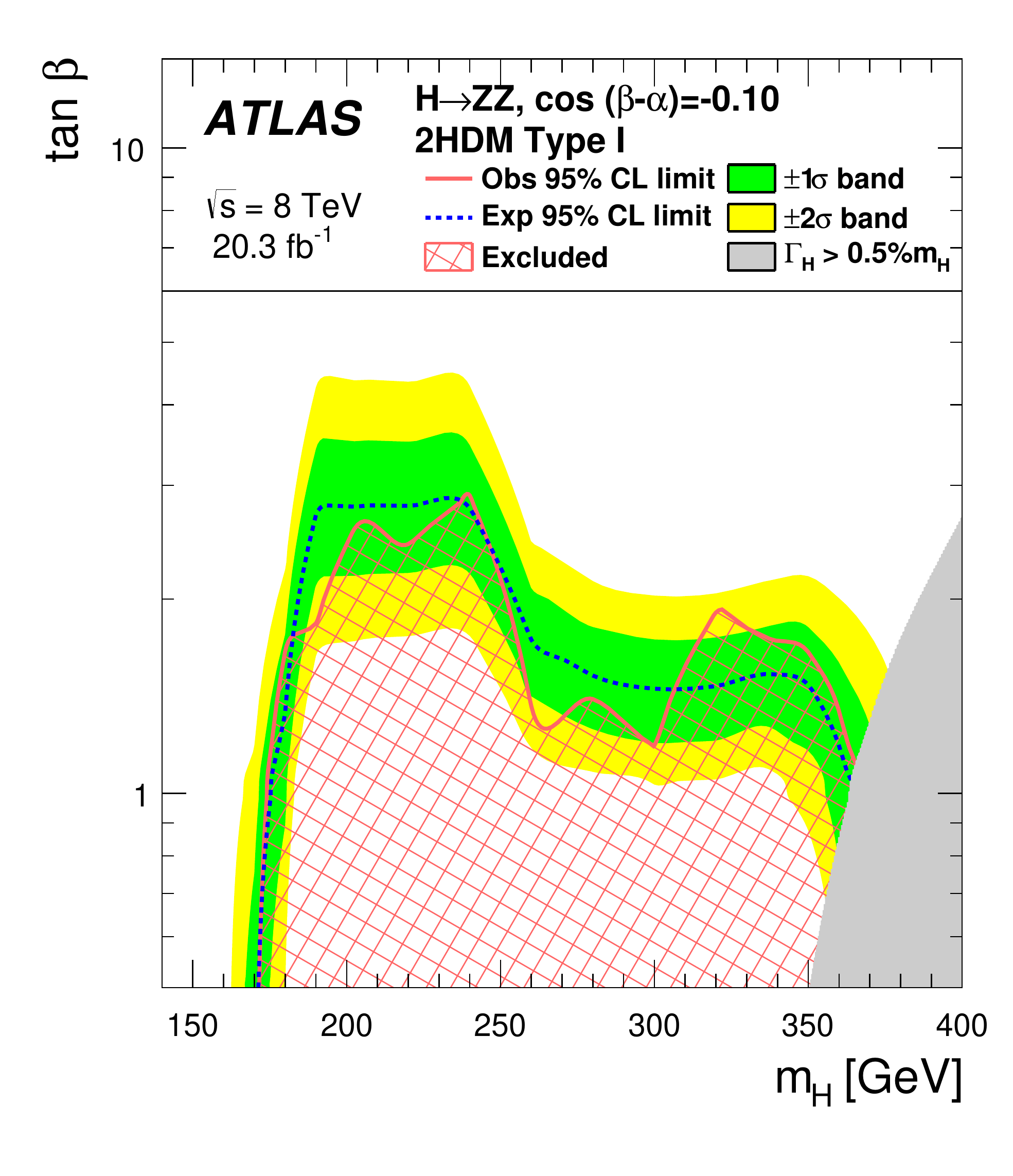}}
\subfloat[Type-II]{\includegraphics[scale=0.35]{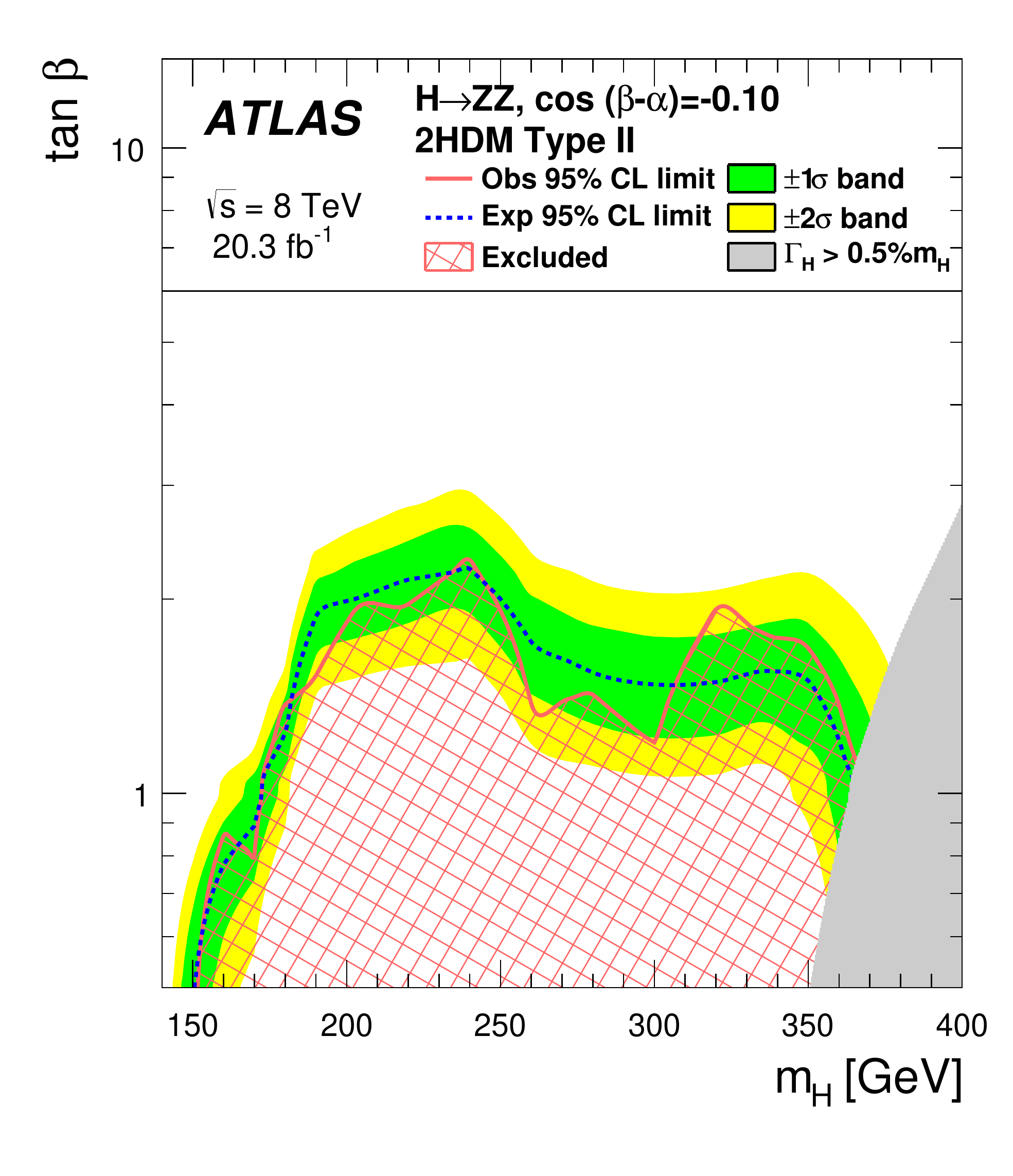}}
\caption{
95\% CL exclusion contours in the 2HDM (a)~Type-I and (b)~Type-II
models for $\cos(\beta-\alpha)=-0.1$, shown as a
function of the heavy Higgs boson mass $\mH$ and the parameter $\tan{\beta}$.
The shaded area shows the observed exclusion, with the black line denoting
the edge of the
excluded region. The blue line represents the expected exclusion contour and the
shaded bands the 1-$\sigma$ and 2-$\sigma$ uncertainties on the
expectation. The grey area masks regions where the
width of the boson is greater than $0.5\%$ of \mH. For the choice of $\cos(\beta-\alpha)=-0.1$ the light Higgs couplings are not
altered from their SM values by more than a factor of two. \label{fig:2hdm_mH}
}
\end{figure}

The previously published ATLAS results using data collected at
$\sqrt{s}=7\tev$~\cite{HIGG-2012-01,HIGG-2012-15,HIGG-2012-14}
assumed a SM Higgs~boson with the relative rate of ggF and VBF production
fixed to the SM prediction.  Thus, they are not directly comparable
with the current results, which assume that the heavy Higgs~boson has
a narrow width but also allow the rates of ggF and VBF production
to vary independently.  These results are also not directly comparable
with the recent results published by the CMS Collaboration~\cite{cms-hzz}
for similar reasons.

\FloatBarrier

\section{Summary}
\label{sec:summary}
  A search is presented for a high-mass Higgs boson in the \htollll, \htollvv, \htollqq, and \htovvqq\  
  decay modes using the ATLAS detector at the CERN Large Hadron Collider. 
  The search uses proton--proton collision data at a
  centre-of-mass energy of $8\tev$ corresponding to an integrated luminosity of 20.3 fb$^{-1}$.
  The results of the search are interpreted in the scenario of a heavy Higgs boson with a width that 
  is small compared with the experimental mass resolution. The Higgs boson mass range considered 
  extends up to $1\tev$ for all four decay modes and down to as low as $140\gev$, depending on 
  the decay mode. No significant excess of events over the Standard Model prediction is found. 
  Limits on production and decay of a heavy Higgs~boson to two $Z$~bosons are set separately for
  gluon-fusion and vector-boson-fusion production modes. For the combination of all decay modes, 95\% CL upper limits
  range from 0.53~pb at $\mH=195$~\gev\ to 0.008~pb at $\mH=950$~\gev\ for the
  gluon-fusion production mode and
  from 0.31~pb at $\mH=195$~\gev\ to 0.009~pb at $\mH=950$~\gev\ 
  for the vector-boson-fusion production mode.
  The results are also interpreted in the context of Type-I and Type-II
  two-Higgs-doublet models, with exclusion contours given in the 
  $\cos(\beta-\alpha)$ versus ~$\tan\beta$ and $\mH$ versus ~$\tan\beta$
  planes for $\mH=200\gev$.  This \mH\ value is chosen so that the assumption of a narrow-width 
  Higgs~boson is valid over most of the parameter space, and so that the experimental 
  sensitivity is at a maximum.
  Compared with recent studies of indirect limits, %
  the two-Higgs-doublet model exclusion presented here is considerably more stringent for
  Type-I with $\cos(\beta-\alpha)<0$ and $1<\tan\beta<2$, and for 
  Type-II with $0.5<\tan\beta<2$.

\section*{Acknowledgements}


We thank CERN for the very successful operation of the LHC, as well as the
support staff from our institutions without whom ATLAS could not be
operated efficiently.

We acknowledge the support of ANPCyT, Argentina; YerPhI, Armenia; ARC,
Australia; BMWFW and FWF, Austria; ANAS, Azerbaijan; SSTC, Belarus; CNPq and FAPESP,
Brazil; NSERC, NRC and CFI, Canada; CERN; CONICYT, Chile; CAS, MOST and NSFC,
China; COLCIENCIAS, Colombia; MSMT CR, MPO CR and VSC CR, Czech Republic;
DNRF, DNSRC and Lundbeck Foundation, Denmark; EPLANET, ERC and NSRF, European Union;
IN2P3-CNRS, CEA-DSM/IRFU, France; GNSF, Georgia; BMBF, DFG, HGF, MPG and AvH
Foundation, Germany; GSRT and NSRF, Greece; RGC, Hong Kong SAR, China; ISF, MINERVA, GIF, I-CORE and Benoziyo Center, Israel; INFN, Italy; MEXT and JSPS, Japan; CNRST, Morocco; FOM and NWO, Netherlands; BRF and RCN, Norway; MNiSW and NCN, Poland; GRICES and FCT, Portugal; MNE/IFA, Romania; MES of Russia and NRC KI, Russian Federation; JINR; MSTD,
Serbia; MSSR, Slovakia; ARRS and MIZ\v{S}, Slovenia; DST/NRF, South Africa;
MINECO, Spain; SRC and Wallenberg Foundation, Sweden; SER, SNSF and Cantons of
Bern and Geneva, Switzerland; NSC, Taiwan; TAEK, Turkey; STFC, the Royal
Society and Leverhulme Trust, United Kingdom; DOE and NSF, United States of
America.

The crucial computing support from all WLCG partners is acknowledged
gratefully, in particular from CERN and the ATLAS Tier-1 facilities at
TRIUMF (Canada), NDGF (Denmark, Norway, Sweden), CC-IN2P3 (France),
KIT/GridKA (Germany), INFN-CNAF (Italy), NL-T1 (Netherlands), PIC (Spain),
ASGC (Taiwan), RAL (UK) and BNL (USA) and in the Tier-2 facilities
worldwide.

\appendix

\section{Flavour tagging in the  $\llqq$ and $\vvqq$ searches}
\label{app:mv1c}
In order to constrain the normalizations of the various flavour components of the $Z$+jets ($Z+jj$,
$Z+cj$, $Z+bj$, and $Z+$hf) and $\Wjets$ ($W+jj$ and $W+cj$) backgrounds in the $\llqq$ and
$\vvqq$ searches, it is necessary to distinguish the different combinations of jet flavour.  This is
achieved by combining the information from the MV1c $b$-tagging discriminant of the two signal jets
in order to disentangle the different light- and heavy-flavour components. 

Besides the MV1c selection criterion described in \secref{sec:objects},
which had an average efficiency of 70\% for jets with $\pt>20\gev$
containing $b$-hadrons ($b$-jets), additional criteria, or operating points,
are defined with average efficiencies of 80\%, 60\%, and 50\%.
The efficiencies for accepting $c$-jets or light-quark jets
for the 50\% (80\%) operating point
are 1/29 (1/3) and 1/1400 (1/30), respectively.
Based on these operating points, five bins
in MV1c are defined:\\
\centerline{%
\begin{tabular}{ll}
  Bin & $b$-tagging efficiency\\
\midrule
Very loose (VL) & $>80\%$\\
Loose (L)       & $80-70\%$\\
Medium (M)      & $70-60\%$\\
Tight (T)       & $60-50\%$\\
Very tight (VT) & $<50\%$\\
\end{tabular}
}

In this analysis, jets selected by the M, T, or VT operating points
(i.e.\ $>70\%$ efficiency for $b$-jets) are considered as $b$-tagged.
Events are then categorized based on the combination of the binned MV1c operating points
for the two signal jets, as shown in \figref{fig:mv1cclassify}, in order to obtain optimal
separation of the flavour components.

\begin{figure}[tbh]
  \centering
\includegraphics[width=0.6\textwidth]{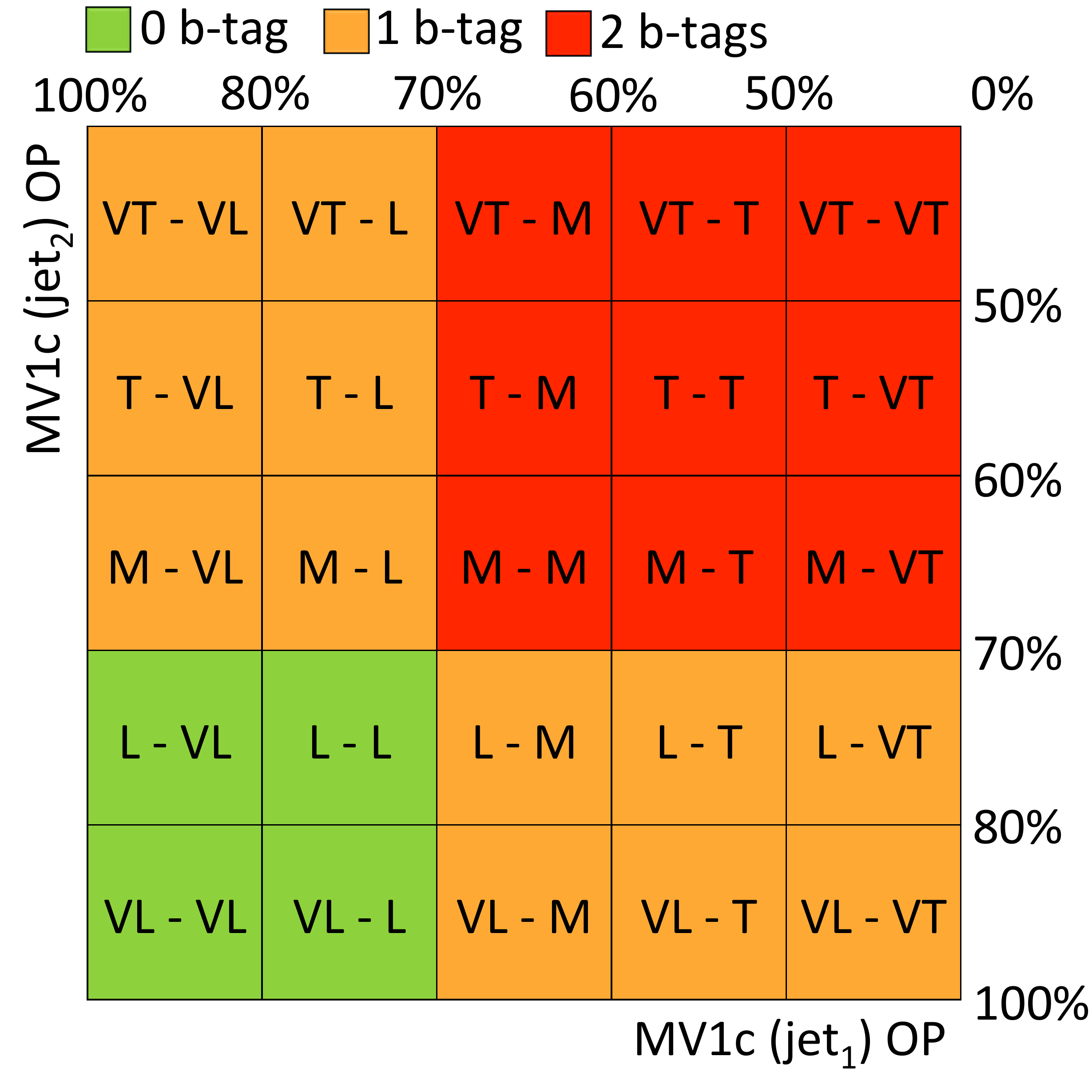}
  \caption{\label{fig:mv1cclassify} Event categorization as a function of the output of the MV1c
    $b$-tagging algorithm for the two signal jets. The bin boundaries correspond
    to the operating points
    (MV1c(jet) OP) giving $b$-tagging efficiencies of 100\%, 80\%, 70\%, and 50\%;
    i.e., the
    $b$-jet purity increases from left (bottom) to right (top). The event categories are labelled
    VL, L, M, T, and VT according to the definitions in the text, and the different colours
  correspond to events with 0, 1, and 2 identified $b$-jets.}
\end{figure}

Distributions of the resulting MV1c event categories are shown
in \figsref{fig:mv1c} and~\ref{fig:vvqq_fit_cr}
for the $\llqq$ $\Zjets$ and $\vvqq$ $\Wjets$ control regions, respectively.  These distributions
are provided as input to the simultaneous profile-likelihood-ratio fit described
in \secref{sec:combination} in order to determine the normalization of the background flavour
components defined above.  Following the fit, the data are well-described by the MC simulation.

\begin{figure}[tbh]
  \centering
\subfloat[Untagged]{\includegraphics[width=0.45\textwidth]{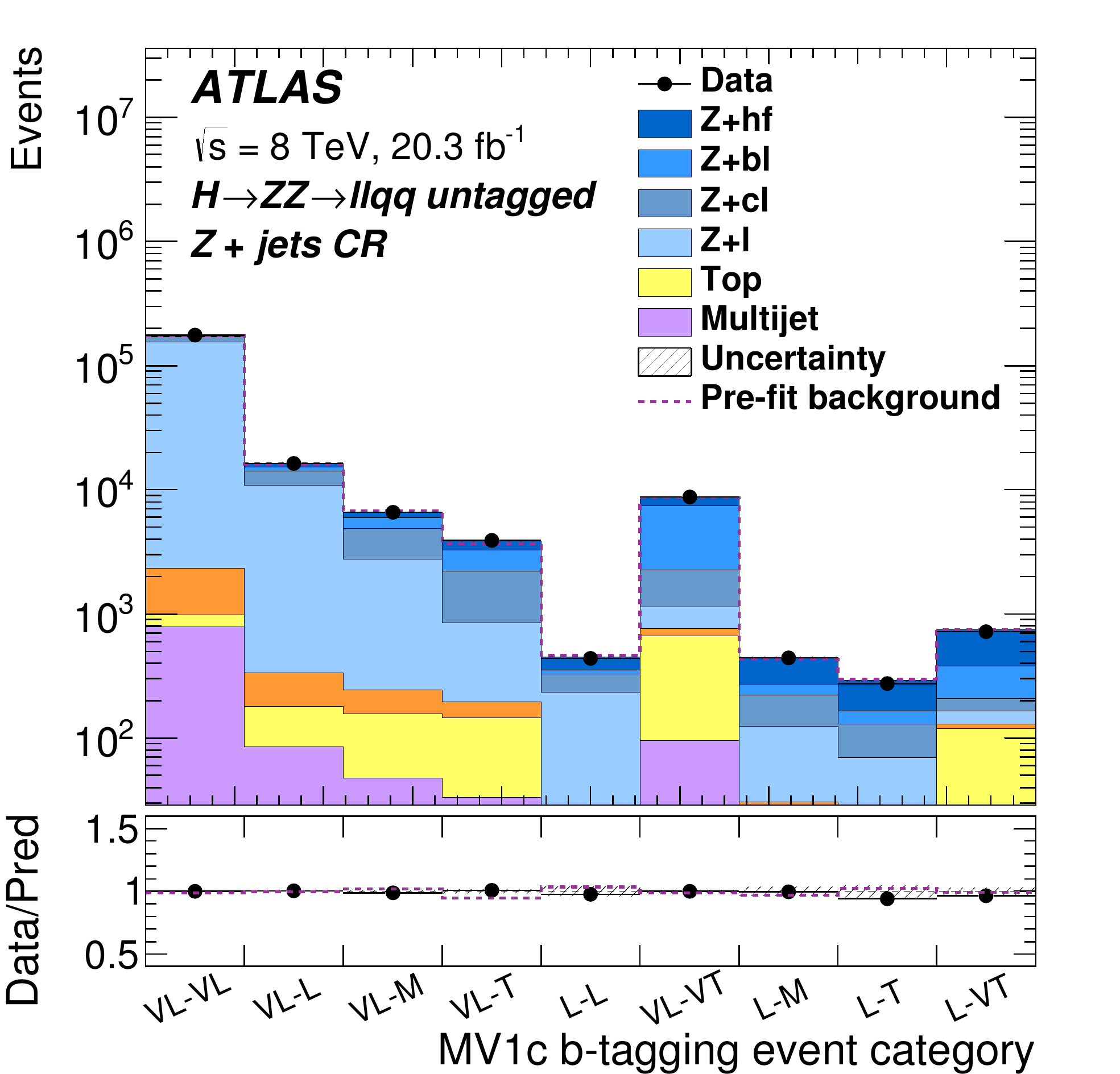}}
\subfloat[Tagged]{\includegraphics[width=0.45\textwidth]{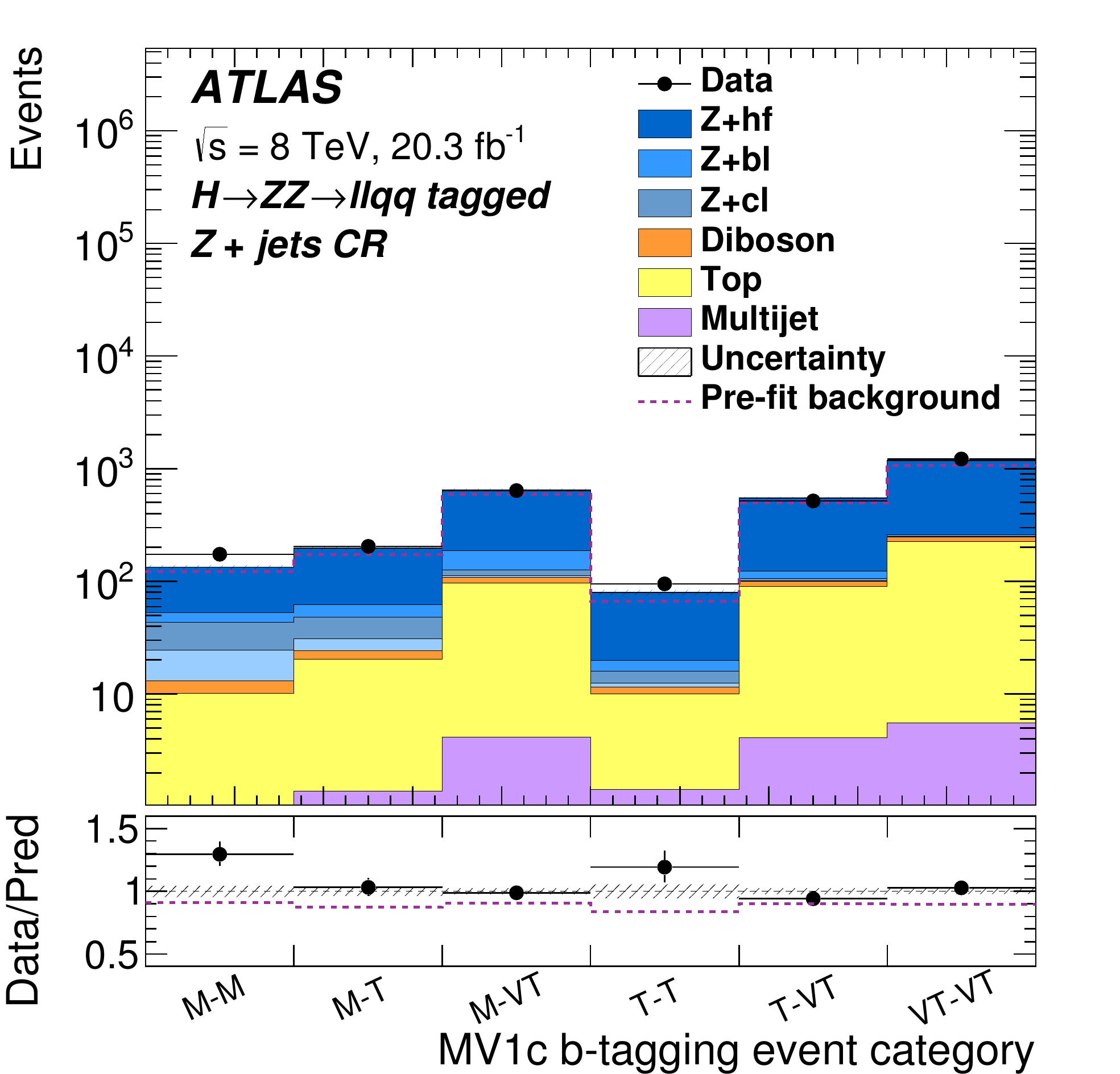}}\\
     \caption{\label{fig:mv1c} The distribution of the MV1c $b$-tagging event categories, based on the two
       signal jets, in the $\Zjets$ control region in the (a)~untagged ggF and (b)~tagged ggF
       channels of the $\htollqq$ search.  The $b$-jet purity generally increases from left to
       right.  The dashed line shows the total background used as input to the fit.  The
       contribution labelled as `Top' includes both the $\ttbar$ and single-top processes.  The
       bottom panes show the ratio of the observed data to the predicted background.  }
\end{figure}

\begin{figure}[htbp]
\centering
\subfloat[0-$b$-tag control region]{\includegraphics[width=0.45\textwidth]{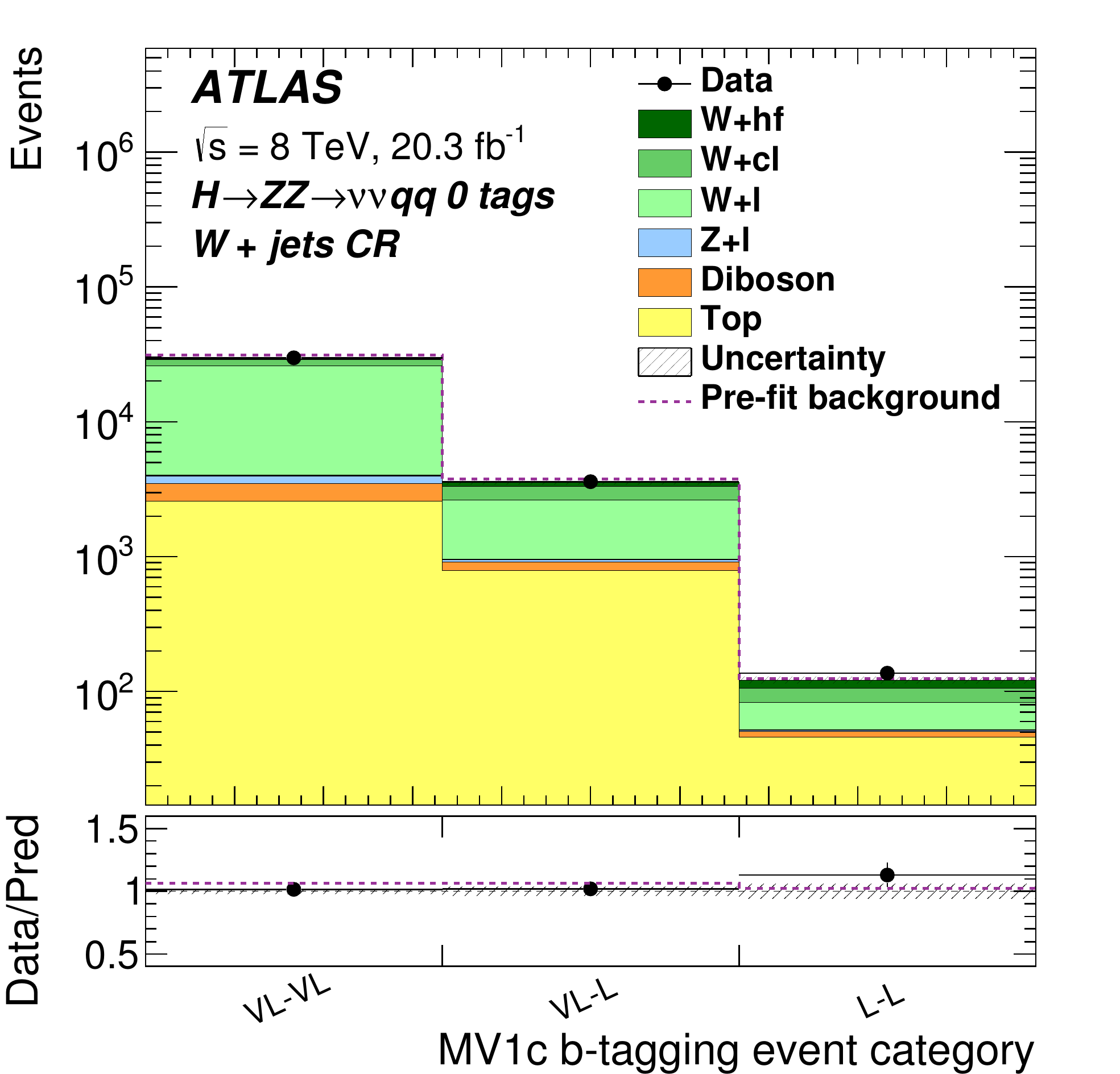}}
\subfloat[1-$b$-tag control region]{\includegraphics[width=0.45\textwidth]{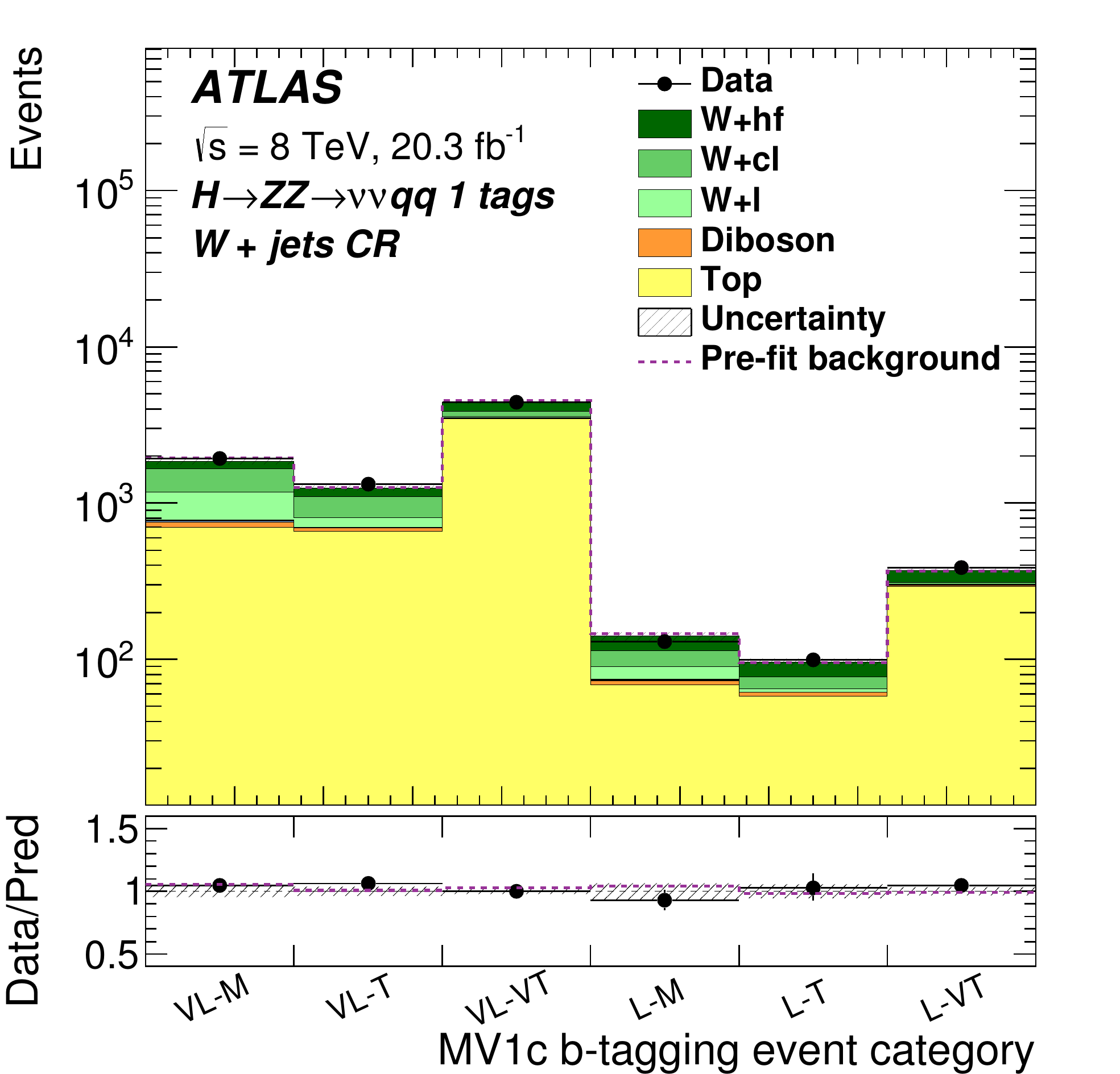}}
\caption{\label{fig:vvqq_fit_cr}
The distribution of the MV1c $b$-tagging event categories, based on the two signal jets, in the $\Wjets$
(a)~0-$b$-tag and (b)~1-$b$-tag control regions of the $\htovvqq$ search.  The $b$-jet purity
generally increases from left to right.  The dashed line shows the total background used as input to
the fit.  The contribution labelled as `Top' includes both the $\ttbar$ and single-top processes.
The bottom panes show the ratio of the observed data to the predicted background.  }
\end{figure}

\section{Corrections to MC simulation for the $\llqq$ search}
\label{app:llqqcorr}
In order to improve the description of the data in the resolved ggF channel, corrections are applied
to the \SHERPA $\Zjets$ simulation (prior to the likelihood fit) as a function of the azimuthal
angle between the two signal jets, $\dphijj$, and the transverse momentum of the leptonic $Z$ boson,
$\ptll$, following Ref.~\cite{HIGG-2013-23}.  The simulation does not model well the observed
$\dphijj$ distribution in the untagged control regions for $\ptll < 120\gev$;
this is not seen at higher $\ptll$ or in the tagged control region.  In order to improve the
modelling, the $Z+jj$ component of the background with $\ptll<120\gev$ is scaled by a linear
function derived from the control region with no $b$-tagged jets at low $\ptll$
with non-$Z$~boson backgrounds subtracted.
Half the value of
the correction is taken as a systematic uncertainty where it is applied.  In the $Z+$hf sample with
$\ptll<120\gev$, the full value of the correction is taken as an uncertainty.  For $\ptll>120\gev$,
no correction is applied for any sample.  In this region, a linear fit is performed to the data/MC
ratio of $\dphijj$ in the untagged subchannel after subtracting the small non-$Z$ background, and 
the uncertainty on the fitted slope taken as an
uncertainty for all $Z+\jets$ samples.  Following this correction, the description of the $\ptll$
distribution in the control region with no $b$-tagged jets also improves, but there is still some
residual discrepancy seen in the control regions that have $b$-tagged jets.  Thus, the $Z+$hf
background component is scaled by a function logarithmic in $\ptll$, determined from the combination
of the control regions with one or more $b$-tagged jets (after subtracting the $Z+jj$ and
non-$\Zjets$ background components).  An uncertainty of half this correction is applied for all
$\Zjets$ channels.  (All these uncertainties are taken to be uncorrelated between the $\Zlight$ and
$Z+$hf samples.)  Following these corrections, the simulation models both the $\dphijj$ and $\ptll$
distributions well in all $\Zjets$ control regions.

For the VBF channel, no significant differences are seen in the $\dphijj$ and $\ptll$ distributions,
but there is a small difference in the $\mlljj$ distribution in the control region.  The simulated
$\Zjets$ background is corrected for this bin-by-bin and the full value of this correction is taken as
an uncertainty, again uncorrelated between light- and heavy-flavour samples.  No corrections are
needed for the merged-jet ggF channel given the small sample size available.

It has been observed in an unfolded measurement of the $\pt$ distribution of
$\ttbar$ quark pairs 
that the simulation does not accurately describe the $\pttt$ distribution~\cite{TOPQ-2012-08}.
To correct for this, $\ttbar$ MC events are
weighed by a function of $\pttt$ taken from $7\tev$ data from Ref.~\cite{TOPQ-2012-08} in order to make the simulation match the data. 
The correction is validated for $8\tev$ data 
using the $e\mu$ top-quark control region, 
and the uncertainty in this correction is estimated by varying it from 50\% to 150\% of its nominal value.

\FloatBarrier
-------------------------------------------------------------------------------
\printbibliography

\clearpage

\clearpage
\begin{flushleft}
{\Large The ATLAS Collaboration}

\bigskip

G.~Aad$^{\rm 85}$,
B.~Abbott$^{\rm 113}$,
J.~Abdallah$^{\rm 151}$,
O.~Abdinov$^{\rm 11}$,
R.~Aben$^{\rm 107}$,
M.~Abolins$^{\rm 90}$,
O.S.~AbouZeid$^{\rm 158}$,
H.~Abramowicz$^{\rm 153}$,
H.~Abreu$^{\rm 152}$,
R.~Abreu$^{\rm 116}$
% ...
\bigskip
\\
% ...
$^{11}$ Institute of Physics, Azerbaijan Academy of Sciences, Baku, Azerbaijan\\
% ...
$^{85}$ CPPM, Aix-Marseille Universit{\'e} and CNRS/IN2P3, Marseille, France\\
% ...
$^{90}$ Department of Physics and Astronomy, Michigan State University, East Lansing MI, United States of America\\
% ...
$^{107}$ Nikhef National Institute for Subatomic Physics and University of Amsterdam, Amsterdam, Netherlands\\
% ...
$^{113}$ Homer L. Dodge Department of Physics and Astronomy, University of Oklahoma, Norman OK, United States of America\\
% ...
$^{116}$ Center for High Energy Physics, University of Oregon, Eugene OR, United States of America\\
% ...
$^{151}$ Institute of Physics, Academia Sinica, Taipei, Taiwan\\
$^{152}$ Department of Physics, Technion: Israel Institute of Technology, Haifa, Israel\\
$^{153}$ Raymond and Beverly Sackler School of Physics and Astronomy, Tel Aviv University, Tel Aviv, Israel\\
% ...
$^{158}$ Department of Physics, University of Toronto, Toronto ON, Canada\\
% ...
\end{flushleft}


\end{document}